\documentclass[
    aps,
    prx,
    superscriptaddress,
    twocolumn,
    a4paper,
    floatfix,
    longbibliography
]{revtex4-2}

\usepackage[american]{babel}
\usepackage[utf8x]{inputenc}

\usepackage{grffile} 

\usepackage{newtxtext,newtxmath}
\usepackage{microtype}

\usepackage{amsmath}
\usepackage{bbm}
\usepackage{bm}
\usepackage{mathtools}
\usepackage{dsfont}
\usepackage{braket}
\usepackage{cancel}
\usepackage{slashed}

\usepackage{physics}
\usepackage{multirow}

\usepackage{graphicx}
\usepackage[svgnames]{xcolor}

\usepackage{siunitx} 

\usepackage{ragged2e}
\usepackage{array}
\usepackage{tabularx}
\usepackage{booktabs}

\definecolor{cset-aps-blueberry}{RGB}{28,128,158}
\definecolor{cset-aps-blue}{RGB}{46,44,184}
\definecolor{cset-aps-turquoise}{RGB}{0,67,88}
\definecolor{cset-aps-limegreen}{RGB}{190,219,67}
\definecolor{cset-aps-green}{RGB}{31,138,112}
\definecolor{cset-aps-yellow}{RGB}{255,225,25}
\definecolor{cset-aps-orange}{RGB}{253,116,0}
\definecolor{cset-aps-red}{RGB}{219,0,43}

\usepackage{tikz}
\usepackage{tikz-feynman}

\usepackage{pgfplots}
\pgfplotsset{%
    every axis legend/.append style={%
        cells={anchor=west},
        at={(0.96,0.04)},
        anchor=south east,
        font=\scriptsize,
        },
    every axis/.append style={%
        yticklabel style={%
            /pgf/number format/fixed zerofill,
            /pgf/number format/precision=2},
        },
    width= \textwidth,
    height=8cm,
    xmajorgrids=true,
    xminorgrids=false,
    minor x tick num=1,
}

\usetikzlibrary{decorations}
\usetikzlibrary{arrows.meta,arrows}

\usepackage{pict2e,picture}

\makeatletter
\DeclareRobustCommand{\Arrow}[1][]{%
\check@mathfonts
\if\relax\detokenize{#1}\relax
\settowidth{\dimen@}{$\m@th\rightarrow$}%
\else
\setlength{\dimen@}{#1}%
\fi
\sbox\z@{\usefont{U}{lasy}{m}{n}\symbol{41}}%
\begin{picture}(\dimen@,\ht\z@)
\roundcap
\put(\dimexpr\dimen@-.7\wd\z@,0){\usebox\z@}
\put(0,\fontdimen22\textfont2){\line(1,0){\dimen@}}
\end{picture}%
}
\makeatother

\usepackage{hyperref}
\hypersetup{%
    colorlinks=true,
    linkcolor={cset-aps-red},
    linkbordercolor={cset-aps-red},
    filecolor={cset-aps-orange},
    filebordercolor={cset-aps-orange},
    citecolor={cset-aps-blue},
    citebordercolor={cset-aps-blue},
    urlcolor={cset-aps-green},
    urlbordercolor={cset-aps-green},
    menucolor={cset-aps-limegreen},
    menubordercolor={cset-aps-limegreen},
    breaklinks=true,
    pdfborderstyle={/S/U/W 2},
    pdfpagemode=UseOutlines,
    pdfstartpage={1},
}

\newcommand{\ee}{e}
\newcommand{\ii}{i}

\usepackage{xspace}

\newcommand{\ie}{i.\,e.,}
\newcommand{\eg}{e.\,g.,}
\newcommand{\vect}[1]{\boldsymbol{#1}}

\newcommand{\pbar}[1]{\,\boldsymbol{\hat{\overline{\!{p}}}}_i^{#1}\,}

\newcommand{\pbarr}[2]{\,\boldsymbol{\hat{\overline{\!{p}}}}_{#1}^{#2}\,}
\newcommand{\bars}[1]{\, \boldsymbol{\hat{\overline{\!{#1}}}} \,}

\usepackage{wasysym}

\usepackage{diagbox}

\usepackage{simplewick}

\usepackage[clock]{ifsym}
\usepackage{lipsum}

\newcommand{\orcid}[1]{\href{https://orcid.org/#1}{\includegraphics[width=7pt]{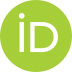}}}

\begin{document}
\title{Quantum Field Theory for Multipolar Composite Bosons with Mass Defect and Relativistic Corrections}
\collaboration{This article has been published in \href{https://doi.org/10.1103/PRXQuantum.5.020322}{PRX Quantum \textbf{5}, 020322 (2024)}; 
licensed under a \href{http://creativecommons.org/licenses/by/4.0/}{Creative Commons Attribution [CC BY]} license.}

\newcommand{\affHAN}{\address{Institut f{\"u}r Quantenoptik, Leibniz Universit{\"a}t Hannover, Welfengarten 1, D-30167 Hannover, Germany}}
\newcommand{\affULM}{\address{Institut f{\"u}r Quantenphysik and Center for Integrated Quantum Science and Technology (IQST), Universit{\"a}t Ulm, Albert-Einstein-Allee 11, D-89081 Ulm, Germany}}
\newcommand{\affTUDa}{\address{Technische Universit{\"a}t Darmstadt, Fachbereich Physik, Institut f{\"u}r Angewandte Physik, Schlossgartenstr. 7, D-64289 Darmstadt, Germany}}

\author{Tobias Asano\,\orcid{0000-0002-6257-8815}\textsuperscript{\affULM}}
\email{tobias.asano@alumni.uni-ulm.de}
\email{tobias-asano@outlook.de}
\affULM
\author{Enno Giese\,\orcid{0000-0002-1126-6352}\textsuperscript{\affTUDa \affHAN}}
\email{enno.giese@tu-darmstadt.de}
\email{enno.a.giese@gmail.com}
\affTUDa
\affHAN
\author{Fabio Di Pumpo\,\orcid{0000-0002-6304-6183}\textsuperscript{\affULM}}
\email{fabio.di-pumpo@uni-ulm.de}
\email{fabio.di-pumpo@gmx.de}
\affULM

\begin{abstract}
\noindent
Atomic high-precision measurements have become a competitive and essential technique for tests of fundamental physics, the Standard Model, and our theory of gravity.
It is therefore self-evident that such measurements call for a consistent relativistic description of atoms that eventually originates from quantum field theories like quantum electrodynamics.
Most quantum-metrological approaches even postulate effective field-theoretical treatments to describe a precision enhancement through techniques like squeezing.
However, a consistent derivation of interacting atomic quantum gases from an elementary quantum field theory that includes both the internal structure as well as the center of mass of atoms, has not yet been addressed.
We present such a subspace effective field theory for interacting, spin carrying, and possibly charged ensembles of atoms composed of nucleus and electron that form composite bosons called cobosons, where the interaction with light is included in a multipolar description.
Relativistic corrections to the energy of a single coboson, light-matter interaction, and the scattering potential between cobosons arise in a consistent and natural manner.
In particular, we obtain a relativistic coupling between the coboson's center-of-mass motion and internal structure encoded by the mass defect.
We use these results to derive modified bound-state energies, including the motion of ions, modified scattering potentials, a relativistic extension of the Gross-Pitaevskii equation, and the mass defect applicable to atomic clocks or quantum clock interferometry. 
\end{abstract}

\maketitle

\section{Introduction}
Quantum field theories (QFTs)~\cite{Wilczek1999} are powerful and successful tools with applications ranging from the field of particle physics described by the Standard Model~\cite{Tanabashi2018}, over quantum electrodynamics (QED)~\cite{Peskin1995} to nonrelativistic (NR) ultracold quantum gases~\cite{Giorgini2008, Baranov2008}.
Because these gases consist of atoms, \ie{} composite particles, and not of elementary particles, they have to be described by an \emph{effective field theory} (EFT)~\cite{Caswell1986, Kinoshita1996, Manohar1997}.
Such EFTs are \emph{the} method of choice, for instance in describing Bose-Einstein condensates (BECs)~\cite{Parkins1998,Kawaguchi2012}, but are usually not derived from an elementary theory.
Hence, they give no direct access to relativistic and further corrections, including radiative corrections~\cite{Wichmann1956,Brown1959,Mohr1974, Jentschura1999}, effects from the composite nature of the nucleus~\cite{Paz2012}, or the mass defect~\cite{Sonnleitner2018,Schwartz2019} relevant for quantum clocks~\cite{Yudin2018}.
In this work, we derive a subspace QFT from QED to describe NR composite particles including relativistic corrections.
As a result, we obtain a field-theoretical description of charged, interacting atomic ensembles including both the coupling of the center-of-mass (c.m.) motion to the internal atomic structure, as well as atom-atom and light-matter interactions with relativistic corrections.

Since in many applications atoms move at NR velocities and pair creation plays no role, the respective EFT leads to nonrelativistic QED (NRQED)~\cite{Caswell1986, Kinoshita1996, Manohar1997, Paz2015}.
It is routinely used to describe an individual neutral, composite particle, where usually the c.m. degrees of freedom are not taken into account and the light-matter interaction is only considered to lowest order.
This approach is suited for studying atomic structures, such as in spectroscopy~\cite{Bings2010}, giving rise to, \eg{} radiative QED corrections. 
On the other hand, atomic scattering experiments imply the presence of more than one particle and rely on the c.m. of atoms, so that the theory mentioned above has to be extended.
Common approaches~\cite{Ueda2010} usually include the c.m. and atom-atom scattering by generalizing single-atom theories to a corresponding effective QFT, so that both coincide in the single-atom limit.
However, fundamental effects from QED remain inaccessible in such a treatment.
By reducing potential NRQED (pNRQED) to a subspace EFT for interacting atomic ensembles and taking their c.m. degrees of freedom into account, we find a description that includes radiative corrections and scattering potentials.
The coupling of the inner-atomic structure to the atom's c.m. motion is a consequence of the relativistic mass defect~\cite{Yudin2018,Sonnleitner2018,Schwartz2019,Martinez2022}, \eg{} used in quantum clock interferometry~\cite{Sinha2011,Zych2011,Pikovski2017,Loriani2019}, and which gives access to quantum tests of fundamental physics~\cite{Derevianko2014,Arvanitaki2015,Arvanitaki2018,Ufrecht2020,Roura2020,DiPumpo2021,DiPumpo2022,DiPumpo2023}.
This coupling has been derived in single-particle quantum mechanics for spin- and chargeless particles with~\cite{Schwartz2019} and without~\cite{Sonnleitner2018} gravitational backgrounds.

In this work, we derive a framework that includes aspects of these concepts.
To this end, we use \smash{pNRQED}~\cite{Pineda1998a, Pineda1998, Pineda1998b}, an EFT of NRQED, to describe two different fermions (constituents of the coboson) on flat spacetime, and project this EFT to  its cobosonic subspace describing charged (ionized) spin carrying atomic clouds, including the mass defect, which are exposed to atom-atom and light-matter interactions.

\begin{figure}
    \centering
    \includegraphics[width=\columnwidth]{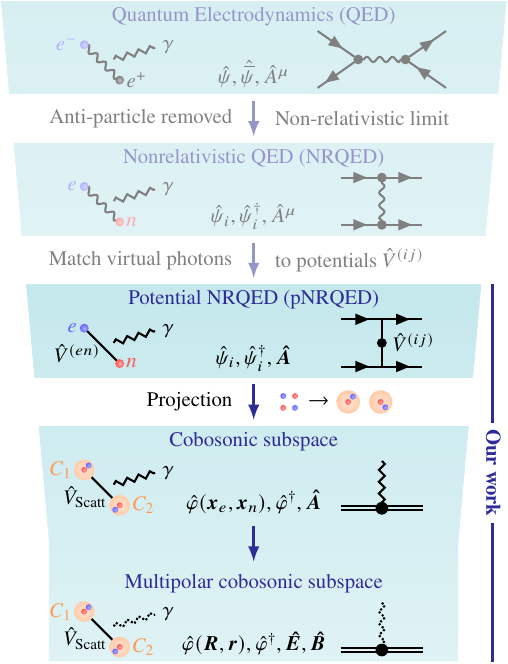}
    \caption{
    Hierarchy of effective field theories: 
    At the top level, quantum electrodynamics (QED) describes the interaction of fermions (dots), in particular electrons (blue) and positrons (black), with photons, where real (measurable) photons are represented by zigzag lines and virtual photons between two fermions by wiggly lines.
    The respective field operators are the four-component spinors $\hat{\psi}$ and $\hat{\overline{\psi}}$ and four potential $\hat{A}^\mu$, whose interaction is described by typical Feynman diagrams such as the vacuum polarization (solid line with arrows are fermions) shown on the right.
    The next refined field theory is nonrelativistic QED (NRQED), where the antiparticle can be removed from the description by the restriction to an NR limit of QED.
    In NRQED fermion field operators are replaced by two-component spinors $\hat{\psi}_i$ creating and annihilating electrons or nuclei, where the latter are assumed as elementary fermions as well. 
    These fermions can interact with each other.
    For example, the Feynman diagram shows two solid lines representing now two fermions scattering via a virtual photon.  
    In a next step potential NRQED (pNRQED) replaces virtual photons in a second-order scattering process by a nonlocal potential $\hat{V}^{(ij)}$ for an effective first-order scattering process, where external photons are completely described by the vector potential $\vect{\hat{A}}$ in Coulomb gauge.
    This level is the starting point of our work.
    The cobosonic subspace follows from a projection, where we restrict the Hilbert space to only electron-nucleus pairs (cobosons) and introduce a separation of scales, \ie{} atomic dimensions and scattering length.
    As a consequence, fermionic field operators are replaced by cobosonic field operators $\hat{\varphi}$ and interactions between cobosons (orange) are mediated by scattering potentials $\hat{V}_\text{scatt}$. 
    Here, Feynman diagrams now feature double solid lines representing a coboson.
    Multipolar cobosonic subspace describes external photons by EM fields $\vect{\hat{E}}$ and $\vect{\hat{B}}$ (dashed zigzag line) interacting with cobosons via multipoles whose degrees of freedom are described by c.m. and relative coordinates.
    }
    \label{fig:intro}
\end{figure}

The paper follows the hierarchical steps presented in Fig.~\ref{fig:intro}, summarizing transitions between different QFTs at each step, where our contribution begins on the level of pNRQED.
This EFT eventually originates from QED as a fundamental starting point for the description of interactions between fermions and photons.
Since we aim for a description of NR interacting composite particles consisting of bound fermions, it is sufficient to work in the established NR limit of QED, where antiparticles are removed from the formalism. 
The resulting model covers the interaction of nuclei and electrons on a field-theoretical level, including relativistic effects.
This interaction is mediated by (virtual) photons that account for the binding potential between the constituents of the composite particle.
Such virtual photons are matched to instantaneous potentials $\hat{V}^{(ij)}$, resulting in the already mentioned potential pNRQED~\cite{Pineda1998a, Pineda1998, Pineda1998b}.
These potentials mediate the electromagnetic (EM) interaction in the spirit of the classical Coulomb problem, but still include relativistic corrections.
Derivations of these EFTs have been discussed previously, which is why we recall in Sec.~\ref{Sec:pNRQED} the pNRQED Hamiltonian that constitutes the starting point for our paper. 
However, in Appendix~\ref{App:A} we discuss NRQED in more detail and present an explicit derivation of pNRQED potentials.

These potentials between elementary fermions in pNRQED do not describe solely attractive potentials between constituents of an atom, but also repulsive interactions in a gas of fermions.
To this end, we reduce pNRQED to its cobosonic subspace via a projection technique on a subspace of paired nuclei and electrons that is based on different length scales (Sec.~\ref{Sec:CbQFT}).
In addition, we introduce field operators for so-called cobosons (composite bosons)~\cite{Combescot2008} whose commutation relation differs from the fundamental bosonic one~\cite{Combescot2010}.
Such a projection was performed for a single atom~\cite{Pineda1998b} without introducing new spatial length scales and consequently scattering plays no role. 
Conversely, an EFT for van der Waals interactions that includes scattering between two atoms was constructed~\cite{Brambilla2017} and matched to pNRQED while we explicitly project pNRQED to the cobosonic subspace of interacting cobosons.
The resulting theory of composite particles has the desired form, but is still given in terms of the degrees of freedom of their constituents.

In Sec.~\ref{Sec:Unitaries} we therefore describe the interaction of such composite particles with light via electric and magnetic fields~\cite{Power1959,Woolley1971} instead of the vector potential.
At the same time, we introduce c.m. and relative coordinates commonly used in the description of bound or composite particles.
This approach clearly distinguishes between the internal structure of composite particles and their c.m. motion, so that it connects to the field-theoretical treatment of quantum gases.

We present the main results in Sec.~\ref{Sec:MpCbQFT}, \ie{} a multipolar cobosonic subspace EFT for atoms, including their c.m. and relative degrees of freedom, a coupling between both, as well as their scattering and interaction with EM fields.

Finally, we put our results in Sec.~\ref{Sec:Discussion} into context with existing approaches in different subfields, and we use this discussion as motivation for sample applications given in Sec.~\ref{Sec:Applications}.
We present the coupling of the atom's energy spectrum to the c.m. motion, reduce the scattering potential to a generalized dipole-dipole potential, derive a QFT for interacting cobosonic quantum gases, and find in a mean-field description a modified Gross-Pitaevskii equation~\cite{Gross1961,Pitaevskii1961} including the mass defect.
We conclude in Sec.~\ref{Sec:Conclusions}.

The detailed discussion of pNRQED from Appendix~\ref{App:A} is followed by Appendix~\ref{App:D} presenting the full transformation from the cobosonic subspace to its multipolar version, while Appendix~\ref{App:E} presents the eigenfunctions for the relative motion of hydrogenlike composite particles.

\section{Potential nonrelativistic QED} \label{Sec:pNRQED}
To create a description of bound-state systems at NR energies characterized by binding potentials, we use \emph{potential nonrelativistic quantum electrodynamics}~\cite{Pineda1998a, Pineda1998, Pineda1998b} as a starting point.
This theory is obtained by taking the NR limit of QED, which leads to NRQED and subsequently matching virtual photons to instantaneous potentials.
For a more detailed discussion of NRQED and the derivation of pNRQED, we refer to Appendix~\ref{App:A}.
Here, we will only present the relevant pNRQED Hamiltonian
\begin{align} \label{eq:Ham_pNRQED2}
\begin{split}
    \hat{H} =& \hat{H}_\text{EM} +  \sum_{i}  \int \dd[3]{x_i} \hat{\psi}^\dagger_i  \hat{h}_i \hat{\psi}_i  \\
    & + \sum_{i,j} \int \dd[3]{x_i} \int \dd[3]{x_j^\prime} \hat{\psi}^\dagger_i \hat{\psi}^{\prime \, \dagger}_j \hat{V}^{(ij)} \hat{\psi}_j^\prime \hat{\psi}_i  
\end{split}
\end{align} 
accounting for photon and spin-1/2 fermion fields.
The former are represented in the free electromagnetic field $\hat{H}_\text{EM} = \varepsilon_0 \int \dd[3]{x} ( \vect{\hat{E}}^2 + c^2 \vect{\hat{B}}^2)/2$ defined through the vacuum permittivity $\varepsilon_0$ as well as electric $\vect{\hat{E}}$ and magnetic field $\vect{\hat{B}}$.
The latter appear in Eq.~\eqref{eq:Ham_pNRQED2} through the single-fermion sector (two field operators) that accounts for the energy of fermion species $i=e,n$, \ie{} we deal with two fermionic species, namely electrons and nuclei.
By that $\hat{h}_i$ represents the (first-quantized) energy of a single fermion $i$.
Note that we treat nuclei as effective fermions, even though they are composite systems as well, which will be accounted for via \textit{Wilson coefficients}~\cite{Wilson1965,Wilson1974}. 
The field operators are two-component spinors obeying anticommutator relations among components $u,v=1,2$ of the same species $\{\hat{\psi}_{i, u} (\vect{x}),\hat{\psi}_{i, v}^\dagger (\vect{x^\prime}) \} = \delta_{uv} \delta(\vect{x}-\vect{x^\prime})$ and $\{\hat{\psi}_{i, u} (\vect{x}),\hat{\psi}_{i, v} (\vect{x^\prime}) \} = 0$.
Simultaneously, electron and nucleus field operators act on different Hilbert spaces implying vanishing commutators $[ \hat{\psi}_{i,u}, \hat{\psi}_{j,v}^\dagger]=0 = [ \hat{\psi}_{i,u}, \hat{\psi}_{j,v}]$ for $i \neq j$ between different particle species.
Next, the multifermion sector (four or more field operators) represents interactions between fermions mediated through instantaneous potentials $\hat{V}^{(ij)}$.

\begin{figure*}
    \centering
        \includegraphics[width=\textwidth]{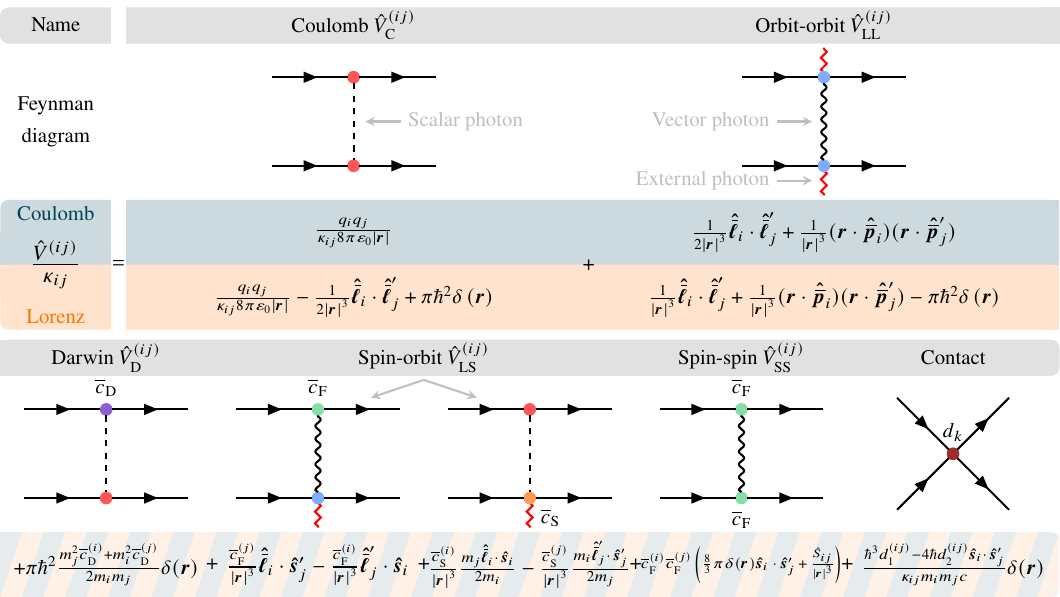}
        \caption{Relevant Feynman diagrams up to order $c^{-2}$.
        The dashed and wiggly lines correspond to scalar photons resulting from a contraction of the scalar potential and vector photons arising from contracting vector potential components, respectively. 
        The vertices are labeled according to the specific term in the Hamiltonian density being contracted:
        electric energy (red), kinetic energy (blue), spin-magnetic field (green, $\overline{c}_\text{F}$), Darwin term (purple, $\overline{ c}_\text{D}$), spin-orbit term (orange, $\overline{c}_\text{S}$), and contact interaction (brown, $d_1$ and $d_2$). 
        Details regarding $c$- and $d$-type Wilson coefficients can be found in Appendix~\ref{App:A}.
        The potential connected to each Feynman diagram is calculated in both Coulomb (top, blue shaded) and Lorenz gauge (bottom, orange shaded). 
        The Feynman diagrams of the first row yield different potentials that depend on the gauge, whereas the potentials of the second row are identical in both gauges. 
        The overall effective potential is the sum of all contributions from the first and second row and is gauge invariant.
        The potential is given in units of $\kappa_{ij} = -q_i q_j/(8 \pi \varepsilon_0 m_i m_jc^2)$ and we introduce abbreviations for individual minimally coupled angular momenta $\, \vect{\hat{ \overline{\! \ell}}}_i= \vect{r} \times \, \vect{\hat{\overline{ \! p}}}_i$ and relative distance $\vect{r} = \vect{x}_i-\vect{x}_j^\prime$ omitting indices $i$ and $j$ for simplicity. 
        In addition, we defined the spin contribution $\hat{S}_{ij} = - \hat{\vect{s}}_i \cdot \hat{\vect{s}}_j^\prime + 3 ( \vect{r} \cdot \hat{\vect{s}}_i ) (\vect{r} \cdot \hat{\vect{s}}_j^\prime)/\abs{\vect{r}}^2$ of the magnetic dipole-dipole potential. }
    \label{fig:potential}
\end{figure*}

In this work, we employ the NR expansion in the order $c^{-2}$ of the speed of light $c$ in which the single-fermion Hamiltonian takes form
\begin{align}
\begin{split}
        \hat{h}_i =& m_i c^2 + \frac{\pbar{2}}{2m_i} - \overline{c}_{\text{F}}^{(i)} q_i \frac{\boldsymbol{\hat{s}}_i\cdot \boldsymbol{\hat{B}}}{m_i} - \frac{\pbar{4}}{8m_i^3c^2} \\ \label{eq:hfirst2}
        & + \overline{c}_{\text{S}}^{(i)} q_i \vect{\hat{s}}_i \cdot \frac{  \pbar{}  \times \vect{\hat{E}} - \vect{\hat{E}} \times \pbar{} }{4 m_i^2c^2}  \\
        &+ \overline{c}_\text{W1}^{(i)} q_i \frac{\left\{ \pbar{2} \!, \vect{\hat{s}}_i \cdot \vect{\hat{B}} \right\}}{4m^3_ic^2} - \overline{c}_\text{A1}^{(i)} q^2_i \hbar^2 \frac{\vect{\hat{B}}^2}{8 m_i^3 c^2}.
\end{split}
\end{align}
The fermions carry rest mass $m_i$ giving rise to rest energy, kinetic energy, and its relativistic correction.
Due to the fermion's spin $\vect{\hat{s}}_i = \hbar \vect{\hat{\sigma}}_i/2$ defined through the Pauli matrix vector $\vect{\hat{\sigma}}=(\hat{\sigma}_x,\hat{\sigma}_y,\hat{\sigma}_z)$ there is also a coupling to the EM fields in the NR expansion.
Note that a contribution $\vect{\nabla} \cdot \vect{\hat{E}}$ (together with a coefficient $\overline{c}_\text{D}$) does not appear explicitly at this point anymore, because we work in Coulomb gauge.
Hence, after accounting for $\overline{c}_\text{D}$ in the potential terms $\hat{V}^{(ij)}$, we can neglect the scalar potential $\hat{\phi}$ for simplicity in this gauge, see also Appendix~\ref{App:A}. 
As a result, also the divergence of the electric field vanishes but both can be reincluded if needed.
The interaction of fermions due to their charge $q_e=-e$ and $q_n=+Ze$ with $Z>0$ and the elementary charge $e$ with external photons is encoded in minimally coupled momentum operators $\pbar{} = \vect{\hat{p}} - q_i \vect{\hat{A}}$. 
This way, both the momentum operator $\vect{\hat{p}} = - \ii \hbar \vect{\nabla}$ and the vector potential $\vect{\hat{A}}$ enter.
The Wilson coefficients $\overline{c}_k^{(i)}$ are discussed in Appendix~\ref{App:A}.
The interaction between an arbitrary number of fermion fields appears also in pNRQED. 
However, contact interactions originating from NRQED that include $N$ fermion fields are at least of order $c^{5-3N/2}$.
Consequently, only the fermion-fermion interaction ($N=4$) presented in Eq.~\eqref{eq:Ham_pNRQED2} with potentials $\hat{V}^{(ij)}$~\footnote{Matching in order $c^{-2}$ can be extended to order $\alpha/c^2$ by including relevant loop corrections~\cite{Pineda1998}.} detailed in Fig.~\ref{fig:potential} is relevant for our purpose and all higher interactions ($N>4$) are suppressed.
In addition, all potentials matched to interactions where more than four fermion fields are present are also beyond the order of $c^{-2}$.
These potentials correspond to one part of the \textit{Breit-Pauli} Hamiltonian~\cite{Breit1929, Breit1930,Breit1932} but also include QED corrections through Wilson coefficients.
The pNRQED Hamiltonian is valid for any Feynman diagram in $c^{-2}$ not yet included in the potentials already.
Even though one can choose an arbitrary gauge, we use Coulomb gauge ($\vect{\nabla} \cdot \vect{\hat{A}} = 0$) for the electromagnetic field for simplicity.
This choice allows us to eliminate $\hat{\phi}$ in $\hat{h}_i$, because all virtual scalar photons are already matched to potentials and the scalar potential of external photons vanishes in Coulomb gauge.
In particular, also the self-energy of the fermions, and by that the Lamb shift~\cite{Bethe1947,Lamb1947}, can still be obtained because only virtual vector photons contribute~\cite{Pineda1998a, Pineda1998c} to this shift.

A derivation of the pNRQED potentials in the order $c^{-2}$ based on the matching of $S$-matrix elements as an alternative to equating Green's functions~\cite{Pineda1998b} is presented in Appendix~\ref{App:A}, also confirming their gauge independence.
In contrast to the usual convention relying only on the relative-coordinate contribution to the potentials, we give the potentials in terms of single-fermion coordinates as we take also the c.m. of our bound-state systems into account.
Moreover, in our description the external photons in the Feynman diagrams are also matched to potentials, which introduces minimally coupled momentum operators.

\section{Cobosonic subspace} \label{Sec:CbQFT}
\begin{figure}[b]
    \centering
    \includegraphics[width=\columnwidth]{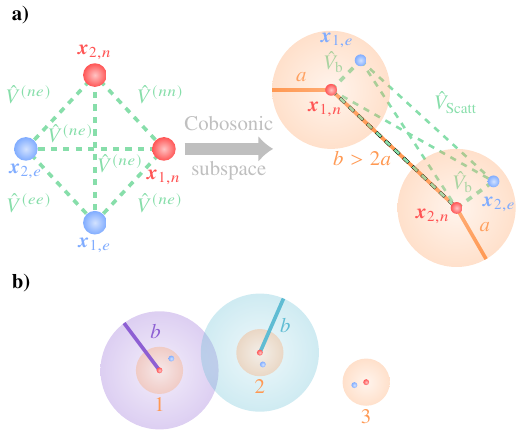}
    \caption{a) The left-hand side shows the situation in pNRQED, where all elementary fermions of the system interact with each other and no interaction is dominant compared to the others. 
    A projection to the coboson subspace introduces length scales, and by that our system is only composed of uniquely assigned electron-nucleus pairs. 
    In particular, we find an atomic length scale $a$ associated with the inner-atomic distance between the constituents and a scattering length scale $b$ associated with the distance between different cobosons. 
    The separation of scales allows for the identification of the dominant binding potential $\hat{V}_\text{b} = \hat{V}^{(ne)}+ \hat{V}^{(en)}$ between the constituents of a cobosons compared to the weaker scattering potential $\hat{V}_\text{scatt}=  \hat{V}^{(nn)}+\hat{V}^{(ne)} +\hat{V}^{(en)}+ \hat{V}^{(ee)}$ composed of attractive and repulsive interactions among constituents of different cobosons.
    b) The separation of scales has also implications for the allowed position of different cobosons. 
    The nucleus of coboson 1 can be placed anywhere, but the corresponding electron only within a vicinity of radius $a$ around it. 
    The nucleus of coboson 2 must not be closer than a distance $b$ from the nucleus of coboson 1. 
    As a consequence, the purple sphere around coboson 1 is excluded for coboson 2. 
    Similarly, coboson 3 must not be located in the purple and green spheres around cobosons 1 and 2, respectively.}
    \label{fig:projection}
\end{figure}

We now move from the single-fermion to a composite-particle description. 
Moreover, we restrict ourselves to ensembles of one atomic species and we reduce the problem to the simplest case of electron-nucleus pairs, forming a composite boson, \ie{} a \textit{coboson}~\cite{Pineda1998b, Combescot2008, Combescot2010}.
The distance between electron and nucleus that form a coboson is given by an atomic length scale, whereas the distance to other cobosons and their constituents is much larger.
Thus, we consider a situation that is sufficiently dilute, such that individual cobosons do not overlap.
Motivated by these different length scales, we describe atoms as \textit{spatially restricted} cobosons resembling \textit{hard-sphere}-based models~\cite{Lee1957,Lee1958,Wu1959}, such that two constituents within a sphere of a certain cutoff radius form a composite particle and are, by definition, free fermions outside of it.
Formally, we achieve such a transition from the fermionic pNRQED to a quantum field theory of its cobosonic subspace by means of a projection $\hat{\pi}_{\text{Cb}}$ of the  Schrödinger equation $\ii \hbar \dd \ket{\Psi} / \dd t = \hat{H} \ket{\Psi}$. 
Here, cobosonic states are part of the general second-quantized state $\ket{\Psi}$, represented in the following by uppercase symbols, in contrast to lowercase symbols that represent first-quantized states. 
Thus, the projection operator is chosen such that only spatially restricted cobosonic states are selected, giving rise to intracobosonic and intercobosonic length scales.
Conversely, observations like atomic decay, free fermions, multielectron atoms, molecules, \textit{etc.}, do not lie within the subspace spanned by this projection. 
Figure~\hyperref[fig:projection]{3a)} shows one exemplary configuration that is ruled out by projection and another one that is selected by the projector. 
Guided by the intuitive picture in the figure, we define the intracobosonic scale $a$ and the length scale $b \gg a$ associated with the distance between different cobosons. 
As a result, the dominant EM interaction between fermions are the attractive binding potentials between atomic constituents.
Contrarily, intercobosonic interactions are based both on attractive and repulsive interactions between the fermionic constituents of \emph{different} cobosons.
Usually, EFTs are derived by directly integrating out certain energy scales such that the EFT is associated with a new energy range. 
By introducing new length scales we do not address the explicit characterization of a new energy scale associated with our theory, but emphasize that such spatial length scales may be identified with a corresponding energy range.

In the spirit of such QFTs for atoms, we expect our cobosonic subspace to be characterized by creation $\hat{\varphi}^\dagger = \hat{\psi}^\dagger_n \hat{\psi}^\dagger_e$ and annihilation operators $\hat{\varphi}=\hat{\psi}_n \hat{\psi}_e$ associated with only the creation and annihilation of whole atoms instead of single-fermion field operators.
To reduce the Hilbert space to the states depicted in Fig.~\hyperref[fig:projection]{3a)} we define a projector $\hat{\pi}_{\text{Cb}} = \sum_{k=0}^{N} \hat{\pi}_k$ that projects on up to $N$ cobosons, where $\hat{\pi}_k$ projects onto a subspace of $k$ cobosons.
As such, the subspace projection
\begin{align} \label{eq:proj}
    \hat{\pi}_k = \frac{1}{k!} \int \limits_{C_1} \dd[6]{x_1} ... \int \limits_{C_k} \dd[6]{x_k} \left( \prod_{\ell=1}^k  \hat{\varphi}^\dagger_\ell \right) \ket{0} \bra{0} \left( \prod_{\ell=1}^k  \hat{\varphi}_\ell \right)
\end{align}
is defined by the cobosonic operator $\hat{\varphi}^\dagger_\ell = \hat{\psi}^\dagger_n ( \vect{x}_{\ell,n}) \hat{\psi}^\dagger_e (\vect{x}_{\ell,e})$, creating a coboson at position $(\vect{x}_{\ell,n}, \vect{x}_{\ell,e})$, with an analogous definition for the annihilation operator.
Moreover, Eq.~\eqref{eq:proj} contains the abbreviation of a six-dimensional integration measure $\dd[6]{x_k} = \dd[3]{x_{k,n}} \, \dd[3]{x_{k,e}} $ and by definition the subspace projectors are orthogonal, \ie{} $\hat{\pi}_k \hat{\pi}_\ell = 0$ for $k\neq \ell$.
As explained in Fig.~\hyperref[fig:projection]{3a)}, such a projection implies that not all fermion coordinates are independent.
In fact, we have to equip the integrals with proper integration regions $C_k = C_{k,n} \cross C_{k,e}$ for the nucleus and electron of the $k$th coboson.
This way, we introduce the internal cobosonic (atomic) length scale $a$ by restricting the electron coordinates $\vect{x}_{k,e}$ of coboson $k$ to $C_{k,e}=B_{a}(\vect{x}_{k,n})$ denoting a spherical volume with radius $a$ around the nucleus of coboson $k$ positioned at $\vect{x}_{k,n}$.
The intercobosonic scale $b$ enters through regions for nuclei coordinates in an iterative manner, see also Fig.~\hyperref[fig:projection]{3b)}. 
The first nucleus may be positioned anywhere, \ie{} in the volume $C_{1,n} = \mathbb{R}^3$.
However, the second nucleus must not be within a sphere of radius $b$ around the first nucleus and is therefore located in a region $C_{2,n} = \mathbb{R}^3 \setminus B_b (\vect{x}_{1,n})$.
The distance $b$ between the cobosons has to be larger than the atomic length scale $a$ such that the pairing of nucleus $k$ and electron $k$ remains unique.
A generalization to the $k$th coboson~\footnote{For simplicity we defined the distance between cobosons with respect to the nucleus coordinate, even though center-of-mass distances are physically more precise. 
However, both are equivalent by replacing $b$ by $b^\prime = b+a$.} yields
\begin{align}
    C_k = C_{k,n} \cross C_{k,e} = \mathbb{R}^3 \setminus \bigcup_{\ell = 1}^{k-1} B_b (\vect{x}_{\ell,n}) \cross B_a ( \vect{x}_{k,n}).
\end{align}
Because of these limits of integration the idempotence of the projector, \ie{} $\hat{\pi}_k^2 = \hat{\pi}_k$ and by that $\hat{\pi}_\text{Cb}^2 = \hat{\pi}_\text{Cb}$, follows from a normalization factor of $1/k!$.
This fact becomes apparent by considering the form of the cobosonic commutator
\begin{align} \label{eq:comm}
    \begin{split}
         \left[ \hat{\varphi}^\prime_{u^\prime v^\prime}, \hat{\varphi}^\dagger_{u v} \right] =& \delta_{u^\prime u} \delta_{v^\prime v} \delta ( \vect{x}^\prime_n - \vect{x}_n) \delta ( \vect{x}^\prime_e - \vect{x}_e ) \\
        &- \delta_{u^\prime u} \delta ( \vect{x}^\prime_n - \vect{x}_n) \hat{\psi}_{e,v}^\dagger \hat{\psi}^\prime_{e,v^\prime} \\
        &- \delta_{v^\prime v} \delta ( \vect{x}^\prime_e - \vect{x}_e) \hat{\psi}_{n,u}^\dagger \hat{\psi}^\prime_{n,u^\prime} ,
    \end{split}
\end{align}
between the cobosonic field operators that is defined through the fermionic anticommutation relations. 
The first term coincides with a fundamental bosonic commutation relation and, thus, corresponds to purely atomic dynamics.
The second and third terms amount to the cobosonic part representing exchange characteristics between fermions of different cobosons.
When we determine $\hat{\pi}_k^2$ by consecutive application of only the bosonic part of the coboson commutator, \ie{} the first line of Eq.~\eqref{eq:comm}, idempotence directly demands a normalization $1/k!$. 
Due to this fact our projector resembles bosonic behavior.
Ignoring the limits of integration, the remaining terms generated by the cobosonic part of the commutator give rise to further contributions, so that  $\hat{\pi}_k$ is not idempotent~\footnote{
For idempotence ingoring the limits of integration, the normalization factor $1/k!$ in \eqref{eq:proj} has to be replaced by $k!^{-2}$, as expected for spatially independent cobosons.
}. 
However, the chosen limits of integration uniquely assign pairwise field operators summarized in $\hat{\varphi}$ and $\hat{\varphi}^\dagger$ each.
Hence, these additional cobosonic terms are vanishing by construction.

Using this time-independent projection operator, we define the projected states via $\ket{\Psi}_\text{Cb} = \hat{\pi}_\text{Cb} \ket{\Psi}$ and their equation of motion
\begin{equation}
    \ii \hbar \frac{\dd }{\dd t} \ket{\Psi}_\text{Cb} = \hat{\pi}_\text{Cb} \hat{H} \hat{\pi}_\text{Cb} \ket{\Psi}_\text{Cb} +   \hat{\pi}_\text{Cb} \hat{H} ( \mathds{1} - \hat{\pi}_\text{Cb}) \ket{\Psi}.
\end{equation}
In the remainder of this paper we focus on the contribution to the motion induced by the cobosonic Hamiltonian $\hat{H}_\text{Cb}=\hat{\pi}_\text{Cb} \hat{H} \hat{\pi}_\text{Cb}$, as well as its eigenvalues and properties.
However, the coupling to states $( \mathds{1} - \hat{\pi}_\text{Cb}) \ket{\Psi}$ that lie outside of our projected Hilbert space leads to additional energy shifts and other effects of the environment in the spirit of open quantum systems~\cite{Rivas2012,Wurtz2020}.  

Projecting the Hamiltonian from Eq.~\eqref{eq:Ham_pNRQED2} yields three contributions.
Since the free EM Hamiltonian does not feature any fermionic field operators, its projection remains invariant and we begin with the single-fermion Hamiltonian $\int \dd[3]{x_i} \hat{\psi}^\dagger_i  \hat{h}_i \hat{\psi}_i \hat{\pi}_{\text{Cb}}$.
Expanding the projector into the subspace projectors $\hat{\pi}_\ell$ and applying the fermionic anticommutation relation $\ell$ times yields $\ell$ terms. 
The integration variables in these terms can be relabeled, but it is not obvious that all terms are identical because of the asymmetry of integration variables also contained in the nested limits of integration.
Nevertheless, the limits of integration are generic in such a way that it is equivalent whether one specific coboson's coordinates are restricted to the whole space except for spheres around all the other cobosons, or the other way around.
As a result, a symmetrized version of the limits of integration exists and allows the exchange of the nested regions.
In the simplest case of two cobosons, this property implies that we may interchange the arguments in any integrand $f$ depending on the six-dimensional vector $\underbar{$\vect{x}$}_k = (\vect{x}_{k,n},\vect{x}_{k,e})$ while preserving the limits of integrations such that
\begin{align} \label{eq:identity1sec3}
    \int \limits_{C_1} \dd[6]{x_1} \int \limits_{C_2} \dd[6]{x_2} f(\mathrm{\underbar{$\vect{x}$}}_1,\mathrm{\underbar{$\vect{x}$}}_2) = \int \limits_{C_1} \dd[6]{x_1} \int \limits_{C_2} \dd[6]{x_2} f(\mathrm{\underbar{$\vect{x}$}}_2,\mathrm{\underbar{$\vect{x}$}}_1) 
\end{align}
holds. 
The case for more than two cobosons follows analogously.
With this consideration, the projection simplifies to
\begin{align}
     \int \dd[3] x \hat{\psi}^\dagger_i \hat{h}_i \hat{\psi}_i \hat{\pi}_{\text{Cb}} = \sum_{\ell=1}^{N} \int \limits_{C_1} \dd[6]{x_1} \hat{\varphi}^\dagger_1 \hat{h}_i \hat{\pi}_{\ell-1 \setminus \mathrm{\underbar{$\vect{x}$}}_1} \hat{\varphi}_1
\end{align}
containing already only cobosonic operators but still the projector
\begin{align} \label{eq:projector_red}
\hat{\pi}_{\ell-1 \setminus \mathrm{\underbar{$\vect{x}$}}_1} = \int_{C_2} \dd[6]{x_2} ... \int_{C_\ell} \dd[6]{x_\ell} \frac{\hat{\varphi}^\dagger_2 ... \hat{\varphi}_\ell^\dagger \ket{0} \bra{0} \hat{\varphi}_\ell ... \hat{\varphi}_2}{(\ell -1)!}
\end{align}
on the subspace of $\ell-1$ cobosons, of which none is located within the sphere associated with the coboson of coordinates $\underbar{$\vect{x}$}_1$.
Hence, this projector begins with $C_2$ instead of $C_1$ and a generalization to $N$ cobosons is defined accordingly.
Finally, the term $\hat{\pi}_{\ell-1 \setminus \mathrm{\underbar{$\vect{x}$}}_1} \hat{\varphi}_1$ indicates the annihilation of a coboson at coordinates $\underbar{$\vect{x}$}_1$ after which we project onto the subspace of $\ell-1$ cobosons that must not coincide with the already annihilated coboson by construction.
This procedure, however, is equivalent to projecting first onto the full subspace of $\ell$ cobosons and then annihilating one coboson at position $\underbar{$\vect{x}$}_1$.
Consequently, the projected single-fermion Hamiltonian 
\begin{align} \label{eq:Ham_atom}
 \int \dd[3]{x_i} \hat{\psi}^\dagger_i  \hat{h}_i \hat{\psi}_i \hat{\pi}_{\text{Cb}} =  \int \limits_{C_1} \dd[6]{x_1} \hat{\varphi}^\dagger_1 \hat{h}_i \hat{\varphi}_1 \hat{\pi}_{\text{Cb}}
\end{align}
has the form of a composite-particle theory as the fermion operators are replaced by coboson operators, while the region accessible to the electron is restricted to the atomic scale around the nucleus.
The projection $\hat{H}_\text{f-f} \hat{\pi}_{\ell}$ of the fermion-fermion Hamiltonian $\hat{H}_\text{f-f} = \sum_{i,j} \int \dd[3]{x_1} \int \dd[3]{x_2} \hat{\psi}_i^\dagger \hat{\psi}_j^\dagger \hat{V}^{(ij)} \hat{\psi}_j \hat{\psi}_i$ comprises the projection of the repulsive ($i = j$) and attractive ($i \neq j$) potentials.
First, the repulsive part requires the commutation of two fermion field operators of the same species with the corresponding $\ell$ operators contained in $\hat{\pi}_\ell$ that yield now $\ell ( \ell -1)$ terms.
With the same argument as before, these terms can be combined to
\begin{align}
\begin{split}
    &\int \dd[3]{x_1} \int \dd[3]{x_2} \hat{\psi}_i^\dagger \hat{\psi}_i^\dagger \hat{V}^{(ii)} \hat{\psi}_i \hat{\psi}_i \hat{\pi}_{\ell} \\
    &= \int \limits_{C_1} \dd[6]{x_1} \int \limits_{C_2} \dd[6]{x_2} \hat{\varphi}_1^\dagger \hat{\varphi}_2^\dagger \hat{V}^{(ii)} \hat{\pi}_{\ell-2 \setminus \mathrm{\underbar{$\vect{x}$}}_1, \mathrm{\underbar{$\vect{x}$}}_2}  \hat{\varphi}_2 \hat{\varphi}_1
\end{split}
\end{align}
with $\hat{\pi}_{\ell-2 \setminus \mathrm{\underbar{$\vect{x}$}}_1, \mathrm{\underbar{$\vect{x}$}}_2}$ defined analogously to Eq.~\eqref{eq:projector_red}.
Similar to the single-fermion Hamiltonian before, it is identical to first annihilate two cobosons and projecting then on the $\ell-2$ subspace or annihilating these two cobosons after we projected onto the subspace of $\ell$ cobosons allowing the interchange of these operations.
When we consider then the attractive part of the fermion-fermion Hamiltonian, we find that this projection follows schematically a combination of the procedures presented with the single-fermion and the repulsive fermion-fermion Hamiltonian.
As a result, the projected fermion-fermion Hamiltonian with both the repulsive and attractive part resolves to
\begin{align} \label{eq:Ham_scatt}
\begin{split}
   \hat{H}_{\text{f-f}} \hat{\pi}_{\text{Cb}} =& \int \limits_{C_1} \dd[6]{x_1}\hat{\varphi}^\dagger_1 \left(\hat{V}^{(ne)} + \hat{V}^{(en)} \right) \hat{\varphi}_1 \hat{\pi}_{\text{Cb}} \\
   &+\int \limits_{C_1} \dd[6]{x_1} \int \limits_{C_2} \dd[6]{x_2} \hat{\varphi}_1^\dagger \hat{\varphi}_2^\dagger \sum_{i,j} \hat{V}^{(ij)} \hat{\varphi}_2 \hat{\varphi}_1 \hat{\pi}_{\text{Cb}}.
   \end{split}
\end{align}
The two different contributions for the attractive potentials can formally be traced back to applying the two different parts of the coboson commutator, where the bosonic part gives rise to the first line and the cobosonic part to the second line.
As also indicated in Fig.~\hyperref[fig:projection]{3a)}, the interaction between the fermions divides into the dominant binding potential given by $\hat{V}_\text{b} =\hat V^{(ne)}+\hat V^{(en)}$ in the single-coboson part (first line), while the intercobosonic scattering potential $\sum_{ij} \hat{V}^{(ij)}$ in the two-coboson part includes attractive and repulsive interactions.
Because of the separation of scales, these interactions are weaker than the binding potential.
Hence, the projected Hamiltonian reads
\begin{align} \label{eq:Ham_cb}
    \hat{H}_{\text{Cb}} =&  \hat{H}_\text{EM} + \int \limits_{C_1} \dd[6]{x_1} \hat{\varphi}^\dagger_1 \hat{h}_{\text{Cb}}  \hat{\varphi}_1 \notag \\
    &+ \int \limits_{C_1} \dd[6]{x_1} \int \limits_{C_2} \dd[6]{x_2} \hat{\varphi}_1^\dagger \hat{\varphi}_2^\dagger  \hat{V}_{\text{scatt}} \hat{\varphi}_2 \hat{\varphi}_1 
\end{align}
with the internal cobosonic energy $\hat{h}_{\text{Cb}} = \hat{h}_n + \hat{h}_e + \hat{V}^{(ne)} + \hat{V}^{(en)}$.
This contribution gives rise to the Breit-Pauli Hamiltonian for an electron and a nucleus~\cite{Breit1929, Breit1930, Breit1932} consisting of the sum of individual fermionic energies together with their total binding potential $\hat{V}^{(ne)} + \hat{V}^{(en)}$ arising from EM interaction between the fermionic constituents.
The scattering potential $\hat{V}_{\text{scatt}} = \hat{V}^{(nn)} + \hat{V}^{(ne)} + \hat{V}^{(en)} + \hat{V}^{(ee)}$ is based on both attractive ($\hat{V}^{(ne)} + \hat{V}^{(en)}$) and repulsive ($\hat{V}^{(nn)} + \hat{V}^{(ee)}$) EM interactions among all fermions of different cobosons. 
In particular, these attractive terms are weaker than the single-particle binding potentials, since the coordinates $\vect{x}_{1,i}$ and $\vect{x}_{2,j}$ of different cobosons are separated by $b \gg a$.
Compared to bosonic field theories for atomic ensembles~\cite{Lewenstein1994, Ueda2010}, our subspace EFT is based on creation $\hat{\varphi}^\dagger$ and annihilation $\hat{\varphi}$ operators whose components $u$,\,$v=1,2$ obey a cobosonic commutation~\cite{Combescot2010} relation from Eq.~\eqref{eq:comm} rather than a purely bosonic one.
Moreover, the projection operator includes integration regions that naturally ensure a bosonlike normalization of $1/k!$ even for this type of commutation relation.
In fact, the cobosonic part of the commutator in the second and third line is responsible for the scattering potential in Eq.~\eqref{eq:Ham_scatt}.
When projecting to a single-coboson subspace~\cite{Pineda1998}, these two aspects, the cobosonic part of the commutator and integration regions ensuring unique electron-nucleus pairs, become irrelevant.
In this case, we recover conventional single-particle quantum mechanics.

So far, we constructed a cobosonic theory from second-order scattering of fermions, where the internal structure is governed by the combined single-fermion energy and where their binding potential results from attractive interactions.
Furthermore, the intercobosonic dynamics arises from the interfermionic interactions between the constituents of different cobosons.
We emphasize that this projection does not exclusively work for the single-fermion Hamiltonian from Eq.~\eqref{eq:hfirst2} and the potential from Fig.~\ref{fig:potential} but rather for arbitrary single-fermion Hamiltonians and potentials.
However, effects of the environment given by states that do not lie within the coboson subspace have been neglected in our treatment.

\section{Second-quantized transformations} \label{Sec:Unitaries}
Although the Hamiltonian from Eq.~\eqref{eq:Ham_cb} has already the desired form of a subspace EFT for cobosons, it still involves the constituents' coordinates.
In the spirit of composite particles, we now move to c.m. and relative coordinates, where the latter take the internal cobosonic, \ie{} atomic, structure into account. 
Also, light-matter interaction enters in lowest order via the vector potential $\vect{\hat{A}}$ contained in the canonical momenta through minimal coupling.
For a description of experiments, it is more convenient to express the coupling by EM fields $\vect{\hat{E}}$ and $\vect{\hat{B}}$.
In this section we derive a method to incorporate the multipolar form of our cobosonic subspace EFT and move to relativistically corrected c.m. and relative coordinates of the second-quantized Hamiltonian $\hat{H}_\text{Cb}$ from Eq.~\eqref{eq:Ham_cb}.

In first-quantized regimes (characterized by lowercase symbols), realizing these operations involves unitary transformations~\cite{Krajcik1974, Liou1974,Power1959, Woolley1971, Woolley2020} $\hat{u}$ that transform a state $\ket{\psi} = \hat{u} \ket{\tilde{\psi}}$, where $\hat{u} = \exp{ \ii \hat{\lambda} /\hbar }$ may be expressed through a (time-independent) single-particle generator $\hat{\lambda}$.
Consequently, the effective Schrödinger equation $\ii \hbar \dd \ket{\psi} / \dd t = \hat{h}_{\text{Cb}} \ket{\psi}$ for the single-coboson Hamiltonian $\hat{h}_{\text{Cb}}$ yields a transformed operator 
\begin{align} \label{eq:unitary}
    \hat{\tilde{h}}_\text{Cb} = \hat{u}^\dagger  \hat{h}_{\text{Cb}} \hat{u}
\end{align}
as long as $\hat{u}$ is time independent.

Guided by this concept, we define for the second-quantized Hamiltonian $\hat{H}_\text{Cb}$ from Eq.~\eqref{eq:Ham_cb} an analogous~\cite{Lin1977,Salam2009} transformation $\ket{\Psi}_\text{Cb} = \hat{U} \ket{\tilde{\Psi}}_\text{Cb}$ on a second-quantized state $\ket{\Psi}_\text{Cb}$ (characterized by uppercase symbols) with a unitary $\hat{U} = \exp{ \ii \hat{\Lambda} / \hbar}$ generated by $\hat{\Lambda}$.
We choose the second-quantized generator $\hat{\Lambda}$ in such a way that the transformation reduces to the single-particle transformation acting on the first-quantized Hamiltonian $\hat{h}_\text{Cb}$.
With the choice 
\begin{align} \label{eq:Lambda}
    \hat{\Lambda} = \int \limits_{C_1} \dd[6]{x_1} \hat{\varphi}^\dagger_1 \hat{\lambda} \hat{\varphi}_1,
\end{align}
where $\hat{\lambda}$ is the generator of the corresponding first-quantized transformation, we achieve the desired behavior.

To see this connection, we determine the transformation
\begin{align}
    \hat{U}^\dagger \hat{\varphi} \hat{U} = \sum_{\ell=0}^\infty \frac{1}{\ell!} \left(- \frac{\ii}{\hbar} \right)^\ell \left[ \hat{\Lambda}, \hat{\varphi} \right]_\ell
\end{align}
with the help of a Baker-Campbell-Hausdorff formula where $\left[ \hat{\Lambda}, \hat{\varphi} \right]_\ell = \left[ \hat{\Lambda}, \left[ \hat{\Lambda}, \hat{\varphi} \right]_{\ell-1} \right]$ and $\left[ \hat{\Lambda}, \hat{\varphi} \right]_0 = \hat{\varphi}$.
Focusing first on the bosonic part of the coboson commutator, it can be shown that
\begin{align}
    \left[ \hat{\Lambda}, \hat{\varphi} \right]_\ell = \left( - \hat{\lambda} \right)^\ell \hat{\varphi} .
\end{align}
Regarding the additional parts of the coboson commutator, they do not vanish trivially but can be resolved by considering only the projected transformation $\hat{\pi}_\text{Cb} \hat{U}^\dagger \hat{\varphi} \hat{U} \hat{\pi}_\text{Cb}$ where these additional terms vanish due to the regions of coboson coordinates.  
Through the cobosonic part of the commutator, we generate terms containing the annihilation of two electrons (nuclei)  within the same sphere around one nucleus (electron), which lies outside of our projected subspace and thus vanishes.
This fact may be made explicit by introducing the projected transformation $\hat{\pi}_\text{Cb} \hat{U}^\dagger \hat{\varphi} \hat{U} \hat{\pi}_\text{Cb}$ as $\hat{\varphi} = \hat{\pi}_\text{Cb} \hat{\varphi}$.
Consequently, the transformation reduces to
\begin{align}
    \hat{U}^\dagger \hat{\varphi} \hat{U} = \hat{u} \hat{\varphi},
\end{align}
\ie{} the second-quantized unitary transformation given by $\hat{U}$ reduces to the first-quantized unitary $\hat{u}$ defined via their respective generator~\cite{Lin1977}.

With this relation, we obtain the transformed second-quantized Hamiltonian
\begin{align} \label{eq:transformation}
\begin{split}
    \hat{U}^\dagger \hat{H}_\text{Cb} \hat{U} =& \hat{U}^\dagger \hat{H}_{\text{EM}} \hat{U} + \int \limits_{C_1} \dd[6]{x_1} \hat{\varphi}^\dagger_1 \hat{u}_1^{\dagger} \hat{h}_\text{Cb} \hat{u}_1 \hat{\varphi}_1 \\
    &+ \int \limits_{C_1} \dd[6]{x_1} \int \limits_{C_2} \dd[6]{x_2} \hat{\varphi}^\dagger_1 \hat{\varphi}^\dagger_2 \hat{u}_1^{\dagger} \hat{u}_2^{\dagger} \hat{V}_{\text{scatt}} \hat{u}_2 \hat{u}_1  \hat{\varphi}_2 \hat{\varphi}_1 
\end{split}
\end{align}
given that $\left[ \hat{\Lambda}, \hat{h}_\text{Cb} \right] = \left[ \hat{\Lambda}, \hat{V}_\text{scatt} \right] =0$.
If $\hat{\lambda}$ contains EM fields, we also need to transform $\hat{H}_\text{EM}$, otherwise it remains invariant.
Here, the first-quantized unitary transformation $\hat{u}_k = \hat{u} (\vect{x}_{k,n}, \vect{x}_{k,e})$ of coboson $k$ acts only on coordinates and operators associated with coboson coordinates $(\vect{x}_{k,n}, \vect{x}_{k,e})$.

In the following, we specify the transformations to describe the multipolar cobosonic subspace by c.m. and relative coordinates including relativistic corrections.

\subsection{Nonrelativistic c.m. and relative coordinates}
First, we move from the set of electron $\{ \vect{x}_{k,e}, \vect{\hat{p}}_{k,e},\vect{\hat{s}}_{k,e} \}$ and nucleus  $\{\vect{x}_{k,n}, \vect{\hat{p}}_{k,n},\vect{\hat{s}}_{k,n} \}$ coordinates to NR c.m. $ \{ \vect{R}_k, \vect{\hat{P}}_k, \vect{\hat{S}}_k \}$ and relative $\{ \vect{r}_k, \vect{\hat{p}}_k, \vect{\hat{s}}_k \}$ coordinates describing coboson $k$.
The connection between the different coordinates are listed in Table~\ref{tab:NRcm} and chosen such that c.m. (position $\vect{R}_k$, momentum $\vect{\hat{P}}_k$) and relative (position $\vect{r}_k$, momentum $\vect{\hat{p}}_k$) share nonvanishing canonical commutators $[R_\ell^{(u)}, \hat{P}_k^{(v)}]= [r_\ell^{(u)} , \hat{p}_k^{(v)}]= \ii \hbar \delta_{u v} \delta_{\ell k} $ where $u,v=x,y,z$. 
\begin{table}[h]
\caption{\label{tab:NRcm}%
Single-fermion coordinates of coboson $k$ expressed through their c.m. and relative counterparts in the NR limit.
The positions, momenta, and spins $\{ \vect{x}_{k,j}, \vect{\hat{p}}_{k,j},\vect{\hat{s}}_{k,j} \}$ of fermion $j=e,n$ are expressed through their respective c.m. and relative coordinates $ \{ \vect{R}_k, \vect{\hat{P}}_k, \vect{\hat{S}}_k \}$ and $\{ \vect{r}_k, \vect{\hat{p}}_k, \vect{\hat{s}}_k \}$. Here, $M= m_e+m_n$ and $m_\text{r} = m_e m_n /M$ describe the total and reduced mass of the coboson, respectively, and $m_j$ is the mass of its constituents.
}
\begin{ruledtabular}
\begin{tabular}{llcc}
 & $j$ & $e$ & $n$ \\
\colrule
Position &$\vect{x}_{k,j}$ & $\vect{R}_k + m_\text{r} \vect{r}_k / m_e$ & $\vect{R}_k - m_\text{r} \vect{r}_k / m_n$   \\
Momentum &$\vect{\hat{p}}_{k,j}$ & $m_e \vect{\hat{P}}_k/M + \vect{\hat{p}}_k$ & $m_n \vect{\hat{P}}_k/M - \vect{\hat{p}}_k$ \\
Spin &$\vect{\hat{s}}_{k,j}$ & $m_e \vect{\hat{S}}_k/M + \vect{\hat{s}}_k$ & $m_n \vect{\hat{S}}_k/M - \vect{\hat{s}}_k$ \\
\end{tabular}
\end{ruledtabular}
\end{table}
These coordinates are defined through the total and reduced mass $M=m_e + m_n$ and $m_\text{r} = m_e m_n/M$ as well as the total spin $\vect{\hat{S}}_k$ and the relative spin $\vect{\hat{s}}_k$.
Changing the coordinates leaves the integration measure invariant and we replace $\dd[6]{x_k} \to \dd[6]{\mathcal{R}_k} =  \dd[3]{R_k} \, \dd[3]{r_k} $ in $\hat{H}_\text{Cb}$ from Eq.~\eqref{eq:Ham_cb} together with single-particle coordinates in $\hat{h}_\text{Cb}$ and $\hat{V}_\text{scatt}$ according to the transformation specified in Table~\ref{tab:NRcm}.
Note that the field operators $\hat{\varphi}_k = \hat{\varphi} ( \vect{R}_k- m_\text{r} \vect{r}_k /m_n, \vect{R}_k + m_\text{r} \vect{r}_k/m_e )$ have thus become a function of c.m. and relative coordinates as well.

\subsection{Relativistic corrections to c.m. and rel. coordinates}
Our description contains relativistic corrections up to the order $c^{-2}$. 
However, the transformation to NR c.m. and relative coordinates from Table~\ref{tab:NRcm} is inconsistent to this order and has to be modified~\cite{Krajcik1970,Close1970, Krajcik1974, Liou1974}.
These corrections can be implemented via a first-quantized unitary transformation~\cite{Krajcik1974,Liou1974}, which circumvents issues regarding the integration measure and the transformation of certain terms in the scattering potential that arise with an actual coordinate transformation.
Such a first-quantized unitary~\cite{Krajcik1974} is generated by
\begin{align} \label{eq:relgen}
    \hat{\lambda}_k^{(\text{rel})} =& \frac{ \vect{r}_k \cdot \bars{P}_k }{4M^2c^2} \left[ \bars{p}_k \cdot \bars{P}_k + \Delta m \left(  \frac{ \bars{p}_k^{2}}{m_{\text{r}}}  + \frac{q_eq_n}{4 \pi \varepsilon_0 \abs{\vect{r}_k}} \right) \right] + \text{H.c.} \notag \\
    &- \frac{1}{4m_{\text{r}} Mc^2} \left(  \bars{p}_k \times \bars{P}_k + \text{H.c.}  \right)  \cdot\vect{\hat{s}}_k.
\end{align}
The Coulomb-potential term proportional to the mass difference $\Delta m = m_n - m_e$ arises due to the internal EM interactions. 
In addition, single-particle masses are contained in the total mass $M=m_e+m_n$ and the reduced mass $m_\text{r}= m_e m_n /M$, and the Hermitian conjugate $\text{H.c.}$ ensures hermiticity of the generator.
We also account for $c^{-2}$ corrections to light-matter interactions by using gauge-invariant minimally coupled momenta~\cite{Krajcik1970} $\, \bars{P}_k = \bars{p}_{k,e} + \bars{p}_{k,n} $ and $ \bars{p}_k = (m_n \, \bars{p}_{k,e} - m_e \bars{p}_{k,n} ) /M$ instead of purely kinetic momentum operators.
To apply this first-quantized transformation to a second-quantized theory, we need to confirm that $[ \hat{\Lambda}_\text{rel}, \hat{h}_\text{Cb} ] = [ \hat{\Lambda}_\text{rel}, \hat{V}_\text{scatt} ]=0$, where $\hat{\Lambda}_\text{rel}$ and $\hat{\lambda}_k^{(\text{rel})}$ are connected through Eq.~\eqref{eq:Lambda}.
Since $\hat{\Lambda}_\text{rel}$ contains an integration over coordinates that are independent of $\hat{h}_\text{Cb}$ and $\hat{V}_\text{scatt}$, the cobosonic operators trivially commute.
While the vector potential commutes in Coulomb gauge with itself, its commutator with the electric fields in $\hat{h}_\text{Cb}$ (spin-orbit term) is nonvanishing.
The resulting additional terms from this commutator, however, are yet ruled out by the limits of integration in $\hat{H}_\text{Cb}$.
This fact is again a consequence of the projection, ensuring that a coboson can contain only one nucleus and electron, and may be made explicit by introducing $\hat{\varphi} = \hat{\pi}_\text{Cb} \hat{\varphi}$, similar to before.
Thus, the transformation reduces to Eq.~\eqref{eq:transformation}.

\subsection{Power-Zienau-Woolley transformation}
We now introduce the interaction of light with matter through electric and magnetic fields  $\vect{\hat{E}}$ and  $\vect{\hat{B}}$ rather than through the vector potential, \ie{} we move to \textit{multipolar cobosonic quantum field theory}.
This transition follows from applying the unitary \textit{Power-Zienau-Woolley} (PZW) transformation~\cite{Power1959, Woolley1971, Woolley2020} defined by its first-quantized generator  $\hat{\lambda}_k^{(\text{PZW})} = \int \dd[3]{y} \vect{\mathcal{P}}_k \left( \vect{y} \right) \cdot \vect{\hat{A}} \left(\vect{y} \right)$ through the polarization field~\cite{Baxter1993}
\begin{align} \label{eq:PZW}
     \vect{\mathcal{P}}_k \left( \vect{y} \right) = \sum_{i} q_i \left(\vect{x}_{k,i} - \vect{R}_k \right) \int \limits_{0}^{1} \dd{\rho} \delta[ \vect{y} - \vect{R}_k - \rho ( \vect{x}_{k,i} - \vect{R}_k )]
\end{align}
of the $k$th coboson.
Here, we choose the coordinate $\vect{R}_k$, which could be arbitrary, in general, to coincide with the c.m. position.
Based on the discussion above, it can be shown that this generator meets the requirements to reduce the second-quantized transformation to the first-quantized one as well.

\subsection{Transformation sequence}
Finally, the total transformation sequence addresses first the generation of relativistic corrections to NR c.m. and relative coordinates, followed by the PZW transformation.
This particular order of transformations is crucial to remain gauge invariant.
The transformations of Eq.~\eqref{eq:Ham_cb} leads to the replacements
\begin{subequations}
\label{eq:trafos}
\begin{align} \label{eq:trafoa}
    \hat{h}_\text{Cb} &\to \hat{u}_1^{(\text{PZW})\, \dagger} \hat{u}_1^{(\text{rel}) \, \dagger} \hat{h}_\text{Cb} \hat{u}_1^{(\text{rel})} \hat{u}_1^{(\text{PZW})} \\ \label{eq:trafob}
    \hat{V}_\text{scatt} &\to \hat{u}_{12}^{(\text{PZW})\, \dagger}  \hat{u}_{12}^{(\text{rel}) \, \dagger} \hat{V}_\text{scatt} \hat{u}_{12}^{(\text{rel})} \hat{u}_{12}^{(\text{PZW})} \\ \label{eq:trafoc}
    \hat{H}_\text{EM} &\to \hat{U}_\text{PZW}^\dagger \hat{U}_\text{rel}^\dagger \hat{H}_\text{EM} \hat{U}_\text{rel} \hat{U}_\text{PZW}
\end{align}
\end{subequations}
where $\hat{u}_{12} = \hat{u}_1 \hat{u}_2$ combines the transformation of both coordinate sets.
For a more detailed discussion on the above transformations, we refer to Appendix~\ref{App:D} and present the major results in the next section.

\section{Multipolar cobosonic subspace} \label{Sec:MpCbQFT}
Together with the transformations from the previous section, our multipolar cobosonic subspace is formulated with respect to c.m. and relative coordinates, while every scale is corrected in $c^{-2}$.
In particular, the theory includes corrections to internal dynamics encoded in $\hat{h}_\text{Cb}$, intercobosonic dynamics in $\hat{V}_\text{scatt}$, and light-matter interactions contained in both terms.
Consequently, the multipolar cobosonic Hamiltonian
\begin{align} \label{eq:Ham_fin}
    \hat{H}_{\text{MpCb}} =&  \hat{H}_\text{EM} + \int \limits_{C} \dd[6]{\mathcal{R}} \hat{\varphi}^\dagger \hat{h}_{\text{MpCb}}  \hat{\varphi} \notag \\
    &+\int \limits_{C_1} \dd[6]{\mathcal{R}_1} \int \limits_{C_2} \dd[6]{\mathcal{R}_2} \hat{\varphi}_1^\dagger \hat{\varphi}_2^\dagger  \hat{\mathcal{V}}_{\text{scatt}} \hat{\varphi}_2 \hat{\varphi}_1 
\end{align}
accounts via $\hat{h}_\text{MpCb}$ for the single-coboson energy, while the scattering potential is accounted for by $\hat{\mathcal{V}}_\text{scatt}$.
In the former, we omit the subscript ``1'' for simplicity and, therefore, whenever there are no interactions between more than one coboson in the following.

\subsection{Single-coboson Hamiltonian}
The explicit form of the single-coboson Hamiltonian
\begin{align}
\begin{split} \label{Eq:Ham_mpcb}
   \hat{h}_{\text{MpCb}} =& Mc^2 + \hat{h}^{(0)}_{\text{rel}} + \hat{h}^{(1)}_{\text{rel}} + \frac{\vect{\hat{P}}_Q^2}{2M} \left( \mathds{1} - \frac{\hat{h}^{(0)}_{\text{rel}}}{Mc^2} \right)- \frac{\vect{\hat{P}}_Q^4}{8M^3c^2}  \\
   &+ \hat{h}_{\text{I}}^{(0)} + \hat{h}_{\text{I}}^{(1)}
\end{split}
\end{align}
consists of the internal structure $\hat{h}^{(0)}_{\text{rel}} + \hat{h}^{(1)}_{\text{rel}}$, c.m. kinetic terms proportional to the minimally coupled momentum $\vect{\hat{P}}_Q^2$, as well as the light-matter interaction $\hat{h}_{\text{I}}^{(0)} + \hat{h}_{\text{I}}^{(1)}$.
The kinetic term couples to the internal structure as a consequence of the mass defect~\cite{Sonnleitner2018,Schwartz2019,Yudin2018}.
The rest energy $Mc^2$ is modified by the relative motion 
\begin{equation}
    \hat{h}^{(0)}_{\text{rel}}= \frac{\vect{\hat{p}}^2}{2m_{\text{r}}} +\frac{ q_e q_n }{4 \pi \varepsilon_0 \abs{\vect{r}}}
\end{equation}
solely given by a hydrogen-type Hamiltonian. 
The next-order correction is given by
\begin{equation} 
\begin{split} 
    \hat{h}^{(1)}_{\text{rel}} =& -\frac{\vect{\hat{p}}^4}{8m_{\text{r}}^3c^2} \frac{m_e^3 + m_n^3}{M^3} - \frac{\kappa}{\abs{\vect{r}}^3} \left( \frac{1}{2} \vect{\hat{\ell}}^2 + \left( \vect{r} \cdot \vect{\hat{p}} \right)^2  \right) \\
    &+ \kappa \alpha_\text{D} \pi \hbar^2 \delta(\vect{r} ) + \kappa \frac{ \alpha_\mathrm{\ell S}}{\abs{\vect{r}}^3} \vect{\hat{\ell}} \cdot \vect{\hat{S}} + \kappa \frac{\alpha_\mathrm{\ell s}}{\abs{\vect{r}}^3} \vect{\hat{\ell}} \cdot \vect{\hat{s}} \\
    &+\kappa \alpha_\mathrm{ss} \pi  \delta(\vect{r}) \vect{\hat{s}}_n \cdot \vect{\hat{s}}_e   + \kappa \frac{ \overline{c}_{\text{F}}^{(n)} \overline{c}_{\text{F}}^{(e)}}{\abs{\vect{r}}^3} \hat{S}_{ne}.
    \end{split}
    \label{eq:Ham_int1}
\end{equation} 
Similar to Sec.~\ref{Sec:pNRQED}, we defined $\kappa=2 \kappa_{ne} = -q_e q_n/(4 \pi \varepsilon_0 m_\text{r} M c^2)$ and the abbreviation $\alpha_v$ summarizes all Wilson coefficients in Table~\ref{tab:Wilson}.
\begin{table}[h]
\caption{\label{tab:Wilson}%
Low-energy effective Wilson coefficients $\alpha_v$ that contribute to the relativistic corrections of the  relative motion, \ie{} to the internal structure. They can be connected to the high-energy Wilson coefficients $\overline{c}_{\text{D}}^{(j)}, \overline{c}_{\text{F}}^{(j)}$, and $\overline{c}_\text{S}^{(j)} $ that correspond to the Darwin, Fermi, and Seagull coefficients of fermion $j=e,n$ as well as to the Wilson coefficients $d_1^{(ij)}$ and $ d_2^{(ij)}$ for $j\neq i$ that originate in the contact interaction.
}
\begin{ruledtabular}
\begin{tabular}{llcc}
&$v$ & $\alpha_v$ \\
\addlinespace\colrule \addlinespace
Darwin &$\text{D}$ &  $ \frac{m_n^2 \overline{c}_{\text{D}}^{(e)}+ m_e^2 \overline{c}_{\text{D}}^{(n)}}{2m_{\text{r}} M} + \frac{d_1^{(en)} + d_1^{(ne)} }{\pi Z \alpha}$  \\ \addlinespace
(Total spin)-orbit &$\mathrm{\ell S}$ & $ \frac{m_e \overline{c}_{\text{F}}^{(e)} + m_n \overline{c}_{\text{F}}^{(n)}}{M} +  \frac{  m_e \overline{c}_{\text{S}}^{(n)} + m_n \overline{c}_{\text{S}}^{(e)}}{2M}$ \\ \addlinespace
(Relative spin)-orbit&$\mathrm{\ell s}$ & $ \overline{c}_\text{F}^{(e)} - \overline{c}_\text{F}^{(n)} + \frac{\overline{c}_\text{S}^{(e)} m_n^2 - \overline{c}_\text{S}^{(n)} m_e^2}{2m_\text{r} M}  $  \\ \addlinespace
Spin-spin & $\mathrm{ss}$ & $  \frac{8}{3} \overline{c}_{\text{F}}^{(n)} \overline{c}_\text{F}^{(e)} - 4  \frac{d_2^{(en)}+d_2^{(ne)}}{ \pi Z \alpha} $  \\ \addlinespace
\end{tabular}
\end{ruledtabular}
\end{table}
These corrections~\cite{Breit1929,Breit1930,Breit1932} give rise to the fine- and hyperfine structure of hydrogenlike atoms and correspond in the same order to the kinetic relative correction, orbit-orbit coupling, Darwin term, spin-orbit coupling of angular momentum $\hat{\vect{\ell}} = \vect{r} \times \vect{\hat{p}}$ to total and relative spin, spin-spin contact coupling, as well as the magnetic dipole-dipole interaction $\hat{S}_{ne} = - \hat{\vect{s}}_n \cdot \hat{\vect{s}}_e + 3 ( \vect{\hat{s}}_n \cdot \vect{r} ) ( \vect{\hat{s}}_e \cdot \vect{r} ) / \abs{\vect{r}}^2 $.
Note that $\vect{\hat{s}}_n$ and $\vect{\hat{s}}_e$ refer to the spin of nucleus and electron and may be expressed through their corresponding total and relative spin.
Here, we reproduce results known from the literature~\cite{Pineda1998b} that are augmented by particle-species dependent Wilson coefficients $\overline{c}_\text{F}^{(i)}$ and $\overline{c}_\text{S}^{(i)}$.

In addition to the internal structure, our results include c.m. degrees of freedom.
The c.m. kinetic energy appears as dominating, lowest-order contribution but is modified by a correction proportional to the relative Hamiltonian $\hat{h}_\text{rel}^{(0)}$ as a consequence of the mass defect~\cite{Sonnleitner2018,Schwartz2019,Yudin2018}, where relative and c.m. degrees of freedom couple to each other.
Relativistic corrections to the relative Hamiltonian $\hat{h}_\text{rel}^{(1)}$ are not included in the mass-defect term for consistency, since these couplings are of order $c^{-4}$.
The mass defect implies an internal-state-dependent c.m. motion that can be identified with a state-dependent mass~\cite{Zych2011,Pikovski2017,Sonnleitner2018,Schwartz2019,Loriani2019,Ufrecht2020,Roura2020,DiPumpo2021,DiPumpo2022,DiPumpo2023}, which we show in detail later.

The fact that our description allows also for charged cobosons (ions) manifests in a monopole coupling where the total charge $Q=q_n + q_e$ and the vector potential evaluated at the c.m. position $\vect{\hat{A}}(\vect{R})$ appear in minimally coupled momenta $\hat{\vect{P}}_Q = \vect{\hat{P}} - Q \vect{\hat{A}} ( \vect{R} )$.
The kinetic c.m. degrees of freedom are completed by the c.m. relativistic kinetic correction proportional to $\hat{\vect{P}}_Q^4$, analogously~\cite{Sonnleitner2018} to the case of neutral atoms with $Q=0$.

Further, external EM fields interact with the coboson in leading order via
\begin{align} \label{eq:externalInteraction}
    \hat{h}_\text{I}^{(0)} = \hat{h}_\text{ME} + \hat{h}_\text{MM} + \hat{h}_\text{R} + \hat{h}_\text{dia} + \hat{h}_\text{self}.
\end{align}
The explicit form of the components are summarized in Table~\ref{tab:Ham_I}, where all EM fields depend on the integration variable $\vect{y}$ if not stated otherwise.
The polarization $\vect{\mathcal{P}}$ and the magnetization $\vect{\hat{M}}$ [see Eq.~\eqref{eq:mag}] depend on $\vect{y}$, as well as on coboson coordinates $\vect{x}_e$ and $\vect{x}_n$ that have to be expressed by c.m. and relative coordinates.
\begin{table}[h]
\caption{\label{tab:Ham_I}%
Contributions to the leading-order light-matter interaction $\hat{h}_\text{I}^{(0)}$.
The individual terms $\hat{h}_v$ describe a coupling of generalized electric (ME) and magnetic (MM) moments as well as the Röntgen (R) and diamagnetic (dia) interaction together with the self-energy (self).
They describe a coupling of the polarization field $\vect{\mathcal{P}}$, the magnetization $ \vect{\hat{M}}$, magnetic moments $\vect{\hat{\mu}}_i$ to c.m. momenta $\vect{\hat{P}}_Q $, magnetic fields $\vect{\hat{B}} $, and the transverse electric field $\vect{\hat{E}}^{\perp}$.   
}
\begin{ruledtabular}
\begin{tabular}{lcc}
$v$ & $\hat{h}_v$ \\
\addlinespace\colrule \addlinespace
$\text{ME}$ &  $ - \int \dd[3]{y} \vect{\mathcal{P}}^{\perp} \cdot \vect{\hat{E}}^{\perp}$  \\ \addlinespace \addlinespace
$\text{MM}$ & $ - \frac{1}{2} \int \dd[3]{y} \left( \vect{\hat{M}} \cdot \vect{\hat{B}} + \text{H.c.} \right) - \sum_{i=e,n} \vect{\hat{\mu}}_i \cdot \vect{\hat{B}} \left( \vect{x}_i \right)$ \\ \addlinespace
$\text{R}$ & $ \frac{1}{2M} \left\{\vect{\hat{P}}_Q , \int \dd[3]{y} \vect{\mathcal{P}} \times \vect{\hat{B}} \right\}   $  \\ \addlinespace
$\text{dia}$ & $\frac{ \left( \int \dd[3]{y} \vect{\mathcal{P}} \times \vect{\hat{B}} \right)^2 }{2M} + \frac{ \Bigg[  \sum_i q_i \frac{m_\text{r}^2}{m_i^2} \vect{r} \times \int \limits_0^1 \dd{\rho} \rho \vect{\hat{B}} \left( \vect{R} - \rho \frac{q_i}{\abs{q_i}} \frac{m_\text{r}}{m_i} \vect{r} \right) \Bigg]^2}{2m_\text{r}}$  \\ \addlinespace
$\text{self}$ & $ \frac{1}{2\varepsilon_0} \int \dd[3]{y} \left( \vect{\mathcal{P}}^\perp \right)^2$  \\ \addlinespace
\end{tabular}
\end{ruledtabular}
\end{table}

The term $\hat{h}_\text{ME}$ couples the transverse electric field $\vect{\hat{E}}^{\perp}$ to the transverse part of the polarization field from Eq.~\eqref{eq:PZW}, giving rise to generalized electric moments (MEs).
For instance, a multipole expansion of the polarization field in $\vect{R}$ implying small relative coordinates, we find in lowest order the dipole moment $\vect{d} = m_\text{r} ( q_e /m_e - q_n / m_n ) \vect{r}$.

Similarly, the magnetic field couples to magnetic moments (MMs) in $\hat{h}_\text{MM}$ and has two contributions:
The particles' spin, \ie{} the magnetic moment $\vect{\hat{\mu}}_i = \overline{c}_\text{F}^{(i)} q_i  \vect{\hat{s}}_i / m_i $, couples to the magnetic field $\vect{\hat{B}} ( \vect{x}_i )$, where the single-fermion coordinates have to be replaced by c.m. and relative coordinates $\vect{x}_i=\vect{R} - \frac{q_i}{\abs{q_i}} \frac{m_{\text{r}}}{m_i} \vect{r}$.
Moreover, constituents of composite particles carry orbital angular momentum $\vect{\hat{\ell}}$ that induces an orbital magnetic moment contained in the quantum magnetization
\begin{align} \label{eq:mag}
    \vect{\hat{M}} ( \vect{y} )  = \sum_{i} \frac{m_{\text{r}}}{m_i} q_i \frac{\hat{\vect{\ell}}}{m_i} \int \limits_0^1 \dd{\rho} \rho \delta [ \vect{y} - \vect{R} - \rho \left( \vect{x}_i - \vect{R} \right) ]
\end{align}
similar to the relation between polarization fields and electric moments.
A multipole expansion of the magnetization and the magnetic field leads in lowest order to the magnetic moment of a coboson $\vect{\hat{\mu}}_\ell + \vect{\hat{\mu}}_n + \vect{\hat{\mu}}_e$, with $\vect{\hat{\mu}}_\ell = m_\text{r} ( q_e / m_e^2 + q_n /m_n^2) \vect{\hat{\ell}}/2$, \ie{} the sum of orbital, electron, and nucleus spin magnetic moments giving rise to the Zeeman shift~\cite{Cohen1986}.

In addition, the c.m. motion of the coboson also yields the c.m. Röntgen Hamiltonian $\hat{h}_\text{R}$~\cite{Wilkens1993, Wilkens1994, Boussiakou2002, Sonnleitner2017, Lopp2021}.
Further, we find that the diamagnetic interaction $\hat{h}_\text{dia}$ with c.m. and relative contribution corresponds to an induced magnetic moment due to the external fields, being part of the quadratic Zeeman effect~\cite{Schiff1939}.
Moreover, the cobosonic self-energy $\hat{h}_\text{self}$ is generally divergent, but can be renormalized~\cite{Vulkics2015} and contributes to the Lamb shift~\cite{Power1959}.

The last contributions to the single-coboson Hamiltonian are relativistic corrections to the light-matter interaction and, in general, depend on the electric or the magnetic field.
In many applications light-matter interactions are dominated by electric fields.
Here, we present these dominant electric terms and  suppress the influence of magnetic fields in $c^{-2}$. 
The full Hamiltonian including magnetic field contributions is given in Appendix~\ref{App:D}, while the electric field contribution resolves to
\begin{align}
\begin{split}
    \hat{h}_\text{I}^{(1)} =&\sum_i \overline{c}_\text{S}^{(i)} q_i \vect{\hat{s}}_i \cdot \frac{\left(\frac{m_i}{M}\vect{\hat{P}}_Q-\frac{q_i}{\abs{q_i}} \vect{\hat{p}}\right) \times \vect{\hat{E}} + \text{H.c.}}{8 m_i^2 c^2} \\
    &+\frac{1}{2} \sum_i \left[ \vect{\hat{E}}^\perp ( \vect{x}_i ) \cdot \vect{\hat{d}}_i^{(1)}+ \vect{\hat{d}}_i^{(1)} \cdot \vect{\hat{E}}^\perp ( \vect{x}_i) \right].
\end{split}
\end{align}
The second line originates from relativistic corrections to c.m. and relative coordinates and describes the coupling of the transverse electric field $\vect{\hat{E}}^\perp$ to dipole-moment corrections
\begin{align} \notag
    \frac{\hat{\vect{d}}_i^{(1)}}{q_i} =& \frac{\vect{r} }{4M^2c^2} \left( \vect{\hat{p}} \cdot \vect{\hat{P}}_Q + \frac{\Delta m}{m_\text{r}} \vect{\hat{p}}^2 + \frac{\Delta m q_e q_n}{4 \pi \varepsilon_0 \abs{\vect{r}}} \right) + \text{H.c.} \\ \notag
    &+ \frac{\vect{r} \cdot \vect{\hat{P}}_Q}{4M^2c^2} \left[ \left( 1- 2 \frac{\Delta m }{m_i} \frac{q_i}{\abs{q_i}} \right) \vect{\hat{p}} - \frac{m_\text{r}}{m_i} \frac{q_i}{\abs{q_i}} \vect{\hat{P}}_Q \right] + \text{H.c.}  \\ \label{eq:Rel_Ecorr}
    &+ \frac{1}{2m_\text{r} M c^2} \left( \frac{m_\text{r}}{m_i} \frac{q_i}{\abs{q_i}} \vect{\hat{P}}_Q + \vect{\hat{p}} \right) \times \vect{\hat{s}}
\end{align}
in accordance~\cite{Krajcik1970} with the limiting case of $Q=0$ and arbitrary loosely bound cobosons.

\subsection{Scattering potential}
The scattering potential has the general form 
\begin{align}
    \begin{split} \label{eq:Scatteringxx}
        \hat{\mathcal{V}}_{\text{scatt}} =& \sum_{i,j} \left[\hat{\mathcal{V}}_{\text{C}}^{(ij)}+ \hat{\mathcal{V}}_{\text{LL}}^{(ij)} + \hat{\mathcal{V}}_{\text{LS}}^{(ij)} + \hat{\mathcal{V}}_{\text{SS}}^{(ij)} \right] + \hat{\mathcal{V}}_\text{self},
    \end{split}
\end{align}
where the components are summarized in Table~\ref{tab:Scattering}.
For simplicity, we include only the most dominant c.m. contribution to the scattering for all $c^{-2}$ terms by omitting terms directly proportional to $\vect{r}_i$, while keeping the general distance between two different constituents $\vect{\chi}_{ij} = \vect{x}_{1,i} - \vect{x}_{2,j}$ and neglecting all terms proportional to the relative momentum $\vect{\hat{p}}_i$.
Besides, we also exclude the influence of light-matter interaction in the scattering processes presented in the table.
The full scattering potential including light-matter interactions and relative contributions is given in Appendix~\ref{App:D}. 
\begin{table}[h]
\caption{\label{tab:Scattering}%
 Contributions to the scattering potential $\hat{\mathcal{V}}_{\text{scatt}}$.
 The individual terms $\hat{\mathcal{V}}_{v}^{(ij)}$ mediate the interaction between two fermions $i,j$ of different cobosons and include a Coulomb (C) interaction, a coupling of orbital magnetic moments through angular momenta including a retardation correction (LL), a spin-orbit interaction (LS), a magnetic dipole potential (SS) and the self-energy (self).
}
\begin{ruledtabular}
\begin{tabular}{lcc}
$v$ & $\hat{\mathcal{V}}_{v}^{(ij)}$ \\
\addlinespace \colrule \addlinespace
C & $\frac{q_iq_j}{8\pi \varepsilon_0 \abs{\vect{\chi}_{ij} }} + \frac{\Delta m }{M} \frac{\vect{e}_{\vect{\chi}_{ij}} \cdot ( \vect{e}_{r_1} - \vect{e}_{r_2} )}{Mc^2} \frac{q_e q_n}{q_i q_j } \left( \frac{q_i q_j}{8 \pi \varepsilon_0 \abs{\vect{\chi}_{ij}}} \right)^2$ \\ \addlinespace
LL & $  -\frac{\mu_0 q_i q_j}{8\pi  M^2} \frac{1}{\abs{\vect{\chi}_{ij}}^3} \left( \frac{1}{2} \vect{\hat{L}}^{(ij)}_1 \cdot \vect{\hat{L}}_2^{(ij)} + \left( \vect{\chi}_{ij} \cdot \vect{\hat{P}}_1\right) \left( \vect{\chi}_{ij} \cdot \vect{\hat{P}}_2\right)  \right) $  \\ \addlinespace
\multirow[b]{2}{*}{LS} & $ -\frac{\mu_0}{8\pi \abs{\vect{\chi}_{ij}}^3} \Big[ \frac{q_i}{M} \left( \vect{\hat{L}}^{(ij)}_1 - \vect{\hat{L}}_2^{(ij)} \right) \left( \frac{q_j}{q_i} \vect{\hat{\mu}}_{1,i} + \vect{\hat{\mu}}_{2,j} \right) \hspace{1cm} $  \\ \addlinespace
& $\hspace{1cm} -\frac{1}{2} \frac{q_j q_i}{M m_i} \vect{\hat{L}}_1^{(ij)} \cdot \vect{\hat{s}}_{1,i} + \frac{1}{2} \frac{q_i q_j}{M m_j} \vect{\hat{L}}_2^{(ij)} \cdot \vect{\hat{s}}_{2,j} \Big]$ \\ \addlinespace
$\text{SS}$ &  $ -\frac{\mu_0}{8 \pi \abs{\vect{\chi}_{ij}}^3} \left[ \vect{\hat{\mu}}_{1,i} \cdot \vect{\hat{\mu}}_{2,j} -3 \left( \vect{\hat{\mu}}_{1,i} \cdot \vect{e}_{\vect{\chi}_{ij}} \right) \left(\vect{\hat{\mu}}_{2,j} \cdot \vect{e}_{\vect{\chi}_{ij}} \right)    \right] $  \\ \addlinespace
$\text{self}$ & $ \frac{1}{2\varepsilon_0} \int \dd[3]{y}  \vect{\mathcal{P}}^\perp_1 \cdot \vect{\mathcal{P}}^\perp_2 $  \\ \addlinespace
\end{tabular}
\end{ruledtabular}
\end{table}
As expected, the leading-order contribution of scattering is the Coulomb (C) interaction between fermion $i$ of coboson 1 and fermion $j$ of coboson 2.
However, we find a second Coulomb-like correction, including the unit vector $\vect{e}_{\vect{w}}$ in $\vect{w}$ direction, where $\vect{w}$ is either the distance $\vect{\chi}_{ij}$ between constituents of different cobosons or relative distances between each coboson's constituents $\vect{r}_1$ and $\vect{r}_2$.
Corrections to the Coulomb potential are interactions between all possible magnetic moments among different cobosons. 
Consequently, in $\hat{V}_\text{LL}^{(ij)}$ we find orbital magnetic moments coupling to each other through orbital angular momenta $\vect{\hat{L}}^{(ij)}_k= \vect{\chi}_{ij} \times \vect{\hat{P}}_k$. 
Moreover, we find also in the scattering potential an additional term containing momentum operators that corresponds to the so-called \emph{retardation correction}~\cite{Breit1929}.
The spin-orbit (LS) interaction includes an interaction of an effective orbital magnetic moment proportional to $\vect{\hat{L}}^{(ij)}_1 - \vect{\hat{L}}_2^{(ij)}$ with an effective spin magnetic moment $q_j \vect{\hat{\mu}}_{1,i}/q_i + \vect{\hat{\mu}}_{2,j}$, complemented by pure spin-orbit coupling.
These potentials are completed by the known~\cite{Goral2000,Olson2013} magnetic dipole-dipole potential $\hat{\mathcal{V}}_\text{SS}^{(ij)}$.

Moreover, one effect of the PZW transformation becomes apparent only in second quantization, which is the scattering self-energy [$\hat{\mathcal{V}}_\text{self}$, Eq.~\eqref{eq:Scatteringxx}], arising analogously (and additionally) to the cobosonic self-energy [$\hat{h}_\text{self}$ in Eq.~\eqref{eq:externalInteraction}] in the single-coboson sector.
Because of the multicoboson nature, this contribution may be associated as one part of the collective (or cooperative) Lamb shift~\cite{Friedberg1973,Scully2009}.

With these results, we have introduced a multipolar cobosonic subspace. 
The single-coboson Hamiltonian includes $c^{-2}$ corrections for the internal structure, the mass defect in the c.m. motion, but also for the light-matter interactions beyond a multipole expansion.
We derived scattering potentials based on lowest-order two-particle scattering, yielding the Coulomb potential with corrections in form of interactions between magnetic moments.

\section{Discussion} \label{Sec:Discussion}
In the following section, we identify different physical systems and issues that can be described or addressed by our multipolar cobosonic subspace, such as models for atomic systems, bound-state energies, the scattering between atoms, as well as ultracold quantum gases.
The examples demonstrate that our theory contributes to the understanding of different subfields, complementing and connecting existing approaches.

\subsection{Models for atomic systems}
First-quantized Lagrangian or Hamiltonian treatments of atomic dynamics restricted to single-particle systems have been studied extensively~\cite{Cohen1986}. 
There are relativistic treatments, \eg{} extending the NR Schrödinger equation for a hydrogen atom to relativistic equations~\cite{Barut1973} or formulating equations of motion for c.m. coordinates of a system of relativistic Dirac particles, which allows for a description of relativistic bound-state systems~\cite{Barut1986}. 
However, the dynamics of atomic ensembles are often studied in NR regimes since atomic quantum gases are mostly restricted to low-energy scales.
One accurate model of bound-state particles that includes relativistic corrections follows from two coupled Dirac equations in the respective NR limit~\cite{Breit1929, Breit1930,Breit1932}, and is known as the Breit-Pauli Hamiltonian.
This Hamiltonian does not account for field-theoretical QED corrections and is derived with respect to single-fermion coordinates.
As a result, additional relativistic corrections enter the Hamiltonian~\cite{Krajcik1970, Close1970} once NR c.m. and relative coordinates are introduced to separate the inner-atomic structure from the c.m. motion.
In the simplified case where one ignores the spin of fermions, such a description of atoms gives rise to a coupling of c.m. to relative degrees of freedom as a consequence of the mass defect~\cite{Sonnleitner2018,Yudin2018, Schwartz2019, Martinez2022}.
In contrast to calculating relativistic corrections to the NR bound-state atoms and their dynamics, there are other models that focus on corrections arising from a finite extension of atoms via hard-sphere models, \ie{} atoms confined in spherical impenetrable boxes, and they have been studied for hydrogen~\cite{Degroot1946,Suryanarayana1976,LeyKoo1979,Marin1991,AquinoA.1995,Marin1995,Varshni1997,Aquino2007,Fernandez2010,Aquino2016}, hydrogenlike~\cite{Laughlin2002}, and many-electron atoms~\cite{Tenseldam1952,Jaskolski1996,Garza1998,Garza2005}.
In this case, different deviations from the standard NR treatment of hydrogenlike atoms than the relativistic ones follow.

Based on such first-quantized models, one usually postulates a corresponding effective field theory~\cite{Boussiakou2002b,Ueda2010}, and rigorous field-theoretical derivations for atoms are not addressed.
Conversely, our work embeds established first-quantized concepts mentioned above into a field-theoretical formulation.
As a result, the Breit-Pauli Hamiltonian is further modified by QED corrections.
Via our projection formalism, we naturally introduce length scales defining the extension of an atom, similar to hard-sphere impenetrable boxes.
Introducing c.m. and relative coordinates leads also to the mass defect, where we extend known derivations~\cite{Sonnleitner2018,Schwartz2019} to arbitrary numbers of spin carrying and charged cobosons in a field-theoretical framework, yet restricted to special relativity.

Models for atomic systems include not only isolated atoms but also the description of their interaction with external fields, \eg{} light and gravity.
In quantum and atom optics for instance, atoms are manipulated via the interaction with light, leading \eg{} to magneto-optical traps~\cite{Raab1987, Steane1991, Steane1992, Gajda1994} for neutral atoms as well as Paul~\cite{Paul1990,Brown1991,Baumann1992, Pedrosa2005} and Penning~\cite{Penning1936, Kretzschmar1991, Blaum2010, Vogel2018} traps for ions.
Instead of trapping atoms, light pulses~\cite{Giltner1995, Mueller2008, Altin2013}, in principle even entangled ones~\cite{Assmann2022}, or Bloch oscillations~\cite{Bloch1929, Wannier1960, BenDahan1996, Wilkinson1996,Clade2009, Pagel2020} are used to manipulate the atoms' momenta and might also induce transitions between internal states~\cite{Hu2017,Rudolph2020}.
In the context of cold atoms, magnetic fields give control over scattering dynamics between atoms via Feshbach resonances~\cite{Feshbach1958,Feshbach1962,Fano1961,Moerdijk1995,Chin2010} but are also crucial to implement, \eg{} $E1M1$~\cite{Labzowsky2005,Labzowsky2006, Janson2021} or magnetically induced single-photon~\cite{Hu2017,Bott2023} pulses.

In many applications~\cite{Jaynes1963, Schleich2013a} and field-theoretical treatments~\cite{Pineda1998b} it is sufficient to take only the lowest-order multipole expansion of the EM field into account.
Further contributions, such as higher-order multipole moments driving transitions~\cite{Seke1994} or respective energy shifts~\cite{Ludlow2015}, are then considered individually to the desired order~\cite{Lopp2021}.
In our present work, we use the generalized polarization-field approach~\cite{Baxter1993} for the PZW transformation~\cite{Power1959, Woolley1971, Woolley2020}.
Moreover, relativistic corrections to light-matter interaction on the level of elementary fermions are known most accurately in the field of NRQED~\cite{Hill2013}.
Once we move to the multipolar form defined with respect to NR c.m. and relative coordinates, additional relativistic corrections arise also for EM fields.
These have been studied for electric field contributions~\cite{Krajcik1970, Grotch1972} but magnetic field contributions have not been discussed explicitly yet.
Our treatment includes relativistic corrections to EM fields appearing in light-matter interactions, in particular, also magnetic field contributions.

Based on operational arguments, particle detectors, also known as Unruh-DeWitt detectors~\cite{Unruh1976,DeWitt1980}, have been postulated~\cite{Huemmer2016} as an alternative to microscopically modeling composite particles and their interaction with, \eg{} EM fields from first principles.
Usually, such models assume an effective first-quantized or two-level system as a detector, either with classical~\cite{Huemmer2016,Ng2018} or quantized~\cite{Sudhir2021,Wood2022,Gale2023} c.m. motion, interacting with an already suitably tailored external field, such as an EM field.
While useful, \eg{} for entanglement-related studies~\cite{Ng2018}, they do not include corrections~\cite{Lopp2021} accessible only by microscopic models like our multipolar cobosonic subspace, such as the full light-matter interaction or many-body effects of the detector itself, although there are attempts to generalize it to second-quantized formulations~\cite{Giacomini2022}.

The simple first-quantized model of atoms falling in gravitational potentials~\cite{Kajari2010}, as another external field, has been extended to a post-Newtonian description for an atomic Hamiltonian.
This description includes relativistic corrections associated with the coupling of gravity to atoms~\cite{Lammerzahl1995,Schwartz2019,Perche2021,Martinez2022} for single, spinless, and neutral atoms.
Some of these works derived the mass defect under gravity~\cite{Schwartz2019}, confirming original ideas for quantum clock interferometry~\cite{Zych2011,Sinha2011,Pikovski2017}.
The mass defect shows a connection to proper time associated with the c.m. of the atom and can be encoded by atomic~\cite{Martinez2022} and quantum~\cite{Loriani2019,Smith2020,DiPumpo2021} clocks in gravitational backgrounds.
Some theories even predict effective gravitational decoherence mechanisms~\cite{Pikovski2015}.  
In addition, quantum clock interferometry allows for tests of general relativity~\cite{Ufrecht2020,Roura2020,DiPumpo2021,DiPumpo2023}.
However, also in the gravity-free case, a coupling of internal degrees of freedom to the atomic c.m. via the mass defect leads to possible measurements of a quantum twin paradox~\cite{Zych2016,Bushev2016,Loriani2019} or allows for dark-matter detection~\cite{Derevianko2014,Arvanitaki2015,Arvanitaki2018,DiPumpo2022}.
With QFT on curved spacetime and a generalization of the respective coordinate transformations, an extension to gravitational fields seems in principle possible.

\subsection{Bound-state energies}
Most models of atoms focus on their internal structure, allowing for calculations of bound-state energies.
Consequently, in most accurate treatments, radiative QED corrections~\cite{Wichmann1956,Brown1959,Mohr1974, Jentschura1999} and effects from the composite-particle nature of the nucleus~\cite{Hill2011, Peset2021} enter bound-state energies~\cite{Pineda1998b} of composite particles.
These can be calculated with relativistic approaches~\cite{Hoyer2021}, \eg{} for hydrogenlike atoms with bound-state QED~\cite{Hoyer1984,Hoyer1989,Jaervinen2005,Lindgren2004} or with the Bethe-Salpeter equation~\cite{Salpeter1951, Salpeter1951a, Cutkosky1954, Nakanishi1969}.
Bound states and their properties for atoms have also been derived from field theory via flow equations and the functional renormalization group~\cite{Floerchinger2010,Alkhofer2019, Jakovac2019}.

In contrast to these fully relativistic treatments, relativistically corrected bound-state energies for hydrogenlike atoms can be obtained from EFTs in the NR regime, \eg{} NRQED~\cite{Kinoshita1996,Nio1997, Pachucki1997,Labelle1997,Czarnecki1999,Kniehl2000,Adkins2018, Haidar2020} or pNRQED~\cite{Pineda1998a, Pineda1998,Pineda1998b,Peset2017}.
These approaches exploit simplifications arising in inherent NR regimes while radiative corrections and effects from the nucleus are still taken into account via Wilson coefficients~\cite{Kinoshita1996, Hill2013}.
However, bound-state calculations within these EFTs are usually restricted to a single atom that is assumed to be trapped.
Consequently, the c.m. motion as well as its relativistic corrections and corrections to the relative coordinates become irrelevant~\cite{Krajcik1970,Close1970,Krajcik1974}.
Naturally, in the single-atom limit no atom-atom interactions occur and usually only basic light-matter interaction is considered, such as electric-dipole coupling for neutral atoms.
These approaches for calculating bound-state energies of trapped atoms may include contributions to the NR Lamb shift~\cite{Pineda1998a, jentschura2005,Peset2015,Peset2015a, Peset2017} but can also be used for fundamental tests~\cite{Lange2021}, the determination of the proton charge radius~\cite{Hill2011,Paz2012, Peset2015,Peset2015a, Peset2017}, and dark-matter searches~\cite{Banerjee2023}.
Since they typically focus only on the atomic spectrum, there are calculations, \eg{} for hydrogen~\cite{Haidar2020}, going beyond the precision of the internal energies derived in our work.
In particular, a pNRQED treatment~\cite{Pineda1998b} may include next-order loop corrections that are omitted in the present paper for simplicity but can be incorporated straightforwardly.
Moreover, these pNRQED derivations focus on positronium (equal-mass case of constituents) or neutral atoms, without taking c.m. degrees of freedom into account, while light-matter interactions enter solely through electric-dipole couplings.
As a result, light-induced internal energy shifts or shifts arising from the interaction with other atoms are not covered in these treatments.
Finally, in pNRQED Wilson coefficients have to be determined for each particular system, which, \eg{} has been carried out explicitly for positronium~\cite{Pineda1998b}, but also for hydrogenlike systems~\cite{Peset2015a}.

In the following applications, we will thus determine bound-state energies, \ie{} the QED-corrected hyperfine structure of hydrogenlike atoms, including parts of the NR Lamb shift~\cite{Lamb1947,Bethe1947,Power1959}, where we keep arbitrary Wilson coefficients such that our results remain valid for generic hydrogenlike atoms~\cite{Peset2015a}.
Because we also extend single-atomic considerations to an arbitrary number of atoms, scattering dynamics arise in addition to the usual pNRQED approaches.
Such two-body scattering dynamics for atoms were previously constructed and matched to pNRQED for the special case of van der Waals interactions~\cite{Brambilla2017}.

\subsection{Scattering between atoms}
Since we aim to describe ultracold quantum gases, these atom-atom interactions become highly relevant.
There are several theoretical models~\cite{Braaten2008} describing NR atomic scattering.
One possible description is based on interaction potentials between two scattering partners, where higher-order scattering events~\cite{Bedaque2000} are neglected.
The NR scattering of neutral atoms is then dominated by van der Waals interactions~\cite{London1937, Holstein2001,Beguin2013,Barcellona2016}.
In this context, theoretical models have been developed to determine van der Waals scattering potentials~\cite{Tang1984,Tang2003} and cover also density-functional-theory approaches~\cite{Kamiya2002, Kurita2001, Sun2008, Berland2015}.
Approximations to the van der Waals interaction are often performed according to an expansion of the form $-C_6/ \Delta R^6- C_7/ \Delta R^7 + ... $ with real constants $C_n$~\cite{Tao2012}, where $\Delta R$ is the distance between the c.m. of two atoms.
Hence, the long-range behavior may be observed in lowest order.
For example, the $C_6$ coefficient for hydrogen~\cite{Koga1985, Ocarroll1968} can be obtained by second-order perturbation theory~\cite{Sakurai2020} of the dipole-dipole potential~\cite{Parker1986} in first-quantized regimes~\cite{Holstein2001}.
Retardation effects may also be taken into account and correspond to the $C_7/\Delta R^7$ term~\cite{Holstein2001,Ishkhanyan2021}.
Another approximation of the van der Waals potential is the Lennard-Jones potential~\cite{LennardJones1931}.
In contrast to such first-quantized approaches, there are also EFTs~\cite{Brambilla2017, Odell2021} dealing with van der Waals interactions directly.
For the case of charged cobosons, ion-ion scattering~\cite{Kim1974,Kim1974b} is characterized by the Coulomb repulsion to lowest order.
We augment these existing approaches for neutral and charged cobosons by deriving relativistic corrections to the lowest-order Coulomb scattering potentials and cover the interactions between magnetic moments associated with orbit-orbit, spin-orbit, and spin-spin (magnetic dipole-dipole potential) interactions.
Spin-orbit and spin-spin magnetic moment interactions are known, \eg{} from magnetic scattering in the context of neutrons~\cite{Bloch1936, Bloch1937, Schwinger1937, Trammell1953, Franco2021}.
Since neutrons are free of charge, no Coulomb interaction is present and such interactions dominate the process.
Magnetic moments coupling in atomic scattering processes are partly discussed in the context of spinor BECs~\cite{Kawaguchi2012, StamperKurn2013}.
In addition to the Coulomb potential and its corrections, we find a scattering self-energy, that is part of the collective Lamb shift and was postulated before by embedding light-matter interaction into a field-theoretical framework~\cite{Boussiakou2002b}.

\subsection{Ultracold quantum gases} \label{Sec:Discussion_uqg}
The combination of bound-state energies with scattering dynamics together allows for a consistent treatment of ultracold quantum gases including their internal structure.
So far, the description of ultracold quantum gases often relies on bottom-up approaches for EFTs based on extensions of NR first-quantized theories.
Consequently, there are successful field-theoretical descriptions of scalar BECs~\cite{Bose1924,Einstein1924,Einstein1925,Ueda2010} as well as spinor BECs including internal states~\cite{Kawaguchi2012, StamperKurn2013}. 
Although their realization is challenging~\cite{Lukin2021,Christensen2021}, due to the Coulomb repulsion among charged bosons, also ionized BECs~\cite{Osborne1949, Foldy1961, Girardeu1962, Ninham1964, Wright1966} have been studied.
These descriptions usually do not address relativistic corrections, the inner-atomic structure is often of minor importance, and light-matter interaction is only partly accounted for.
Our basic assumption for cobosonic subspace EFT, to introduce different length scales, enters our description of an ultracold quantum gas in terms of hard-sphere atoms, and naturally the scattering dynamics in our model remain perturbations to the single-coboson contribution.

Moreover, this scattering dynamics is usually treated with approximations, leading to effective scattering lengths from $s$-wave scattering~\cite{Ueda2010} as well as introducing effective pseudopotentials for scattering from hard-sphere interactions~\cite{Huang1957,Lee1957} instead of the full scattering potential.
Within these approximations, we may derive the Gross-Pitaevskii equation (GPE)~\cite{Gross1961,Pitaevskii1961, Gross2004} that describes a Schrödinger-type equation complemented by a nonlinear collision term corresponding to the lowest-order effects of the condensate mean-field contribution~\cite{Proukakis2008,Ueda2010, Kawaguchi2012}.
Here, the field operator can be approximated by a wave function of the condensate by symmetry-breaking~\cite{Fetter1972} or number-conserving approaches~\cite{Koashi2000,Ueda2002}.
Higher-order corrections such as fluctuations~\cite{Proukakis2008, Salasnich2018} arising from the coupling of the condensate to a noncondensed thermal cloud may also be taken into account.
There are extensions to coupled GPEs, both for different modes~\cite{Ho1996, Pu1998} and quantized light fields~\cite{Boussiakou2000, Boussiakou2002b}, which are usually postulated extensions of first-quantized considerations.
Some studies~\cite{Anandan1981,Matos2011} generalize the GPE to a relativistic equation by postulating an invariant Klein-Gordon-type equation~\cite{Klein1926,Gordon1926,Peskin1995} to account for relativistic effects.
The modified GPE then follows in these approaches in the NR limit by separating a rest-energy phase from the condensate function~\cite{Gabel2019}, resulting into a relativistic correction proportional to a second derivative in time of the condensate function.
However, such a treatment does not include relativistic effects and the mass defect.
Consequently, as another application, we will derive a GPE including the mass defect, relativistic corrections, also for light-matter interactions, and a coupling of different internal states of the coboson.
This modified GPE differs significantly from previous Klein-Gordon-type derivations and might lead to fundamentally different predictions.
The deviation originates from the fact that atoms, as composite particles, are not fundamental bosons but rather cobosonic in their nature and, thus, they do not obey a Klein-Gordon equation describing spin-0 particles.

\section{Applications} \label{Sec:Applications}
Following the discussion above, we aim to derive the dynamics of interacting quantum gases and their internal structure encoded in the coboson field operator $\hat{\varphi}$.
This includes modified bound-state energies associated with the fine and hyperfine structure of the coboson and a coupling via the mass defect to its c.m. motion.
We determine the scattering potentials between two internal states of the coboson with respect to internal degrees of freedom giving access to generalized van der Waals potentials.
The mean-field contribution of the field operator gives rise to a GPE modified by relativistic corrections and the mass defect.

\subsection{Modes of relative motion}
In the spirit of composite particles, we introduced c.m. and relative coordinates for the multipolar cobosonic subspace.
As a next step, we explicitly describe the equation of motion of cobosons and separate between the c.m. and modes for the relative motion between constituents.
In contrast to the Schrödinger equation, where the complete time dependence enters through the state vector $\ket{\Psi(t)}$ while all field operators $\hat{\varphi}$, $\vect{\hat{A}}$, $\vect{\hat{E}}$, $\vect{\hat{B}}$ are explicitly time independent, $\hat{\varphi}$ and all other field operator become time dependent once we consider the equation of motion, effectively corresponding to a change into the Heisenberg picture.

\subsubsection{Cobosonic equation of motion}
First, we derive the equation of motion for the cobosonic field operator $\hat{\varphi}$ based on the Heisenberg equation $\ii \hbar \text{d} \hat{\varphi} / \text{d} t = [ \hat{\varphi}, \hat{H}_\text{MpCb} ]$, neglecting the influence of the environment that lies outside of our cobosonic subspace.
We recall that the equation of motion follows from the cobosonic commutation relation, generating additional terms compared to a purely bosonic field operator.
However, these additional terms correspond to processes that lie outside of the projected Hilbert space, such as the annihilation of an electron and a nucleus of different cobosons.
To derive the effective equation of motion, we rely on the projected equation of motion $\ii \hbar \hat{\pi}_\text{Cb} \text{d} \hat{\varphi} / \text{d} t \hat{\pi}_\text{Cb}  = \hat{\pi}_\text{Cb}  [ \hat{\varphi}, \hat{H}_\text{MpCb} ] \hat{\pi}_\text{Cb} $ that resolves to
\begin{align} \label{eq:mass_motion}
    \ii \hbar \dv{t} \hat{\varphi} = \Theta ( a - \abs{\vect{r}}) \left( \hat{h}_\text{MpCb}  + \int \limits_{C_2} \dd[6]{\mathcal{R}}_2 \hat{\varphi}_2^\dagger 2 \hat{\mathcal{V}}_\text{scatt} \hat{\varphi}_2 \right) \hat{\varphi}.
\end{align}
The Heaviside step function $\Theta ( x )$ accounts for creation and annihilation of only such cobosons whose constituents posses relative distances $\abs{\vect{r}} \leq a$.
The equation of motion yields a Schrödinger-like equation for the single-coboson energy governed by $\hat{h}_\text{MpCb}$, while the second term accounts for the influence of all other cobosons in the system via scattering.
Equation~\eqref{eq:mass_motion} does not exhibit general analytical solutions due to the combination of the relativistic corrections and the coupling to arbitrary light fields contained in the Breit-Pauli Hamiltonian $\hat{h}_\text{MpCb}$ as well as due to the scattering dynamics $\hat{\mathcal{V}}_\text{scatt}$.
For specific initial states, the time evolution of the associated $N$-particle Hamiltonian can be solved numerically. 
Since both the relativistic corrections and the scattering act perturbatively in such an atomic model, they can be approached with perturbative methods.
In particular, the relativistic corrections are dealt with an expansion into unperturbed relative hydrogen modes while the influence of scattering is resolved via a delta potential~\cite{Huang1957} together with mean-field and subsequent beyond-mean field approaches~\cite{Proukakis2008}. 
Within these mean-field approaches, Bogoliubov theory~\cite{Bogoliubov1947,Kawaguchi2012} may allow for an analytical solution.
Special cases like a single-mode or few-mode approximation for the coboson field operator~\cite{Anglin2001} allow further simplifications of the problem.
Solving for the field operator in the presence of light-matter interactions often requires further approximations like a pure dipole coupling or single-mode approximations.
Thus, in the following we present such a perturbative treatment.

\subsubsection{Expansion into unperturbed hydrogenlike modes}
While the dynamics implied by Eq.~\eqref{eq:mass_motion} is involved, the limits of integration restrict the relative distances between constituents of different cobosons to $\abs{\vect{x}_{1,i}- \vect{x}_{2,j}} > b \gg a$, and allow for a perturbative treatment of the scattering potentials.
The remaining dominant term denotes the single-coboson contribution associated with $\hat{h}_\text{MpCb}$, where the leading-order contribution $\hat{h}_\text{rel}^{(0)} = \vect{\hat{p}}^2/(2 m_\text{r}) +q_e q_n /  (4\pi \varepsilon_0 \abs{\vect{r}})$ is followed by other perturbative terms contained in $\hat{h}_\text{rel}^{(1)}$.
Consequently, we use an expansion into eigenmodes of $\hat{h}_\text{rel}^{(0)}$, \ie{} into hydrogenlike modes of the relative motion, and find
\begin{align} \label{eq:field_op}
    \hat{\varphi} = \sum_\beta \psi_\beta (\vect{r}) \hat{\Psi}_\beta ( \vect{R},t ),  
\end{align}
where $\psi_\beta$ is the (first-quantized) wave function of the relative motion associated with internal state $\beta$. 
The field operator $\hat{\Psi}_\beta ( \vect{R},t)$ annihilates a coboson in state $\beta$ at c.m. position $\vect{R}$.
The commutation relation of the remaining field operator $\hat{\Psi}_\beta ( \vect{R},t)$ is completely defined through the original cobosonic commutator from Eq.~\eqref{eq:comm}.
Furthermore, the cobosonic equation of motion requires the wave functions to vanish at $\abs{\vect{r}}=a$, similar to the case of atoms in an impenetrable spherical box~\cite{Degroot1946,Suryanarayana1976,LeyKoo1979,Marin1991,AquinoA.1995,Marin1995,Varshni1997,Aquino2007,Fernandez2010,Aquino2016}.
This condition is numerically solvable, with an energy depending on the particular choice of $a$ and converging to the known energies of hydrogen-type atoms for $a \to \infty$.
In the following, we choose the standard hydrogenlike wave functions for the relative motion, because for suitable values of $a$ the probability density is exponentially suppressed in regions $\abs{\vect{r}} > a $.
However, a numerical treatment is possible as well~\cite{AquinoA.1995}.
Hence, the hydrogenlike wave function $\psi_\beta ( \vect{r} )$ is associated with a generalized quantum number $\beta$ encompassing all quantum numbers, \ie{} the principal quantum number $n$ with energy eigenvalues $E_n^{(0)} = - m_\text{r} (Z\alpha c)^2/(2n^2)$, the quantum number $j$ associated with total angular momentum $\vect{\hat{j}} = \hat{\vect{\ell}} + \vect{\hat{S}}$, its projection to the $z$ axis ($m_j$), the orbital angular momentum ($\ell$) and the quantum number of the total spin $S$ associated with the spin $\vect{\hat{S}} = \vect{\hat{s}}_e + \vect{\hat{s}}_n$.
We present the angular momentum basis in Appendix~\ref{App:E}, where $\psi_\beta$ takes also the spin degrees of freedom of the coboson into account, \ie{} it contains in total four spinor components from the two spin-1/2 fermions.

By inserting the field-operator expansion from~\eqref{eq:field_op} into the equation of motion, multiplying with the conjugate wave function $\psi_\alpha^*$, and using the orthonormality of the relative modes when integrating over the relative coordinate, the equation of motion for the c.m. field operator of mode $\alpha$ resolves to
\begin{align} \label{eq:Ham_eigen}
\begin{split}
    \ii \hbar \dv{t} \hat{\Psi}_\alpha =& \hat{h}_{\text{MpCb}, \alpha} \hat{\Psi}_\alpha + \sum_{\beta \neq \alpha} \hat{T}_{\alpha \beta} \hat{\Psi}_{\beta} \\
    &+ \sum_{\beta \nu \mu} \, \int \limits_{\abs{\vect{\Delta R}} > b^\prime} \dd[3]{R^\prime} \hat{\Psi}^{\prime \, \dagger}_{\nu} \hat{\mathcal{V}}_{\alpha \nu ;\beta \mu} \hat{\Psi}^\prime_{\mu} \hat{\Psi}_{\beta}.
\end{split}
\end{align}
It contains only an integration with respect to c.m. coordinates.
The intercobosonic scale was introduced through the nucleus coordinate.
Thus, for consistency with the previous definition, we replace $b$ with $b' = b+a$ for the distance $\Delta{\vect{R}} = \vect{R} - \vect{R}^\prime$ between the c.m. positions of two cobosons.
Next, we present the internal Hamiltonian $\hat{h}_\mathrm{MpCb,\alpha}$, the transition elements $\hat{T}_{\alpha \beta}$ between internal states, and scattering matrix elements $\hat{\mathcal{V}}_{\alpha \nu ;\beta \mu}$ in the following two subsections.

\subsection{Modified bound-state energies}
The equation of motion for the field operator $\hat{\Psi}_\alpha$ associated with the annihilation of a coboson in mode $\alpha$ at c.m. position $\vect{R}$ includes the bound-state energy
\begin{align} \label{eq:approx_mass}
   \hat{h}_{\text{MpCb}, \alpha} = M_\alpha c^2 + E_\alpha^{(1)} + \frac{\vect{\hat{P}}_Q^2}{2M_\alpha}- \frac{\vect{\hat{P}}_Q^4}{8M^3c^2} +  \langle \hat{h}_\text{I} \rangle_\alpha,
\end{align}
of a coboson in internal state $\alpha$. 
\begin{figure}
    \centering
    \includegraphics[width=\columnwidth]{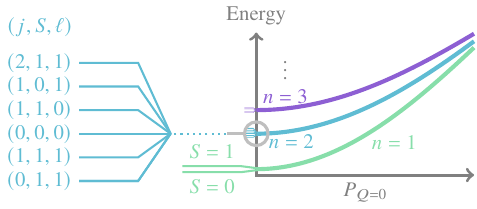}
    \caption{
    Due to the mass defect, the energy-momentum dispersion of a neutral coboson (with $Q=0$) depends on the unperturbed internal state defined by the principal quantum number $n=1,2,3,...$\,. 
    Additional relativistic corrections cause a splitting into sublevels that correspond to the fine and hyperfine structure given by the set of quantum numbers $(j,S,\ell)$.
    These structures do not enter the modified energy-momentum dispersion at our level of approximation in $c^{-2}$.
    The quantum number $j$ coincides with the quantum number $f$ that is the standard form for the hyperfine splitting in atomic physics and results from coupling the electron's spin with the relative angular momentum, before coupling the nuclear spin.}
    \label{fig:levels}
\end{figure}
Figure~\ref{fig:levels} shows the energy-momentum dispersion for the Hamiltonian $\hat{h}_{\text{MpCb}, \alpha}$ for different modes $\alpha$.
By introducing an internal-state-dependent rest mass $M_\alpha = M [ 1 + E^{(0)}_\alpha/(Mc^2)]$, the spectrum of the atom enters the rest energy $M_\alpha c^2$ and, through the relativistic mass defect, the minimally coupled kinetic energy $\vect{\hat{P}}_Q^2/(2M_\alpha)$, where the latter implies a lowest-order Taylor expansion.
As a result, the energy-momentum dispersion depends on the internal energy of the coboson.
Fine and hyperfine splittings~\cite{Pineda1998} enter through relativistic internal corrections $E^{(1)}_\alpha= \langle \hat{h}_{\text{rel}}^{(1)} \rangle_\alpha$, where $\langle \hat{o} \rangle_\alpha = \int \dd[3]{r} \psi_\alpha^* \hat{o} \psi_\alpha$ denotes the expectation value with respect to internal state $\alpha$ of an arbitrary operator $\hat{o}$.
These first-order corrections are of the order $E^{(1)}_\alpha/E^{(0)}_\alpha \sim 10^{-5}$ of the unperturbed energy $E^{(0)}_\alpha$, when $\alpha$ corresponds to the ground state of a hydrogen atom.
A detailed expression for these corrections is given in Appendix~\ref{App:E}.
Due to these corrections, there is a splitting of the unperturbed hydrogenlike energy levels also presented in Fig.~\ref{fig:levels}.
In the presence of EM fields, energy shifts occur in the form of $\langle \hat{h}_\text{I} \rangle_\alpha = \langle  \hat{h}_\text{I}^{(0)} \rangle_{\alpha}  + \langle \hat{h}_\text{I}^{(1)} \rangle_\alpha$, accounting for first-order perturbative shifts such as, \eg{} the linear Zeeman shift~\cite{Cohen1986}. 
Contrarily, second-order effects like the quadratic Stark effect~\cite{Schiff1939} are not explicitly accounted for in the diagonal matrix elements.
Further nonperturbative EM fields, giving rise to shifts such as ac-Stark~\cite{Delone1999} and other light shifts~\cite{Gauguet2008,Giese2015a} are also not solely represented by these diagonal elements.
To cover such additional effects, the second term in Eq.~\eqref{eq:Ham_eigen}, including all off-diagonal transition elements
\begin{equation}
    \hat{T}_{\alpha, \beta} = \int \dd[3]{r} \psi^*_\alpha \left( \hat{h}_\text{rel}^{(1)} + \hat{h}_\text{I} \right) \psi_{\beta}
\end{equation}
from internal state $\beta$ to $\alpha$, cannot necessarily be treated perturbatively.

In summary, using the expansion into relative hydrogenlike modes, we find both the bound-state energy of a coboson, including energy shifts due to internal relativistic corrections, as well as transitions between different internal-coboson states driven by both internal interactions and light fields.

\subsection{Modified scattering potentials}
The multicoboson aspect of our theory enters via the scattering matrix elements
\begin{align} \label{eq:gen_Scatt}
    \hat{\mathcal{V}}_{\alpha \nu; \beta \mu} = \int \dd[3]{r} \int \dd[3]{r^\prime} \psi_\alpha^* \psi_{\nu}^{\prime \, *} 2 \hat{\mathcal{V}}_\text{scatt} \psi_{\mu}^\prime \psi_{\beta},
\end{align}
describing the scattering from internal modes $\beta \mu$ into $\alpha \nu$, where $\hat{\mathcal{V}}_{\alpha \nu; \beta \mu}$ is a function of both $\vect{R}$ and $\vect{R}^\prime$.
Similar to the splitting of the single-coboson energy into the bound-state energies and internal transitions, we divide the scattering matrix elements into one part without transitions, \ie{} $\alpha = \beta$, and a part including actual transitions, \ie{} $\alpha \neq \beta$, that corresponds to internal state changing collisions.
As a result, the equation of motion 
\begin{align} \label{eq:Full_EQ}
\begin{split}
    \ii \hbar \dv{t} \hat{\Psi}_\alpha =& \left( \hat{h}_{\text{MpCb}, \alpha} + \sum_{\nu \mu} \int \limits_{\abs{\vect{\Delta R}} > b^\prime} \hspace{-0.4cm} \dd[3]{R^\prime} \hat{\Psi}^{\prime \, \dagger}_{\nu} \hat{\mathcal{V}}_{\alpha \nu; \alpha \mu} \hat{\Psi}^\prime_{\mu} \right)\hat{\Psi}_\alpha \\
    &+ \sum_{\beta \neq \alpha} \left(  \hat{T}_{\alpha \beta} + \sum_{\nu  \mu} \int \limits_{\abs{\vect{\Delta R}} > b^\prime} \hspace{-0.4cm} \dd[3]{R^\prime} \hat{\Psi}^{\prime \, \dagger}_{\nu} \hat{\mathcal{V}}_{\alpha \nu; \beta \mu} \hat{\Psi}^\prime_{\mu} \right)  \hat{\Psi}_{\beta}
\end{split}
\end{align}
for internal state $\alpha$ includes the single-coboson energy $\hat{h}_{\text{MpCb},\alpha}$ and is augmented by the scattering accounting for the mean field created by all other cobosons interacting with the coboson of mode $\alpha$.
Transitions from mode $\beta$ to $\alpha$ are either induced via internal or light-matter interactions but also by scattering with other cobosons that change its internal state from $\mu$ to $\nu$.
By integrating over relative degrees of freedom to obtain the scattering matrix elements from Eq.~\eqref{eq:gen_Scatt}, we gain via Eq.~\eqref{eq:Full_EQ} access to exact scattering potentials predicted by our model.

\begin{figure}
    \centering
    \includegraphics[width=\columnwidth]{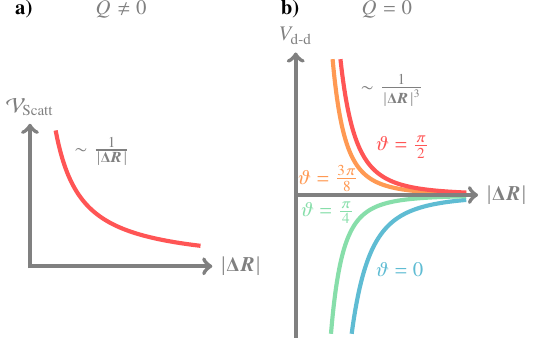}
    \caption{
    Form of the scattering potential.
    a) Two ionized cobosons with $Q\neq 0$ experience a repulsive Coulomb potential as the dominant contribution of their interaction.
    b) For neutral cobosons with $Q= 0$ higher-order dipole terms are dominating the interaction. The panel shows the dipole-dipole potential $V_\text{d-d} \propto (Z-1)^2/\abs{\vect{\Delta R}} + Z a^2 (1- 3 \mathrm{cos}^2 \vartheta)/\abs{\vect{\Delta R}}^3$ for the simplified case of $\vect{r} = \vect{r}^\prime = a \vect{e}_r$, where $\vartheta$ is the angle $\vect{r}$ and $\vect{\Delta R}$. 
    Depending on the orientation, the potential is either attractive or repulsive. 
    }
    \label{fig:dipdip}
\end{figure}

We obtain analytic expressions for the potentials approximated order by order, at least for the regime where $b' \gg a$, via the Taylor expansion of $\hat{\mathcal{V}}_\text{scatt}$  around  $\vect{x}_{i} - \vect{x}^\prime_{j}  \cong \vect{\Delta R}$ in Eq.~\eqref{eq:gen_Scatt}.
The dominant contribution in this regime follows from the Coulomb potential.
We find the generalized electric dipole-dipole potential 
\begin{widetext}
\begin{align}\label{eq:dipdip}
        \hat{\mathcal{V}}_{\text{scatt}} \approx \frac{1}{8\pi \varepsilon_0} \Bigg\{& \frac{Q^2}{\abs{\vect{\Delta R}}}  +Q \frac{\vect{e}_{\Delta R} \cdot \left( \vect{d} - \vect{d}^\prime \right)}{\abs{\vect{\Delta R}}^2} + Q\frac{ \sum\limits_u (\mathcal{Q}_{uu} + \mathcal{Q}^\prime_{uu}) - 3 \sum\limits_{u,v}  \vect{e}_{\Delta R}^{(u)} \left( \mathcal{Q}_{uv} + \mathcal{Q}^\prime_{uv} \right) \vect{e}_{\Delta R}^{(v)} }{\abs{\vect{\Delta R}}^3}+ \frac{ \vect{d} \cdot \vect{d}^\prime - 3 \left( \vect{e}_{\Delta R} \cdot \vect{d} \right) \left( \vect{e}_{\Delta R} \cdot \vect{d}^\prime \right)}{\abs{\vect{\Delta R}}^3} \Bigg\}
\end{align}
\end{widetext}
that accounts in general for cobosonic ions~\cite{Salam2009}.
For $Q\neq 0$, the leading order corresponds to a repulsive Coulomb potential proportional to $Q^2$ as indicated in Fig.~\hyperref[fig:dipdip]{5a)}. 
It is followed by corrections in which the difference of generalized dipole moments $\vect{d}= m_\text{r} ( q_e /m_e - q_n / m_n ) \vect{r}$ enters as well as the quadrupole-moment tensor $\mathcal{Q}_{uv} = -r_u r_v m_\text{r}^2 (q_e/m_e^2 + q_n/m_n^2)/2$ with components $u,v=x,y,z$.
The last term, the only one remaining in the limit of neutral cobosons with $Q=0$, corresponds to the standard electric dipole-dipole potential whose  dipole moment simplifies to $\vect{d} = q_e \vect{r}$ for $q_e=-q_n$.
Such a potential is the starting point to describe interatomic interactions in dipolar quantum gases~\cite{Craig1998,Kurizki2004,Lahaye2009}. 
We plot it in Fig.~\hyperref[fig:dipdip]{5b)} for parallel $\vect{r}$ and $\vect{r}^\prime$, as well as for different values of the angle between $\vect{\Delta R}$ and $\vect{r}$.
For instance, using second-order perturbation theory in first quantization, the dipole-dipole potential gives rise to the energy shift associated with the van der Waals potential~\cite{London1937, Moerdijk1995,Holstein2001} of the form $-C_6/\abs{\vect{\Delta R}}^6$ with a real constant $C_6$.
As a consequence, using the full Coulomb potential together with all relativistic corrections and explicitly integrating over relative degrees of freedom gives access to generalized van der Waals scattering potentials between cobosonic modes.
This approach can serve as the cobosonic model prediction to van der Waals potentials that may be compared with experimental results.
In addition, as we derived scattering dynamics with respect to internal states of the coboson, we are able to model cobosonic entanglement through scattering.
Since scattering can also be used for squeezing of internal states of atoms~\cite{Ma2011,Pezze2018,Salvi2018,Anders2021} and its description requires a field-theoretical formulation, our subspace EFT can be embedded into the field of quantum metrology.  
  
\subsection{Modified Gross-Pitaevskii equation}
The derivation of approximate solutions to the equation of motion from Eq.~\eqref{eq:Ham_eigen} for the c.m. field operator often follows a mean-field approach~\cite{Bogoliubov1947}.
Such a treatment leads to the celebrated \emph{Gross-Pitaevskii equation} (GPE)~\cite{Gross1961,Pitaevskii1961}, which we derive below in favor of approaches following, \eg{} density-functional theory~\cite{Malet2015}.

The scattering potentials in Eq.~\eqref{eq:Full_EQ} have the form of a hard-sphere interaction~\cite{Huang1957, Lee1957} characterized by a nonvanishing potential only at distances $\abs{\vect{\Delta R}} > b^\prime$, which is ensured by integration regions in our model. 
In this case and for low temperatures as well as weakly interacting, dilute gases, such hard-sphere potentials can be replaced by a pseudopotential~\cite{Huang1957} of the form $\eta_{\alpha \nu; \alpha \mu} \delta( \vect{R} - \vect{R}^\prime )$, where no integration region appears~\footnote{The replacement may imply a restriction to quantum numbers $\beta$ whose angular momentum $\ell=0$ vanishes.}.
Instead, we find an effective, renormalized~\cite{Beliaev1958, Popov1988, Proukakis2008} scattering length~\footnote{As a consequence the effective scattering length is not only the width of a narrow dipole-dipole potential but contains numerical contributions as well.} $\eta_{\alpha \nu, \alpha \mu}$, mediating scattering between cobosons of mode $\alpha$ with that of mode $\mu$ transitioning into mode $\nu$.
An analogous replacement in the collision-induced coupling between modes $\alpha$ and $\beta$ in the second line of Eq.~\eqref{eq:Full_EQ} with an effective scattering length $\eta_{\alpha \nu, \beta \mu}$ can be made.
Within this approximation the equation of motion for the field operator takes the form
\begin{align} \label{eq:final}
\begin{split}
    \ii \hbar \dv{t} \hat{\Psi}_{\alpha} =& \left( \hat{h}_{\text{MpCb},\alpha}  + \sum_{\nu \mu} \eta_{\alpha \nu; \alpha \mu} \hat{\Psi}^\dagger_\nu \hat{\Psi}_{\mu} \right) \hat{\Psi}_\alpha \\
    &+ \sum_{\beta \neq \alpha} \left(  \hat{T}_{\alpha, \beta} + \sum_{\nu \mu} \eta_{\alpha \nu;\beta \mu} \hat{\Psi}_\nu^\dagger \hat{\Psi}_{\mu} \right) \hat{\Psi}_{\beta}.
    \end{split}
\end{align}
Such an approximation is often applied in the context of ultracold quantum gases~\cite{Ueda2010, Pethick2008} and corrections may also be taken into account~\cite{Fu2003,Veksler2014}.

However, already in a mean-field theory we observe a difference to the conventional treatment, \ie{} we have access to relative and c.m. relativistic corrections, as well as to the full coupling to external EM fields.
To this end, we approximate Eq.~\eqref{eq:final} by moving to a first-quantized equation of motion $\hat{\Psi}_\alpha \to \Psi_\alpha$ where $\Psi_\alpha$ represents the mean field of the condensate~\cite{Bogoliubov1947}.
There are several ways to introduce the mean field as lowest-order contribution of the equation of motion~\cite{Nandi2007,Proukakis2008,Ufrecht2019}.
Extending the lowest-order contribution to beyond mean-field theory~\cite{Braaten1997,Anglin2001,Lima2011,Lima2012} may be achieved by including also an operator-valued noncondensate part of the field operator in terms of a thermal cloud that couples to the mean field~\cite{Proukakis2008}.
Within the mean-field approach, we find new effective scattering lengths $\tilde{\eta}_{\alpha \beta}$ and $\tilde{\eta}_{\alpha \beta; \alpha^\prime \beta^\prime}$ that may differ from the previous values.
These approximations result in the modified GPE
\begin{widetext}
\begin{align}
\begin{split} \label{eq:GPE}
    \ii \hbar \dv{t} \Psi_{\alpha} =&\bigg( M_\alpha c^2 + E_\alpha^{(1)} + \langle \hat{h}_\text{I} \rangle_\alpha + \frac{\vect{\hat{P}}_Q^2}{2M_\alpha}- \frac{\vect{\hat{P}}_Q^4}{8M^3c^2}
     + \sum_{\nu \mu} \tilde{\eta}_{\alpha \nu; \alpha \mu} \Psi_\nu^* \Psi_{\mu} \Bigg) \Psi_\alpha 
    + \sum_{\beta \neq \alpha} \left(\hat{T}_{\alpha, \beta} + \sum_{\nu \mu} \tilde{\eta}_{\alpha \nu;\beta \mu} \Psi^*_\nu \Psi_{\mu} \right) \Psi_{\beta}
\end{split}
\end{align}
\end{widetext}
that contains, compared to the NR bosonic GPE~\cite{Gross1961,Pitaevskii1961} , first-order relativistic corrections. 
It is valid for spinor Bose-Einstein condensates~\cite{Kawaguchi2012} and has a state-dependent mass $M_\alpha$ differing from previous derivations~\cite{Anandan1981,Matos2011,Gabel2019}.
Moreover, the description applies also to cobosonic ions (coupling via $\vect{\hat{P}}_Q$), as long as the gas can still be treated as weakly interacting.
In addition, we find the energy shift $E_\alpha^{(1)}$ from the internal cobosonic structure and we account for light-matter interaction in $\langle \hat{h}_\text{I} \rangle_\alpha$.
Moreover, we observe that the GPE for mode $\alpha$ may couple to other modes through $\Psi_\nu^* \Psi_{\mu}$ terms~\cite{Gupta2009}, where usually only the contributions proportional to $\abs{\Psi_\mu}^2$ are taken into account.
The coupling to other modes enters via nonvanishing internal transition elements $\hat{T}_{\alpha, \alpha^\prime}$, as well as via a scattering element including transitions from mode $\mu$ to $\nu$.

To our knowledge, a modified GPE for c.m. degrees of freedom, taking into account internal degrees of freedom and incorporating the mass defect, has not yet been derived from first principles.
In contrast to previous derivations of relativistically corrected GPEs~\cite{Anandan1981,Matos2011,Gabel2019}, discussed in more detail in Sec.~\ref{Sec:Discussion_uqg}, our results include the mass defect expressed through the internal-state-dependent total mass $M_\alpha$, and by that a state-dependent dispersion relation. 
Moreover, we include also the first-order internal-energy shift $E^{(1)}_\alpha$, the kinetic correction $\vect{\hat{P}}_Q^4$, and a complete treatment of light-matter interaction.
In addition, the equation is usually derived only for neutral and spinless particles.
References~\cite{Anandan1981,Matos2011,Gabel2019} aim at describing relativistic Bose-Einstein condensates starting from a nonlinear generalization of the Klein-Gordon equation.

In experiments with cold atoms, actual interactions not approximated as contact potentials and beyond-mean-field effects may dominate over these relativistic corrections and Eq.~\eqref{eq:GPE} needs to be extended to account also for these effects.
Such a procedure can be applied to Eq.~\eqref{eq:final} and gives rise to Bogoliubov-des Gennes equations~\cite{Bogoliubov1947, deGennes1966, Fetter1972}, for example, as well as couplings of the mean field to the noncondensate part of the field proportional to the density of the noncondensate in lowest order~\cite{Proukakis2008}.

However, high-precision experiments relying on atom interferometry are currently discussed for fundamental-physics tests, such as Einstein-equivalence-principle violations~\cite{Ufrecht2020,DiPumpo2021}, or dark-matter~\cite{Arvanitaki2018,DiPumpo2023b} and gravitational-wave~\cite{Dimopoulos2008Atomic_gravitational} detection.
Since these campaigns aim for the detection of perturbations, an analysis of contributions, such as the special-relativistic effects derived in this work are required for a consistent treatment.
For example, proposals for fountain geometries, ranging from ten~\cite{Schlippert2020} to several hundred meters~\cite{Badurina2020,Mitchell2022,Abend2023}, give rise to velocities $v/c$ that are in the order $10^{-7}$ when the atoms are accelerated for a hundred meters by Earth's gravitational field. 
Due to the $\vect{\hat{P}}^4$ term, the kinetic energy experiences then a correction proportional to $(v/c)^2 \sim 10^{-14}$ that, in principle, has to be included for the analysis of high-precision tests targeting at such sensitivities, depending on the observable. 
For atomic clocks in a storage ring moving with a c.m. velocity with $v=0.03 c$~\cite{Reinhardt2007}, this correction would even increase to $(v/c)^2 \sim 10^{-3}$, becoming important for already performed experiments.

Moreover, the relativistic mass defect corrects the kinetic energy by a term of the order $10^{-11}$ during optical clock transitions, for example, through the $^{1}S_0 \text{-} ^{3}P_0$ transition in $^{88} \text{Sr}$~\cite{Hu2017}. 
This correction is key to geometries for ambitious tests of the Einstein equivalence principle as it induces differential recoil shifts~\cite{Loriani2019} and has to be considered in quantum clock interferometry~\cite{Zych2011}.
Even though the most precise measurements of the fine-structure constant were achieved with atom interferometers generated by a combination of Raman pulses and Bloch oscillations~\cite{Morel2020}, similar setups based on large-momentum transfer through optical single-photon transitions might have to consider a modified recoil velocity. 
Furthermore, the mass defect was also explicitly measured in atomic clock setups~\cite{Daams1974,Brewer2019,Miao2022} as part of the relativistic or second-order Doppler shift. 
Thus, the analysis of state-of-the-art atomic clocks has to consider the mass defect, \ie{}  second-order Doppler shifts, which we discuss in this context explicitly in the next subsection.

\subsection{Reduction to mass defect}
With the modified GPE we reproduce two special cases: 
(i) We find the typical atomic physics NR GPE by neglecting all relativistic contributions. 
(ii) By restricting the treatment for $Q=0$ to two modes, \ie{} ground ($g$) and excited ($e$) state, and by neglecting the $\vect{\hat{P}}^4$ term, internal relativistic corrections in $E_\alpha^{(1)}$, as well as the influence of any scattering, we reproduce a Hamiltonian~\cite{Sonnleitner2018,Schwartz2019} that is relevant in an atomic~\cite{Yudin2018,Martinez2022} and quantum clock context~\cite{Zych2011,Sinha2011,Loriani2019,Ufrecht2020,Roura2020,DiPumpo2021,DiPumpo2022,DiPumpo2023}.
For the sake of presentation, we neglect light-matter interactions for the moment.
In this limit, the equation of motion for both, ground and excited state, reduces to $\ii \hbar \dd \ket{j} / \dd{t} = \hat{h}_j \ket{j}$ with a first-quantized Hamiltonian $\hat{h}_j = M_jc^2 + \vect{\hat{P}}^2/(2M_j)$, including the abstract form of the wave function in position representation $\Psi_j = \braket{\vect{R}}{j}$, with $j=g,e$.
Since the differential equations for the internal states are now decoupled, we find a Schrödinger equation for the general state $\ket{\Psi} = \Psi_g \ket{g} + \Psi_e \ket{e}$ with $\abs{\Psi_g}^2+ \abs{\Psi_e}^2=1$. 
After Taylor expanding the state-dependent mass $M_j$, the system Hamiltonian, \ie{} the sum of the two Hamiltonians $\hat{h}_j$, takes the form
\begin{align}
    \hat{h} = Mc^2 \mathds{1} + \hat{h}_\text{rel}^{(0)} + \frac{\vect{\hat{P}}^2}{2M} \left( \mathds{1} - \frac{\hat{h}_\text{rel}^{(0)}}{Mc^2} \right),
\end{align}
which is the limit of addressing only two internal states of $\hat{h}_\text{rel}^{(0)} = E_g \ket{g} \! \bra{g} + E_e \ket{e} \! \bra{e}$ in Eq.~\eqref{Eq:Ham_mpcb}, as expected.
This Hamiltonian can be recast into the form
\begin{equation}
\label{eq.Mass_defect}
    \hat{h} = \, \overline{\! M} \, c^2 \mathds{1} + \hat{h}_\text{cl} + \frac{\vect{\hat{P}}^2}{2 \, \overline{\! M} \,}\left(\mathds{1}-\frac{\hat{h}_\text{cl}}{\, \overline{\! M} \,c^2}\right)
\end{equation}
by introducing a new mean mass $\, \overline{\! M} \, = M + (E_e^{(0)}+E_g^{(0)}) /(2c^2)$ together with replacing the unperturbed Hamiltonian of the relative degrees of freedom $\hat{h}_\text{rel}^{(0)}$ by the clock Hamiltonian
\begin{equation} \label{eq:clock}
    \hat{h}_\text{cl} = \frac{E_e^{(0)}-E_g^{(0)}}{2} [\ket{e}\!\bra{e}-\ket{g}\!\bra{g}].
\end{equation}
This clock Hamiltonian describes the internal (relative) dynamics and constitutes the basis of atomic and quantum clocks~\cite{Zych2011, Sinha2011, Sonnleitner2018, Yudin2018, Schwartz2019, Loriani2019, Ufrecht2020, Roura2020, DiPumpo2021, Martinez2022, DiPumpo2022, DiPumpo2023}, in our case without gravity.
In particular, the preceding two equations are related by the fact that in the order $c^{-2}$ the equivalence 
\begin{align}
\hat{h} (M, \hat{h}_\text{rel}^{(0)})=\hat{h} (\, \overline{\! M} \,, \hat{h}_\text{cl} )
\end{align}
holds.
Moreover, the energy difference $E_e^{(0)}-E_g^{(0)}$ can be associated with the transition frequency of a clock as well as the mass difference between both internal states.
This equivalence can be extended in the order $c^{-2}$ to the case where the total momentum $\vect{\hat{P}}$ is replaced by its minimally coupled version $\vect{\hat{P}}_Q$.
Similarly, the equivalence holds in the order $c^{-2}$ also for the corrected relative degrees of freedom $\hat{h}^{(1)}_{\text{rel}}$, and the corrected EM interaction $\hat{h}_{\text{I}}^{(1)}$ in Eq.~\eqref{Eq:Ham_mpcb}.
However, replacing $M$ by $\, \overline{ \! M}$ would lead to additional relativistic modifications for some parts of the NR EM interaction $\hat{h}_{\text{I}}^{(0)}$, especially when considering magnetic fields, while its leading-order NR contributions maintain the same form.

\begin{figure}
    \centering
    \includegraphics[width=\columnwidth]{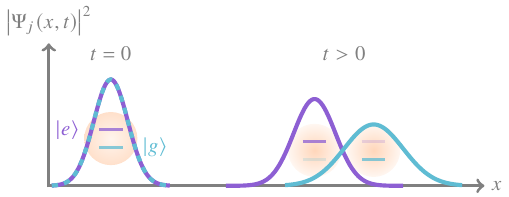}
    \caption{Sketch of the time evolution of noninteracting cobosons in a superposition of ground $\ket{g}$ and excited $\ket{e}$ state.
    The c.m. probability distribution $\abs{\Psi_j(x,t)}^2$ of the atom in state $\ket{j}$ is shown by different wave packets.
    Their form depends on the internal states due to different dispersion relations, which depend on the masses $M_g$ and $M_e$ due to the mass defect.
    Moreover, if both wave packets have the same initial momentum, the ground state propagates with a higher velocity.}
    \label{fig:dispersion}
\end{figure}

The Hamiltonian accounts for a modified c.m. motion and dispersion relation for atoms in different internal states through the mass defect. 
To underline the implications of the mass defect, we observe that wave packets associated with the ground and excited state of a free coboson disperse and propagate differently over time, as indicated in Fig.~\ref{fig:dispersion}.
Due to the state-dependent mass, the amplitude and the uncertainty differ.
Since both wave packets share the same initial momentum they evolve with a different velocity.

The c.m. of clocks moving at a velocity $v$ experiences a time dilation $t^\prime = t (1-[v/c]^2)^{-1/2}$, which, in turn, implies, after Taylor expansion, a second-order Doppler shift $\Omega^\prime = \Omega [1- (v/c)^2/2]$ of any frequency $\Omega$ as a first-order correction, and by that also the transition frequency $\Omega = (E_e^{(0)}-E_g^{(0)}) / \hbar$ of atomic clocks~\cite{Chou2010}.
A semiclassical treatment of such frequency shifts leads to relative values up to order $(v/c)^2\sim10^{-12}$ and has to be taken into account for the analysis of high-accuracy frequency standards~\cite{Kajita2018}.
These shifts, which are nonneglibile for high-precision clocks~\cite{Martinez2022,Miao2022}, are, for example, caused by micro- or macromotion of ions in traps, where the ion is not prepared in the c.m. ground state, but has some thermal excitation, and is not located at the exact minimum of the trap, which has a complex and time-dependent structure~\cite{Leibfried2003}. 
To estimate these shifts, usually an ensemble average over the induced classical motion of a thermal ensemble is performed.
However, Eq.~\eqref{eq.Mass_defect} incorporates the time-dilaton shift on a quantum level, since it can be recast in the form
\begin{equation}
    \hat{h} = \left[ \overline{\! M} \, c^2 + \frac{\vect{\hat{P}}^2}{2 \, \overline{\! M} \,} \right] + \frac{\hbar \Omega}{2} [\ket{e}\!\bra{e}-\ket{g}\!\bra{g}]  \left[1- \frac{1}{2}\left(\frac{\vect{\hat{P}}}{\, \overline{\! M} c}\right)^2\right].
\end{equation}
A generalization to ions and the inclusion of trapping potentials caused by electromagnetic fields is straightforward within our framework.
Tracing out the c.m. degrees of freedom of the bound systems, we find a relative shift $- \langle (\vect{\hat P} / \, \overline{\! M} c)^2\rangle/2$ of the transition frequency.
It can be associated with the ensemble average of the semiclassical approach and is determined by the specific parameters of the trap and the associated micro- and macromotion~\cite{Miao2022}.
Moreover, combining our description with a perturbative treatment~\cite{Ufrecht20202} also allows for the explicit description of second-order Doppler effects in Ramsey sequences or for probing transitions, where the atoms undergo a complex motion in time-dependent traps.
Here, one distinguishes between a momentum distribution caused by finite temperatures and the associated uncertainties in momentum, but also the excess micromotion when the clock is initially displaced from the trap minimum~\cite{Zhukas2021}.
Hence, our results can be directly applied for the analysis of time-dilation effects in atomic or ion clocks, complementing semiclassical approaches that so far are the backbone of the analysis of relativistic motional effects of atomic clocks.
For the order of magnitude of such effects, see Refs.~\cite{Chou2010,Ludlow2015}.

Moreover, in contrast to previous works discussing Hamiltonians of the form of Eq.~\eqref{eq.Mass_defect}, our framework allows for the treatment of spin-dependent interactions such as in spinor ion traps~\cite{Leibfried2003} or spinor atom interferometry~\cite{Hu2017} through $\hat{h}_{\text{I}}$ in Eq.~\eqref{Eq:Ham_mpcb}.
Besides well-known Zeeman-like effects~\cite{Ludlow2015}, including the coboson's spin also enables the treatment of, \eg{} magnetically induced single-photon transitions~\cite{Bott2023} and beyond-Standard-Model effects such as axion couplings~\cite{Graham2016}.
Additionally, such spin- and internal-state-dependent interactions allow for addressing each spin state with a different trapping potential, leading to guided quantum clock interferometry~\cite{DiPumpo2021}, where the clock's c.m. state is being brought into spatial delocalization.
Such schemes can be used for novel internal sensors, \eg{} for rotation sensing~\cite{Stevenson2015}, as well as for fundamental tests such as the Einstein equivalence principle~\cite{DiPumpo2021}.
Hence, an extension of our framework to gravity would allow for a detailed study of gravitational-redshift tests across single wave packets or on small scales~\cite{Bothwell2021,Zheng2022}, exceeding so far semiclassical descriptions of clocks.
For such tests, reaching an uncertainty level of $10^{-21}$, the effects discussed in our treatment are crucial.
More generally, such an extension could be used to study the gravitational redshift itself as a further relativistic correction, augmenting the $\vect{\hat{P}}^4$ kinetic correction in similar orders of magnitude.

\section{Conclusions} \label{Sec:Conclusions}
By projecting existing NR EFTs, we derived in this paper a subspace EFT for (possibly) charged, spin carrying, and interacting composite bosons based on their constituents.
Our approach includes relativistic contributions such as radiative corrections, mass defects, as well as atom-atom scattering, and light-matter interactions.
We therefore combined low-energy aspects of particle physics, quantum optics, and atomic physics in one multipolar cobosonic subspace EFT with a range of applications, \eg{} to scattering experiments~\cite{Taylor2006, Braaten2008}, ultracold quantum gases~\cite{Giorgini2008,Baranov2008,Ueda2010}, and high-precision measurements based on quantum clocks~\cite{Zych2011,Sinha2011,Derevianko2014,Arvanitaki2015,Arvanitaki2018,Loriani2019,Ufrecht2020,Roura2020,DiPumpo2021,DiPumpo2023,Yudin2018, Martinez2022}.

By considering their c.m. motion, we observed intercobosonic interactions via relativistically corrected scattering potentials and a coupling between c.m. and relative degrees of freedom that arises from the relativistic mass defect.
Moreover, our projection technique can be applied to other single-fermion Hamiltonians $\hat{h}_i$ and potentials $\hat{V}^{(ij)}$.
Our second-quantized transformations led to relativistic corrections to c.m. and relative coordinates together with the multipolar version of our subspace EFT.
The extension to spin carrying charged cobosons confirms that the coupling between c.m. and relative degrees of freedom induced by the mass defect holds also in this framework.
To the best of our knowledge, relativistically corrected scattering potentials to this extent are given for the first time, including a scattering self-energy term.
In addition, we presented a modified, coupled GPE including light-matter interactions, other relativistic corrections, and the mass defect.

The projection formalism introduces length scales associated with atoms composed of electron-nucleus pairs.
Introducing further different length scales may result into subspace EFTs for other types of composite particles, \eg{} multielectron atoms and molecules.
A subspace EFT for molecules would directly connect to and extend existing approaches~\cite{Brambilla2018} and lead to a field-theoretical description of interacting ultracold molecules.
Such an effective theory revolves around established concepts such as the Born-Oppenheimer approximation~\cite{Born1927} and other bound-state calculations for many-body bound systems, such as density-functional theory~\cite{Parr1995}.
Furthermore, our model describing single-species ensembles may be extended to mixtures, \eg{} of cobosons and free fermions, different species, isotopes, as well as ions within neutral quantum gases, the latter giving rise to respective spinor quantum gases.
Moreover, effects of the environment that lie outside of the cobosonic subspace could, in principle, be incorporated by techniques known from open quantum systems~\cite{Rivas2012,Wurtz2020}, and will lead to additional energy shifts as well as to decoherence mechanisms.

Including external nonelectromagnetic fields, such as gravity or violation fields in a similar fashion would set a quantum-field-theoretical foundation for established single-particle descriptions, being of essence for quantum clock interferometry but also for atomic clocks exposed to micromotion~\cite{Yudin2018, Martinez2022}, tests of special and general relativity~\cite{Ufrecht2020, Roura2020, DiPumpo2021, DiPumpo2023}, as well as dark-matter detection~\cite{Derevianko2014,Arvanitaki2015,Arvanitaki2018,DiPumpo2022}.
By determining c.m. scattering potentials between two cobosons by integrating over relative degrees of freedom numerically, we expect to find corrected \textit{van der Waals} scattering potentials~\cite{Tang1984,Tang2003}.
Moreover, our results facilitate a field-theoretical description of both the c.m. motion as well as the internal states of atomic quantum gases. 
Since quantum-metrological methods enhancing the precision through techniques like squeezing rely on such a treatment and might be even generated through scattering, our results lay the basis for the description and modeling of supersensitive measurements below the shot-noise limit and can be applied to spin-squeezed experiments~\cite{Ma2011,Pezze2018} or momentum-squeezed atom interferometry~\cite{Salvi2018,Anders2021}.

In summary, the presented multipolar cobosonic subspace EFT can be applied to a large class of atomic ensembles, \eg{} Bose-Einstein condensates~\cite{Goral2000,Kurizki2004,Ueda2010,Kawaguchi2012,Olson2013}, ionized quantum gases~\cite{Osborne1949, Foldy1961, Girardeu1962, Ninham1964, Wright1966}, and thermal clouds~\cite{Koehl2002,Miller2005} that may be exposed to light-matter interactions including trapping potentials~\cite{Fetter2009} and light pulses~\cite{Giltner1995, Mueller2008, Altin2013}. 
It also includes relativistic corrections to the relative Hamiltonian, the mass defect, light-matter interaction in its most general form, and scattering potentials.
Therefore, our results are a basis for studies of composite particles, both for fundamental physics but also for the application of quantum systems in a vast area of different subfields.

\begin{acknowledgments}
We are grateful to W. P. Schleich for his stimulating input and continuing support.
Moreover, we are thankful to A. Friedrich for helpful support and instructive feedback throughout the whole project, as well as proofreading our manuscript.
We also thank O. Buchmüller, M. A. Efremov, C. Kiefer, R. Lopp, C. Niehof, G. Paz, R. Walser, A. Wolf, as well as the QUANTUS and INTENTAS teams for fruitful and interesting discussions.
The projects ``Building composite particles from quantum field theory on dilaton gravity'' (BOnD) and ``Metrology with interfering Unruh-DeWitt detectors'' (MIUnD) are funded by the Carl Zeiss Foundation (Carl-Zeiss-Stiftung).
The QUANTUS and INTENTAS projects are supported by the German Space Agency at the German Aerospace Center (Deutsche Raumfahrtagentur im Deutschen Zentrum f\"ur Luft- und Raumfahrt, DLR) with funds provided by the Federal Ministry for Economic Affairs and Climate Action (Bundesministerium f\"ur Wirtschaft und Klimaschutz, BMWK) due to an enactment of the German Bundestag under Grants No. 50WM2250D-2250E (QUANTUS+), No. 50WM2450D-2450E (Quantus-VI), as well as 50WM2177-2178 (INTENTAS).
The Qu-Gov project in cooperation with the ``Bundesdruckerei GmbH'' is supported by the Federal Ministry of Finance (Bundesministerium der Finanzen, BMF).
E.G. thanks the German Research Foundation (Deutsche Forschungsgemeinschaft, DFG) for a Mercator Fellowship within CRC 1227 (DQ-mat).
F.D.P. is grateful to the financial support program for early career researchers of the Graduate \& Professional Training Center at Ulm University and for its funding of the project ``Long-Baseline-Atominterferometer Gravity and Standard-Model Extensions tests'' (LArGE).
\end{acknowledgments}

\begin{appendix}
\section{Potential NRQED revisited} \label{App:A}
We assume that composite particles are built from two fermionic particle species, namely electrons ($e$) and nuclei ($n$)~\footnote{
We omit the composite-particle nature of the nucleus in its dynamical description by treating it as a single fermionic field. 
The composite-particle nature of the nucleus may be taken into account by deriving an effective QFT in the same spirit.
However, nondynamical composite-particles effects are included in prefactors, as detailed later.
}. 
Since NR effects are primarily responsible for the bound-state dynamics between the constituents, one usually relies on 
NRQED~\cite{Caswell1986, Kinoshita1996, Nio1997, Manohar1997, Paz2015, Haidar2020} for a field-theoretical description.
NRQED is an established EFT of QED valid in NR regimes of both nucleus and electron momenta where antiparticles are of no relevance.
This assumption requires sufficiently low photon energies such that the particle-antiparticle dynamics remains negligible.

Because of the absence of antiparticles, the constituents are simply described by two-component field operators $\hat{\psi}_i(\vect{x})$ and $\hat{\psi}_i^\dagger (\vect{x})$, associated with the annihilation and creation of particle $i=e,n$ at position $\vect{x}$, rather than four-component spinors containing both particle and antiparticle field operators.
Thus, the components $u,v=1,2$ of field operators of the same species obey anticommutation relations $\{\hat{\psi}_{i, u} (\vect{x}),\hat{\psi}_{i, v}^\dagger (\vect{x^\prime}) \} = \delta_{uv} \delta(\vect{x}-\vect{x^\prime})$ and $\{\hat{\psi}_{i, u} (\vect{x}),\hat{\psi}_{i, v} (\vect{x^\prime}) \} = 0$.
Simultaneously, electron and nucleus field operators act on different Hilbert spaces implying vanishing commutators $[ \hat{\psi}_{i,u}, \hat{\psi}_{j,v}^\dagger]=0 = [ \hat{\psi}_{i,u}, \hat{\psi}_{j,v}]$ for $i \neq j$ between different particle species.

The Lagrangian density governing the dynamics of the fermionic field operators may be constructed~\cite{Paz2015} by considering all possible operator combinations that preserve the symmetries (namely hermiticity, as well as gauge, rotational, parity, and time-reversal invariance).
Each combination is then equipped with a coefficient determined by a matching of cross sections in the low-energy limit of QED~\cite{Kinoshita1996,Nio1997}. 
Alternatively, the NRQED Lagrangian follows directly from the QED Lagrangian by applying the Foldy-Wouthuysen transformation~\cite{Foldy1950}, where the matching coefficients have to be added manually.
These so-called \textit{Wilson coefficients}~\cite{Wilson1965,Wilson1974} partly account for QED effects that are no longer accessible in NRQED, such as the anomalous magnetic moment~\cite{Schwinger1948, Laporta1996, Aoyama2012}.
In the spirit of EFTs, they also account for composite-particle aspects of the nucleus, loop corrections, or radiative effects.

After a Legendre transformation of the NRQED Lagrangian density~\cite{Hill2013} up to order $c^{-2}$ of the speed of light $c$, the NRQED Hamiltonian density
\begin{align} \label{eq:NRdensity}
        \hat{\mathcal{H}} =& \hat{\mathcal{H}}_\text{EM} +\displaystyle \sum_{i=e,n} \hat{\psi}^\dagger_i  \hat{h}_i \hat{\psi}_i + \hat{\mathcal{H}}_\text{cont} 
\end{align}
contains three contributions~\footnote{The following form of the Hamiltonian density follows from the flat spacetime Minkowski metric $\eta_{\mu,\nu}=\operatorname{diag}(+1,-1,-1,-1)$.}.
The first one corresponds to the free energy density of the EM field $\hat{\mathcal{H}}_\text{EM} = \varepsilon_0 ( \vect{\hat{E}}^2 + c^2 \vect{\hat{B}}^2)/2$.
The second term accounts for the energy density of electrons and nuclei, where the single-fermion Hamiltonian
\begin{align}
\begin{split}
        \hat{h}_i =& m_i c^2 + q_i \hat{\phi} + \frac{\pbar{2}}{2m_i} - \overline{c}_{\text{F}}^{(i)} q_i \frac{\boldsymbol{\hat{s}}_i\cdot \boldsymbol{\hat{B}}}{m_i} - \frac{\pbar{4}}{8m_i^3c^2} \\ \label{eq:hfirst}
        & - \overline{c}_{\text{D}}^{(i)} q_i \hbar^2 \frac{ \vect{\nabla} \cdot \vect{\hat{E}}}{8 m_i^2c^2} + \overline{c}_{\text{S}}^{(i)} q_i \vect{\hat{s}_i} \cdot \frac{  \pbar{}  \times \vect{\hat{E}} - \vect{\hat{E}} \times \pbar{} }{4 m_i^2c^2}  \\
        &+ \overline{c}_\text{W1}^{(i)} q_i \frac{\left\{ \pbar{2} \!, \vect{\hat{s}}_i \cdot \vect{\hat{B}} \right\}}{4m^3_ic^2} - \overline{c}_\text{A1}^{(i)} q^2_i \hbar^2 \frac{\vect{\hat{B}}^2}{8 m_i^3 c^2}
\end{split}
\end{align}
corresponds to the energy of a single fermion of species $i$.
The Hamiltonian constitutes the basis for the Schrödinger equation of first-quantized systems. 
It is sandwiched between the field operators $\hat{\psi}_i^\dagger$ and $\hat{\psi}_i$ and creates a weighted particle-number density in a field-theoretical treatment.

In leading order, the energy of particle species $i$ is the sum of rest energy due to its rest mass $m_i$, energy caused by the scalar potential $\hat{\phi}$ because of its charge $q_i$, kinetic energy, as well as energy due to the coupling of spin $\vect{\hat{s}}_i = \hbar \vect{\hat{\sigma}}_i/2$ with Pauli-matrix vector $\vect{\hat{\sigma}}_i$ of particle $i$ to a magnetic field $\vect{\hat{B}}$.
The kinetic energy associated with the particle's minimally coupled momentum $\pbar{} = \vect{\hat{p}} - q_i \vect{\hat{A}}$, with momentum operator $\vect{\hat{p}} =- \ii \hbar \vect{\nabla}$, is modified by the vector potential $\vect{\hat{A}}$, and obeys the commutation relation $[ x_u, \hat{p}_v]= \ii \hbar \delta_{u v}$ for $u,v=x,y,z$.
Note that we rely on positions $\vect{x}$ as integration variables and consequently they are not operator valued.
In contrast to that, we introduced the momentum operator, similarly to the covariant derivative, with a hat to emphasize the particular fact that $\vect{\hat{p}}$ and $\vect{x}$ do not commute and $\vect{\hat{p}}$ has to be used in position representation.

The first-order relativistic corrections are the kinetic ($\pbar{4}$) and electric field corrections, covering the Darwin term ($\vect{\nabla} \cdot \vect{\hat{E}}$) and spin-orbit term ($\pbar{} \times \vect{\hat{E}}$), which give rise to a corresponding hydrogen fine-structure contribution.
While $\vect{\hat{p}}_i$ acts on the field operator, $\nabla \cdot \vect{\hat{E}}$ is only a spatial derivative of $\vect{\hat{E}}$. 
The last line of Eq.~\eqref{eq:hfirst} contains relativistic corrections to light-matter interaction in the form of general magnetic moment and diamagnetic corrections. 
All light fields are functions of position $\vect{x}$ and are connected to $\hat{\phi}$ and $\hat{\vect{A}}$ via $\vect{\hat{E}} = - \vect{\nabla} \hat{\phi} - \partial_t \vect{\hat{A}}$ and $\vect{\hat{B}} = \vect{\nabla} \times \vect{\hat{A}}$.

The Wilson coefficients $\overline{c}_k^{(i)}$ in Eq.~\eqref{eq:hfirst} are determined from tree-level QED matching~\cite{Kinoshita1996}, and particular subscripts stand for \textit{Fermi}, \textit{Darwin}, and \textit{Seagull}.
Because we also allow for different charge numbers $Z_i$ of nucleus and electron, we include $Z_i$ explicitly into Eq.~\eqref{eq:hfirst}, so that our coefficients $\overline{c}_k^{(i)}$ and $\overline{c}_\text{A1}^{(i)}$ are connected to the conventional Wilson coefficients~\cite{Hill2013,Peset2015a} through $c_k^{(i)} = Z_i \overline{c}_k^{(i)}$ and $c_\text{A1}^{(i)} = Z_i^2 \overline{c}_\text{A1}^{(i)}$.
This factor of $Z_i$ is directly connected to the charges $q_i = Z_i e$.
In particular, $\overline{c}_\text{F}^{(i)} = 1 + a_i/Z_i $ is related to the anomalous magnetic moment $a_i$ of particle $i$ and its charge number $Z_e= -1$ and $Z_n = Z$.
For instance, we can relate $\overline{c}_\text{F}^{(e)}=g_e /2$ to the $g$ factor of the electron. 
Some Wilson coefficients are defined completely through other coefficients~\cite{Hill2013}; and specific values for electrons~\cite{Kinoshita1996} or protons~\cite{Peset2021} have been determined.

The third term  
\begin{align}
     \hat{\mathcal{H}}_{\text{cont}} =\hbar^3 \sum_{i,j} \frac{d_1^{(ij)} \hat{\psi}^\dagger_i \hat{\psi}_i \hat{\psi}^\dagger_j \hat{\psi}_j - d_{2}^{(ij)} \hat{\psi}^\dagger_i \boldsymbol{\hat{\sigma}}_i \hat{\psi}_i \cdot \hat{\psi}^\dagger_j \vect{\hat{\sigma}}_j \hat{\psi}_j}{m_i m_j c}
\end{align}
of the Hamiltonian density from Eq.~\eqref{eq:NRdensity} describes contact interactions through which fermions couple directly (Darwin-like contact interaction) and through their spin (spin-spin contact interaction).
The Wilson coefficients $d_1^{(ij)}$ and $d_2^{(ij)}$ are in lowest order proportional to the fine-structure constant $\alpha = e^2/(4 \pi \varepsilon_0 \hbar c)$ with vacuum permittivity $\varepsilon_0$.
These $d$-type Wilson coefficients are given in standard form, \ie{} factors of $Z_i Z_j$ are included in their definition.
However, compared to common definitions~\cite{Pineda1998b}, our coefficients differ by a factor of $1/2$ in the case of the interaction between two different particle species due to the summation over all values of $i$ and $j$.
For example, the interaction between electron and positron~\cite{Pineda1998b} corresponds to $d_1^{ee^{+}} = d_1^{e^{+}e} = d_s/2 = 3 \pi \alpha / 4 + \mathcal{O} (\alpha^2)$.
A similar identification can be made with our $d_2$ and $d_v$.
These contact terms solely arise from loop corrections~\cite{Pineda1998c}, such that they cannot be obtained from a pure tree-level treatment.
As a result, $\hat{\mathcal{H}}_\text{cont}$ is of order $\alpha/c$ and by that of $c^{-2}$~\footnote{We use for convenience $c$ to specify the order, but since $c$ is connected to other fundamental constants, it is more precise to fix it to $\alpha/c$ or equivalently $1/(\varepsilon_0 c^2)$.}.
The Hamiltonian neglects loop corrections of the order $c^{-2}$, which are suppressed by another $d$-type Wilson coefficient and are in fact of order $\alpha/c^2$.
Moreover, the NRQED Hamiltonian generally features also terms proportional to more than four fermionic field operators and there are also further contributions to the EM Hamiltonian which we both omitted here as they are beyond our order.

\subsection{Potential matching} \label{App:A1}

\begin{figure}[h]
    \centering
    \includegraphics[width=\columnwidth]{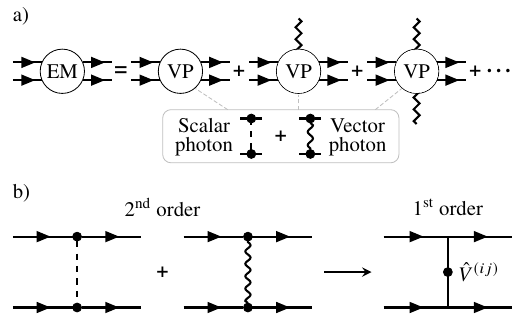}
    \caption{a) Set of all possible Feynman diagrams of two real fermions (which model in our treatment the constituents of a composite particle), interacting with EM fields.
    It can be described by the exchange of purely virtual photons (VP) and real photons (zigzag lines).
    The former can be any combination of scalar photons (dashed line), and vector photons (wiggly line) mediating the interaction between both fermions.
    This set of Feynman diagrams can be expanded into an increasing amount of scattering processes with real photons that account for external EM fields.
    b) The lowest-order nontrivial set of virtual photons includes all possible Feynman diagrams with one virtual scalar or vector photon (second-order scattering). 
    All diagrams are reduced to an effective first-order scattering by matching to an effective potential $\hat{V}^{(ij)}$ in the framework of pNRQED.}
    \label{fig:Feynman}
\end{figure}

The Hamiltonian from Eq.~\eqref{eq:NRdensity} allows a description of composite particles based on their fermionic constituents. 
However, the defining property of composite particles, \ie{} a bound-state potential due to EM interactions between fermions, does not appear explicitly yet.
Instead, it is contained in the EM fields which give rise to all allowed NRQED Feynman diagrams involving photons.
These photons may be categorized into real (external lines in Feynman diagrams) and virtual (internal lines in Feynman diagrams) photons.
The former describe all photons from external fields that scatter with the composite particle, the latter are virtual mediating EM interaction between the fermionic constituents of the composite particle. 
Such a separation is sketched in Fig.~\hyperref[fig:Feynman]{7a)} where all possible Feynman diagrams between two constituents (solid lines) may be written as a sum of all virtual photon diagrams (dashed and wiggly lines), that scatter an increasing number of real photons (zigzag lines). 

Integration over all positions and performing such a separation leads to the Hamiltonian
\begin{align} \label{eq:sepa}
    \hat{H} =& \hat{H}_\text{EM} + \displaystyle \sum_{i}  \int \dd[3]{x_i} \hat{\psi}^\dagger_i   \hat{h}_i \hat{\psi}_i 
     + \hat{H}_{\text{f-f}},
\end{align}
where the EM interaction between two fermions now explicitly appears in a modified fermion-fermion interaction Hamiltonian $\hat{H}_{\text{f-f}}$ absorbing also $\hat{H}_\text{cont}$, while the EM fields in the original single-fermion Hamiltonian [first term in Eq.~\eqref{eq:sepa}] are now only associated with real photons scattering with fermions (light-matter interaction).

To determine $\hat{H}_\text{f-f}$, one considers the first type of Feynman diagram in Fig.~\hyperref[fig:Feynman]{7a)} involving solely virtual photons.
According to Fig.~\hyperref[fig:Feynman]{7b)}, the interaction or scattering between two real fermions $i$ and $j$ follows to lowest order from a second-order scattering process, \ie{} two vertices in a Feynman diagram connected by a virtual scalar (dashed line) or vector photon (wiggly line).
These second-order interactions are reduced to effective first-order scattering processes with one vertex containing an instantaneous potential $\hat{V}^{(ij)}$.
Consequently, the resulting Hamiltonian takes the form
\begin{align} \label{eq:potentialNRQED}
    \hat{H}_{\text{f-f}} = \sum_{i,j} \int \dd[3]{x_i} \int \dd[3]{x_j^\prime} \hat{\psi}^\dagger_i \hat{\psi}^{\prime \, \dagger}_j \hat{V}^{(ij)} \hat{\psi}_j^\prime \hat{\psi}_i + \mathcal{O} \left(c^{-3}\right)
\end{align}
and gives rise to the effective field theory of \textit{potential nonrelativistic quantum electrodynamics}~\cite{Pineda1998a, Pineda1998, Pineda1998b, Brambilla2000, Adkins2018, Peset2021}.
In Eq.~\eqref{eq:potentialNRQED} we use the abbreviations $\hat{\psi}_i = \hat{\psi}_i ( \vect{x}_i )$, $\hat{\psi}_j^\prime = \hat{\psi}_j ( \vect{x}_j^\prime)$, and $\hat{V}^{(ij)} = \hat{V}^{(ij)} ( \vect{x}_i, \vect{x}_j^\prime, \vect{\hat{p}}_i, \vect{\hat{p}}_j^\prime, \vect{\hat{s}}_i, \vect{\hat{s}}_j^\prime )$.
The potential itself is determined by considering all possible virtual photons as indicated in Fig.~\hyperref[fig:Feynman]{7b)}, leading to potentials of order $c^{-2}$.

In the following, we review the actual matching procedure for the potentials~\footnote{The renormalization of the resulting theory is similar to the transition from QED to NRQED~\cite{Pineda1998}.} based on the reduction of scattering-matrix elements represented in Fig.~\hyperref[fig:Feynman]{7b)} giving rise to the Hamiltonian from Eq.~\eqref{eq:potentialNRQED}.
To this end, we move to an interaction picture in the Hamiltonian associated with Eq.~\eqref{eq:NRdensity} with respect to the Hamiltonian
\begin{align}
    \hat{H}_0 = \sum_i \int \dd[3]{x_i} \hat{\psi}_i^\dagger \left( m_i c^2 + \frac{\vect{\hat{p}}_i^2}{2m_i} \right) \hat{\psi}_i + \hat{H}_\text{EM}
\end{align} 
accounting for the free EM and fermion fields.
The resulting interaction Hamiltonian density
\begin{align} \label{eq:effective_matching_den}
    \hat{\mathcal{H}}_I =& \sum_i \hat{\psi}_i^\dagger \Bigg( q_i \hat{\phi} + \frac{ -\left\{ \vect{\hat{p}}, q_i \vect{\hat{A}} \right\} + q_i^2 \vect{\hat{A}}^2}{2m_i} - \overline{c}_\text{F}^{(i)} q_i \frac{\vect{\hat{s}}_i \cdot \vect{\hat{B}}}{m_i} \notag \\
    &- \overline{c}_\text{D}^{(i)} q_i \hbar^2 \frac{ \vect{\nabla} \cdot \vect{\hat{E}}}{8m_i^2c^2} - \overline{c}_\text{S}^{(i)} q_i \vect{\hat{s}}_i \cdot \frac{\pbar{} \times \vect{\hat{E}} - \vect{\hat{E}} \times \pbar{}}{4m_i^2c^2} \Bigg) \hat{\psi}_i
\end{align}
shows only those terms relevant for the matching in the order $c^{-2}$.
In particular, the kinetic correction and the last two terms of $\hat{h}_i$ in Eq.~\eqref{eq:NRdensity} give rise only to Feynman diagrams of orders higher than $c^{-2}$.
Moreover, the contact interaction has already the form of first-order scattering so that it does not give rise to additional second-order Feynman diagrams.
In this interaction picture, all field operators $\hat{\psi}_i$, $\hat{\phi}$, and $\vect{\hat{A}}$ depend on $x^\varrho = (ct, \vect{x})$ with $\varrho=0,1,2,3$, \ie{} they become explicitly time dependent.
The EM fields are connected to the  four potential $\hat{A}^\varrho = ( \hat{\phi}/c, \vect{\hat{A}})$ via $\vect{\hat{E}} = - \partial \vect{\hat{A}} / \partial t - \vect{\nabla} \hat{\phi} $ and $\vect{\hat{B}} = \vect{\nabla} \times \vect{\hat{A}}$.
With the interaction Hamiltonian density from Eq.~\eqref{eq:effective_matching_den} and the time-ordering operator $\hat{\mathcal{T}}$, we define the scattering matrix
$\hat{\mathcal{S}} = \hat{\mathcal{T}}  \exp{ - \frac{\ii}{\hbar c}\int \dd[4]{x} \hat{\mathcal{H}}_I (x) }$.
The actual matching corresponds to determining scattering-matrix elements up to a given order for the desired interactions~\cite{Pineda1998b}.
In our case, we match up to the order $c^{-2}$ for virtual photons between two fermions.
Hence, second-order scattering
\begin{align}
    \hat{\mathcal{S}}^{(2)} = \frac{1}{2!} \left( - \frac{\ii}{\hbar c} \right)^2  \int \dd[4]{x} \int \dd[4]{x^\prime} \hat{\mathcal{T}} \left\{ \hat{\mathcal{H}}_I (x) \hat{\mathcal{H}}_I (x^\prime) \right\}
\end{align}
is the first and only relevant scattering-matrix element whose time ordering is resolved by Wick's theorem~\cite{Wick1950, Peskin1995}, giving rise to only normally ordered contractions.
As a result, we select all contractions with two (real) fermions entering and exiting the scattering (constituents of the cobosons), \ie{} only contractions of EM fields, scalar photons $\bcontraction{}{\hat{A}}{{}_0}{\hat{A}} \hat{A}_0 \hat{A}_0^\prime$, and vector photons $\bcontraction{}{\hat{A}}{{}_r}{\hat{A}} \hat{A}_r \hat{A}_s^\prime$ with $r,s=1,2,3$, are involved.
After power counting~\cite{Pineda1998b}, we find that consistent matching up to the order $c^{-2}$ requires the contraction of all scalar and vector photons between terms in the first line of Eq.~\eqref{eq:effective_matching_den}.
The terms in the second line can only be contracted via a scalar photon with the Coulomb potential (compare with the Feynman diagrams depicted in Fig.~\ref{fig:potential}). 

\subsubsection{Coulomb gauge} \label{App:A1C}
To determine the matrix elements, a choice of gauge is required to resolve the contractions.
First, we move to Coulomb gauge where contractions of the scalar and vector photon take the form~\cite{Berestetskii1982}
\begin{subequations}
\begin{align} \label{eq:Coulomb_scalar}
    \bcontraction{}{\hat{A}}{{}_0}{\hat{A}} \hat{A}_0 \hat{A}_0^\prime=& \ii \hbar  \frac{\delta ( x_0 -x_0^\prime ) }{ 4 \pi \varepsilon_0 c \abs{\vect{x} - \vect{x}^\prime }}
\end{align}
and
\begin{align}
\begin{split}
        \bcontraction{}{\hat{A}}{{}_r}{\hat{A}} \hat{A}_r \hat{A}_s^\prime =& \frac{\hbar \mu_0 c}{2} \frac{1}{\left( 2\pi \right)^{3}} \int \dd[3]{k} \ee^{ \ii \vect{k} \cdot \left( \vect{x} - \vect{x}^\prime \right) } \left( \frac{\delta_{rs}}{\abs{\vect{k}}} - \frac{k_r k_s}{\abs{\vect{k}}^3} \right) \\
        &\times \left( \ee^{- \ii \abs{\vect{k}} ( x_0 - x_0^\prime )}  \Theta ( x_0 - x_0^\prime ) + \left( x_0 \leftrightarrow x_0^\prime \right) \right)
\end{split}
\end{align}
\end{subequations}
with the Heaviside step function $\Theta(x_0 - x_0^\prime)$. 
Further, contractions between the vector and scalar potential $\bcontraction{}{\hat{A}}{{}_r}{\hat{A}} \hat{A}_r \hat{A}_0^\prime =0$ vanish.
Multiplying all elements of $\hat{\mathcal{H}}_I ( x ) \hat{\mathcal{H}}_I ( x^\prime )$ yields a term proportional to $\hat{\phi} \hat{\phi^\prime}$, whose contraction leads to the Coulomb potential and corresponds to the first Feynman diagram in Fig.~\ref{fig:potential}.
This particular second-order matrix element
\begin{align} \label{eq:Coulomb:second}
    \hat{S}^{(2)}_{\text{C}} = \left( - \frac{ \ii }{\hbar} \right)^2 \sum_{i,j} \frac{q_i q_j}{2!} \int \dd[4]{x} \int \dd[4]{x^\prime} \hat{\psi}_{i}^\dagger \hat{\psi}_{j}^{\prime \, \dagger }\bcontraction{}{\hat{A}}{{}_0}{\hat{A}} \hat{A}_0 \hat{A}_0^\prime\hat{\psi}_{j}^\prime\hat{\psi}_{i}
\end{align}
contains abbreviations $\hat{\psi}_i= \hat{\psi}_i \left( x \right)$ and $\hat{\psi}_j^\prime = \hat{\psi}_j \left( x^\prime \right)$, where the order of fermion operators takes both commuting and anticommuting field operators into account.
Inserting the contraction of the scalar potential from Eq.~\eqref{eq:Coulomb_scalar} yields
\begin{align}
     \hat{S}^{(2)}_{\text{C}} =  - \frac{ \ii }{\hbar} \int \dd{t} \int \dd[3]{x} \int \dd[3]{x^\prime} \sum_{i,j=e,n} \hat{\psi}_{i}^\dagger \hat{\psi}_{j}^{\prime \, \dagger} V^{(ij)}_{\text{C}} \hat{\psi}_{j}^\prime \hat{\psi}_{i}
\end{align}
and the corresponding Coulomb potential $\hat{V}^{(ij)}_{\text{C}} = q_i q_j/(8 \pi \varepsilon_0 \abs{\vect{x} - \vect{x}^\prime })$.
Thus, we reduced the second-order scattering process associated with the Coulomb interaction to effective first-order scattering whose corresponding Hamiltonian density takes the same form as was presented in the fermion-fermion Hamiltonian in Eq.~\eqref{eq:potentialNRQED}.

Moving first to the contribution proportional to $\left\{ \hat{p}_r , \hat{A}^{(r)} \right\} \left\{ \hat{p}^\prime_s , \hat{A}^{\prime \, (s)} \right\} $, \ie{} the case where external photons (red zigzag lines in Fig.~\ref{fig:potential} are not yet accounted), we find the matrix element
\begin{align} 
\begin{split}
    \hat{S}^{(2)}_{\text{LL}} =& \left( - \frac{ \ii }{\hbar c} \right)^2 \frac{1}{2!} \sum_{i,j=e,n} \frac{q_i q_j}{m_i m_j} \int \dd[4]{x} \int \dd[4]{x^\prime} \\
    &\times \hat{\psi}_{i}^\dagger \hat{\psi}_{j}^{\prime \, \dagger} \hat{p}_r \hat{p}_s^\prime \bcontraction{}{\hat{A}}{{}_r}{\hat{A}} \hat{A}_r \hat{A}_s^\prime \hat{\psi}_{j}^\prime \hat{\psi}_{i}
\end{split}
\end{align}
with the help of $\vect{\hat{p}}_\ell \cdot \hat{\vect{A}}_\ell = \hat{\vect{A}}_\ell \cdot \vect{\hat{p}}_\ell $ in Coulomb gauge.
The contraction of the vector photon is not an exact delta function in time, \ie{} not instantaneous.
However, by partial integration with respect to one temporal coordinate we extract from the instantaneous part of the matrix element 
\begin{align}
         \hat{S}^{(2)}_{\text{LL}} =   \frac{- \ii }{\hbar} \int \dd{t} \int \dd[3]{x} \int \dd[3]{x^\prime} \sum_{i,j} \hat{\psi}_{i}^\dagger \hat{\psi}_{j}^{\prime \, \dagger} \hat{V}^{(ij)}_{\text{LL}} \hat{\psi}_{j}^\prime \hat{\psi}_{i} 
\end{align}
the potential  
\begin{align}
    \hat{V}^{(ij)}_{\text{LL}} = \frac{4\pi \kappa_{ij}}{\left( 2\pi \right)^{3}} \int \dd[3]{k} \ee^{ \ii \vect{k} \cdot \left( \vect{x} - \vect{x}^\prime \right) } \left( \frac{\delta_{rs}}{\abs{\vect{k}}^2} - \frac{k_r k_s}{\abs{\vect{k}}^4} \right) \hat{p}_r \hat{p}_s^\prime
\end{align}
associated with the orbit-orbit coupling, while the remainder of the integral is of higher order and thus neglected.
Here, we introduce $\kappa_{ij}=-q_i q_j /(8 \pi \varepsilon_0 m_i m_j c^2)$.
After performing the Fourier transform we find
\begin{align}
     \hat{V}^{(ij)}_{\text{LL}} = \frac{\kappa_{ij}}{2} \left( \frac{1}{ \abs{\vect{r}}} \vect{\hat{p}} \cdot \vect{\hat{p}}^\prime + \frac{1}{\abs{\vect{r}}^3} \vect{r} \cdot \left( \vect{r} \cdot \vect{\hat{p}} \right) \vect{\hat{p}}^\prime \right).
\end{align}
It is straightforward to show the equivalence to the form given in Fig.~\ref{fig:potential}.

The remaining potentials in relevant order are derived in a completely analogous procedure and are therefore summarized by Fig.~\ref{fig:potential} showing all relevant Feynman diagrams and their corresponding terms contributing to $\hat{V}^{(ij)}$.
The full potential $\hat{V}^{(ij)}$ is given with respect to single-fermion coordinates up to order $c^{-2}$.
In addition to the explicitly derived lowest-order Coulomb interaction $\hat{V}^{(ij)}_{\text{C}}$ and orbit-orbit $\hat{V}^{(ij)}_{\text{LL}}$ interaction, the potential is completed by spin-orbit $\hat{V}^{(ij)}_{\text{LS}}$, spin-spin $\hat{V}^{(ij)}_{\text{SS}}$, Darwin $\hat{V}^{(ij)}_{\text{D}}$, and the contact interaction.
The last term already had the form of a fermion-fermion interaction.
These potentials are also known as part of the \textit{Breit-Pauli} Hamiltonian~\cite{Breit1929, Breit1930,Breit1932}, however, augmented by QED corrections in our description.

\subsubsection{Lorenz gauge} \label{App:A2}
Figure~\ref{fig:potential} presents also the potentials calculated in Lorenz gauge~\footnote{In Lorenz gauge the EM four potential may be quantized with respect to a weaker Lorenz-gauge condition in the spirit of Gupta and Bleuler~\cite{Gupta1950,Bleuler1950} or with the help of BRST quantization~\cite{Becchi1975, Tyutin1975}.}. When we use this gauge instead of Coulomb gauge to determine the potentials, the general procedure remains identical but we need to take into account that $\vect{\hat{p}}_\ell \cdot \hat{\vect{A}}_\ell \neq \hat{\vect{A}}_\ell \cdot \vect{\hat{p}}_\ell $.
The contraction then reads
\begin{align}
\begin{split}
    \bcontraction{}{\hat{A}}{{}_\mu}{\hat{A}} \hat{A}_\mu \hat{A}_\nu^\prime =& - \frac{\hbar \mu_0 c}{2} \frac{ \eta_{\mu \nu} }{\left( 2\pi \right)^3} \int \dd[3]{k} \frac{ \ee^{ \ii \vect{k} \cdot \left( \vect{x} - \vect{x}^\prime \right) }}{\abs{\vect{k}}} \\
    &\times \left( \ee^{- \ii \abs{\vect{k}} ( x_0 - x_0^\prime)} \Theta( x_0 - x_0^\prime) + \left( x_0 \leftrightarrow x_0^\prime \right)  \right).
\end{split}
\end{align}
All potentials are identical to the ones obtained in Coulomb gauge except for the orbit-orbit and Coulomb potentials.
For the first, we find
\begin{align}
    \hat{V}^{(ij)}_{\text{LL}} = \frac{\kappa_{ij}}{\abs{\vect{r}}^3} \left( \vect{\hat{\ell}} \cdot \vect{\hat{\ell}}^\prime + \left( \vect{r} \cdot \vect{\hat{p}} \right) ( \vect{r} \cdot \vect{\hat{p}}^\prime) \right) - \kappa_{ij} \pi \hbar^2 \delta ( \vect{r} ).
\end{align}
The Coulomb potential is of order $c^0$, but in Lorenz gauge the scalar photon propagator is not instantaneous in time.
As a consequence, there is a non-negligible remaining integral after a partial integration.
The instantaneous part of this matrix element corresponds to the Coulomb potential and the remainder yields a term of order $c^{-2}$.
We resolve the second part as well with the help of partial integration in time and use consecutively the continuity equation
\begin{align}
    \pdv{t} \left( \hat{\psi}_i^\dagger \hat{\psi}_i \right) = \frac{ \ii \hbar}{2m_i}  \vect{\nabla} \cdot \left( \hat{\psi}_i^\dagger \left[\vect{\nabla} \hat{\psi}_i \right] + \left[\vect{\nabla} \hat{\psi}_i^\dagger \right] \hat{\psi}_i \right) 
\end{align}
to remove partial derivatives in time.
This procedure leads to the potential
\begin{align}
    \hat{V}^{(ij)}_{\text{C}} = \frac{q_i q_j}{8 \pi \varepsilon_0 \abs{\vect{r}}} - \kappa_{ij} \frac{1}{2\abs{\vect{r}}^3} \vect{\hat{\ell}} \cdot \vect{\hat{\ell}}^\prime + \kappa_{ij}  \pi \hbar^2 \delta ( \vect{r} )
\end{align}
showing that the sum of all potentials is identical in both gauges.

\subsection{Matching potentials with external photons}
\begin{figure}[h]
    \centering
        \includegraphics[width=\columnwidth]{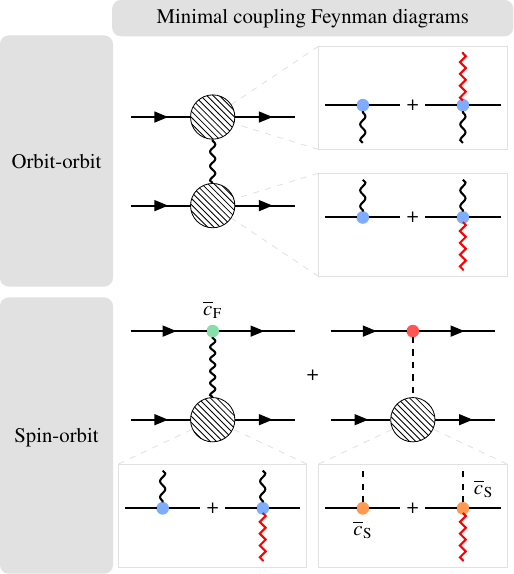}
        \caption{Canonical momenta appear in the effective potentials of the orbit-orbit (LL) coupling $\hat{V}^{(ij)}_{\text{LL}}$ and the spin-orbit (LS) coupling $\hat{V}^{(ij)}_{\text{LS}}$.
        To replace these by minimally coupled momentum operators, we augment the orbit-orbit Feynman diagram by three additional diagrams: two fermions $i$ and $j$ exchanging a virtual photon while they scatter a real photon, respectively, together with the diagram where simultaneously both fermions scatter a  real photon.
        Both spin-orbit diagrams exchanging a virtual vector or a virtual scalar photon, respectively, have to be augmented by one additional diagram describing the scattering of a real photon during this process.}
    \label{fig:mincoup}
\end{figure}

So far we discussed the potential between two fermions due to EM interactions mediated by virtual photons that will eventually give rise to the binding potential of composite particles.
In addition, we also aim to describe light-matter interaction between composite particles and external light fields.
The scattering process of a real photon with a composite particle contains fermion-photon interactions that correspond to the EM fields appearing in the single-fermion Hamiltonian $\hat{h}_i$ from Eq.~\eqref{eq:sepa}.
Since the constituents form a bound system, we have to include also the case of a real photon that scatters from two fermions exchanging a virtual photon.
This process is not yet accounted for in $\hat{H}_\text{f-f}$, since no minimally coupled momentum operators appear in the potential that originates in the first term of Fig.~\hyperref[fig:Feynman]{7a)}.
We incorporate this case in $\hat{H}_\text{f-f}$ by including Feynman diagrams according to the remaining two terms in Fig.~\hyperref[fig:Feynman]{7a)}.
The relevant additional Feynman diagrams with external photons (depicted in red) are summarized in Fig.~\ref{fig:mincoup}.
Together with these additional diagrams, we find potentials in which the momentum operators are replaced by minimally coupled momenta to arrive at the potentials presented in Fig.~\ref{fig:potential}.
Here, we use Coulomb gauge ($\vect{\nabla} \cdot \vect{\hat{A}} = 0$) to determine the matching, and consequently our remaining external photons are also fixed to this gauge from now on. 
In this case, the vector potential can be decomposed as
 \begin{align} \label{Eq:vecpotC}
 \vect{\hat{A}} = \sum_{r=1}^{2} \frac{1}{(2\pi)^{3/2}} \int \dd[3]{k} \mathcal{A}_{\vect{k}} \vect{e}_{r} \left( \vect{k} \right) \left[ \hat{a}_r \left(\vect{k} \right) \ee^{\ii \vect{k} \cdot \vect{x}} + \text{H.c.} \right].
 \end{align}
We introduced the vacuum amplitude $\mathcal{A}_{\vect{k}} =\sqrt{\hbar/2 \varepsilon_0 c \abs{\vect{k}}} $ and unit polarization vectors $\vect{e}_{r} \left( \vect{k} \right)$ corresponding to the wave vector $\vect{k}$.
In the interaction picture used in Appendix~\ref{App:A1C} the vector potential is time dependent and its nonvanishing commutator is responsible for the propagator of the electromagnetic field. 
However, since we work in the Schrödinger picture here, all field operators including the vector potential given in Eq.~\eqref{Eq:vecpotC} are time independent.
Thus, the commutators of the field-operator components commute such that $\left[ \hat{A}_u ( \vect{x} ), \hat{A}_v ( \vect{x}^\prime) \right] =0$ for $u,v=x,y,z$.

The scalar potential $\hat{\phi}$ contains both real and all virtual photons arising from contractions, such as the Coulomb potential $\hat{V}^{(ij)}_{\text{C}}$.
In Coulomb gauge, there are no real scalar photons in the absence of a free charge density sourcing the external field, and up to order $c^{-2}$ we determined all possible virtual scalar-photon contributions by collecting them in $\hat{H}_{\text{f-f}}$.
Consequently, we set $\hat{\phi} = 0$~\footnote{Dropping the scalar potential implies that higher order scalar photons become inaccessible. 
But when moving to the next order, $\hat{H}_\text{f-f}$ has to be redetermined anyway for a consistent treatment.}.
The self-energy of the fermions, being the main contribution to the Lamb shift~\cite{Bethe1947, Lamb1947}, can still be determined since only the virtual vector photon contributes in Coulomb gauge to it. 

The result is a full minimally coupled Hamiltonian
\begin{align} \label{eq:Ham_ele}
\begin{split}
    \hat{H} =& \hat{H}_\text{EM} +  \sum_{i}  \int \dd[3]{x_i} \hat{\psi}^\dagger_i  \hat{h}_i \hat{\psi}_i  \\
    & + \sum_{i,j} \int \dd[3]{x_i} \int \dd[3]{x_j^\prime} \hat{\psi}^\dagger_i \hat{\psi}^{\prime \, \dagger}_j \hat{V}^{(ij)} \hat{\psi}_j^\prime \hat{\psi}_i  
\end{split}
\end{align} 
where the potential $\hat{V}^{(ij)} = \hat{V}^{(ij)} ( \vect{x}_i, \vect{x}_j^\prime, \pbar{}, \pbarr{j}{\prime}, \vect{\hat{s}}_i, \vect{\hat{s}}_j^\prime)$ is now a function of minimally coupled momenta and all EM fields are given in Coulomb gauge.

\section{Multipolar cobosonic Hamiltonian from unitary transformations} \label{App:D}
In this Appendix, we present the transformation of the coboson Hamiltonian into its multipolar form including relativistic corrections of c.m. and relative coordinates, and provide the full expressions omitted in the main body of the paper.
In Sec.~\ref{Sec:Unitaries} we demonstrated that the transformation of the field-theoretical Hamiltonian  can be reduced to its single-particle counterpart, summarized in Eq.~\eqref{eq:trafos}.
This Appendix gives details on the respective single-particle transformations.

As discussed in Sec.~\ref{Sec:Unitaries}, we first perform a unitary transformation to introduce relativistic corrections to c.m. and relativistic coordinates, before we transform in a second step the resulting operators with the help of the PZW transformation.
Because only the lowest-order relativistic correction is of relevance, we find for any single-particle operator $\vect{\hat{O}}$ the transformation $ \hat{u}^{(\text{rel}) \, \dagger} \vect{\hat{O}} \hat{u}^{(\text{rel})} =   \vect{\hat{O}}  - \ii \left[ \hat{\lambda}^{(\text{rel})}, \vect{\hat{O}} \right] / \hbar= \vect{\hat{O}}+ \vect{\hat{O}}^{(1)} $.
The generator $\hat{\lambda}^{(\text{rel})}$ already given in Eq.~\eqref{eq:relgen} has the form
\begin{align}
    \hat{\lambda}_k^{(\text{rel})} =& \frac{ \vect{r}_k \cdot \bars{P}_k }{4M^2c^2} \left[ \bars{p}_k \cdot \bars{P}_k + \Delta m \left(  \frac{ \bars{p}_k^{2}}{m_{\text{r}}}  + \frac{q_eq_n}{4 \pi \varepsilon_0 \abs{\vect{r}_k}} \right) \right] + \text{H.c.} \notag \\
    &- \frac{1}{4m_{\text{r}} Mc^2} \left(  \bars{p}_k \times \bars{P}_k + \text{H.c.}  \right)  \cdot\vect{\hat{s}}_k.
\end{align}
The subsequent PZW transformation is performed with the help of the generator   $\hat{\lambda}_k^{(\text{PZW})} = \int \dd[3]{y} \vect{\mathcal{P}}_k \left( \vect{y} \right) \cdot \vect{\hat{A}} \left(\vect{y} \right)$, where $\vect{\mathcal{P}}_k$ is a polarization field of the $k$th coboson and is defined in Eq.~\eqref{eq:PZW}. 
According to Eq.~\eqref{eq:trafos}, we discuss in the following the individual transformations of the single-coboson Hamiltonian, the coboson scattering potential, and the EM Hamiltonian.

\subsection{Single-coboson Hamiltonian}
The single-coboson Hamiltonian from Eq.~\eqref{eq:trafoa} includes the operators $\bars{P}$, $\bars{p}$, $1/\abs{\vect{r}}$, $\vect{\hat{B}}$, $\vect{\hat{s}}_i$, and $\vect{\hat{E}}$~\footnote{We drop the subscript ``1'' for operators in the single-coboson sector} that have to be transformed.

\subsubsection{Relativistic corrections}
To compute the relativistic corrections $\vect{\hat{O}}^{(1)}$, we first calculate the commutators between the components $l,m,n=x,y,z$ of minimally coupled momenta to
\begin{subequations}
\begin{align}
    \left[ \, \hat{\overline{\! P}}_\ell \, , \, \hat{\overline{\! P}}_m \right] &= \ii \hbar \varepsilon_{\ell m n} Q \hat{B}_{Q,n}, \\
    \left[ \, \hat{\overline{\! p}}_\ell \, , \, \hat{\overline{\! P}}_m \right] &= \ii \hbar \varepsilon_{\ell m n} q_1 \hat{B}_{q_1,n}, \, \text{and} \\
        \left[ \, \hat{\overline{\! p}}_\ell \, , \, \hat{\overline{\! p}}_m \right] &= \ii \hbar \varepsilon_{\ell m n} q_2 \hat{B}_{q_2,n}.
\end{align}
\end{subequations}
Here, we introduce the abbreviation $\vect{\hat{B}}_{q_r} = \sum_i \text{sgn} (-q_i)^{r} q_i ( m_\text{r} / m_i )^r \vect{\hat{B}}(\vect{x}_i ) / q_r$ with $r=0,1,2$ that contains the weighted charge $q_r =  \sum_i \text{sgn} (-q_i)^{r} q_i ( m_\text{r} / m_i )^r $ and is chosen such that the lowest-order multipole expansion of $\vect{\hat{B}}_{q_r}$ coincides with $ \vect{\hat{B}}(\vect{R})$.
In the particular case $q_0 = Q$ we write the total charge $Q$ and $\text{sgn}$ is the sign function.
With the commutators of minimally coupled momenta we determine all corrections, which are presented in Table~\ref{tab:relO}.
\begin{table*}
\caption{\label{tab:relO}%
Lowest-order relativistic corrections obtained by the transformation $ \hat{u}^{(\text{rel}) \, \dagger} \vect{\hat{O}} \hat{u}^{(\text{rel})} = \vect{\hat{O}}+ \vect{\hat{O}}^{(1)} $ for the c.m. momentum, the relative momentum, Coulomb-like terms, the magnetic field at the position of a fermion, and its spin. 
They are obtained by calculating $\vect{\hat{O}}^{(1)} = - \ii \left[ \hat{\lambda}^{(\text{rel})}, \vect{\hat{O}} \right] / \hbar$. 
They depend on the operators $\vect{\hat{\delta}}_{\vect{r}} ,\vect{\hat{\delta}}_{\vect{R}} $ and $\vect{\hat{\delta}}_{\vect{B}} $ given in Eqs.~\eqref{eq:operators_delta} and~\eqref{eq:delta_B} as well as the terms $\vect{\hat{B}}_{q_r} = \sum_i \text{sgn} (-q_i)^{r} q_i ( m_\text{r} / m_i )^r \vect{\hat{B}}(\vect{x}_i ) / q_r$ and $\vect{\hat{B}}_{rs} =  q_r \vect{\hat{B}}_{q_r} + \frac{\Delta m}{m_\text{r}} q_s \vect{\hat{B}}_{q_s} $ with a weighted charge  $q_r =  \sum_i \text{sgn} (-q_i)^{r} q_i ( m_\text{r} / m_i )^r $.
}
\begin{ruledtabular}
    \begin{tabular}{cc}
        $\vect{\hat{O}}$ & $\vect{\hat{O}}^{(1)}$ \\
        \addlinespace \colrule \addlinespace
       $\bars{P}$  & $-\vect{\hat{\delta}}_{\vect{R}} \times Q \vect{\hat{B}}_Q - \vect{\hat{\delta}}_{\vect{r}} \times q_1 \vect{\hat{B}}_{q_1} + \text{H.c.} - \frac{\hbar^2}{4M^2c^2} ( \vect{r} \cdot \vect{\nabla} ) \vect{\nabla} \times \vect{\hat{B}}_{12} $  \\ \addlinespace
         $\bars{p}$ & $ -\vect{\hat{\delta}}_{\vect{R}} \times q_1 \vect{\hat{B}}_{q_1} - \vect{\hat{\delta}}_{\vect{r}} \times q_2 \vect{\hat{B}}_{q_2} + \text{H.c.} - \frac{\hbar^2}{4M^2c^2} ( \vect{r} \cdot \vect{\nabla} ) \vect{\nabla} \times \vect{\hat{B}}_{23} + \frac{1}{4M^2c^2} \left[ \bars{P} \left( \bars{p} \cdot \bars{P} + \frac{\Delta m}{m_\text{r}} \bars{p}^{2} + \frac{\Delta m q_e q_n}{4 \pi \varepsilon_0 \abs{\vect{r}}} \right) - \frac{\Delta m q_e q_n}{4\pi \varepsilon_0} \vect{r} \cdot \bars{P} \frac{\vect{r}}{\abs{\vect{r}}^3} + \text{H.c.}\right] $ \\ \addlinespace
        $1/\abs{\vect{r}}$ & $ \frac{1}{\abs{\vect{r}}^3} \frac{\big( \, \bars{P} \cdot \vect{r} \big)^2}{2M^2c^2} + \frac{\Delta m}{2m_\text{r} Mc^2} \left( \frac{1}{\abs{\vect{r}}^3} \left( \bars{P} \cdot \vect{r} \right) \left( \vect{r} \cdot \bars{p} \right) + \text{H.c.} \right) - \frac{1}{2 m_\text{r}Mc^2} \frac{1}{\abs{\vect{r}}^3} \vect{r} \times \bars{P} \cdot \vect{\hat{s}}$ \\ \addlinespace
        $\vect{\hat{B}} ( \vect{x}_i )$ & $ \left(\vect{\hat{\delta}}_{\vect{\hat{B}}} \cdot \vect{\nabla} \right) \vect{\hat{B}} + \text{H.c.} + \frac{\hbar^2}{4M^2c^2} \frac{q_i}{\abs{q_i}} \frac{m_\text{r}}{m_i} \left( 1 - \frac{\Delta m}{m_i} \frac{q_i}{\abs{q_i}} \right) \left( \vect{r} \cdot \vect{\nabla} \right) \vect{\nabla}^2 \vect{\hat{B}}$ \\ \addlinespace
        $\vect{\hat{s}}_i$ & $\vect{\hat{s}}_i \times \frac{q_i}{\abs{q_i}} \frac{ \bars{p} \times \, \bars{P} + \text{H.c.}}{4m_iMc^2}$ \\
    \end{tabular}
    \label{tab:my_label}
\end{ruledtabular}
\end{table*}
We see that c.m. and relative momentum are modified by light-induced corrections with the general $\vect{r} \times \vect{\hat{B}}$ structure.
In Table~\ref{tab:relO} we introduced the operators
\begin{subequations}
\label{eq:operators_delta}
\begin{align}
    \vect{\hat{\delta}}_{\vect{r}} =&  \frac{ \vect{r} \cdot \bars{P}}{4M^2c^2}  \left( \bars{P} + 2 \frac{\Delta m}{m_\text{r}} \bars{p} \right) - \frac{ \bars{P} \times \vect{\hat{s}}}{4 m_\text{r} M c^2} \\
    \vect{\hat{\delta}}_{\vect{R}} =& \frac{ \vect{r} \cdot \bars{P} }{4M^2c^2} \bars{p}    + \frac{ \bars{p} \cdot \bars{P} + \Delta m \Big( \frac{ \bars{p}^{2}}{m_\text{r}} + \frac{q_e q_n}{4 \pi \varepsilon_0 \abs{\vect{r}}} \Big)}{4M^2c^2} \vect{r} + \frac{ \bars{p} \times \vect{\hat{s}}}{4m_\text{r}Mc^2} 
\end{align}
\end{subequations}
where $\vect{\hat{\delta}}_{\vect{r}}$ and $\vect{\hat{\delta}}_{\vect{R}}$ follow from a commutator involving the relative and  c.m. momentum, respectively.
Moreover, terms arise that are of the form of a second-order multipole expansion of the magnetic field $\vect{\hat{B}}_{rs} =  q_r \vect{\hat{B}}_{q_r} + \frac{\Delta m}{m_\text{r}} q_s \vect{\hat{B}}_{q_s} $. 

While the c.m. momentum contains only light-induced corrections, the relative momentum has an additional correction that is not induced by magnetic fields.
The correction to the Coulomb potential coincides~\cite{Sonnleitner2018} with the one if light-field corrections are neglected, but the canonical c.m. and relative momenta are exchanged by minimally coupled ones.
Finally, the correction to the magnetic field may be rewritten into a second-order multipole expansion form but with the operator
\begin{align}
\label{eq:delta_B}
\begin{split}
    \vect{\hat{\delta}}_{\vect{\hat{B}}} =&  - \frac{\bars{p} \cdot \bars{P} + \frac{\Delta m}{m_\text{r}} \bars{p}^{\,2} + \frac{\Delta m q_e q_n}{4 \pi \varepsilon_0 \abs{\vect{r}}}}{4M^2c^2} \vect{r} + \frac{q_i}{\abs{q_i}} \frac{m_\text{r}}{m_i} \frac{ \bars{P} \cdot \vect{r}}{4M^2c^2} \bars{P}  \\ 
    &-  \left( 1 - 2 \frac{q_i}{\abs{q_i}} \frac{\Delta m}{m_i} \right) \frac{ \bars{P} \cdot \vect{r} }{4M^2c^2} \bars{p} - \frac{ \bars{p} + \frac{q_i}{\abs{q_i}} \frac{m_\text{r}}{m_i} \bars{P}  }{4m_\text{r}Mc^2} \times \vect{\hat{s}}
\end{split}
\end{align}
together with a contribution proportional to the Laplacian of the magnetic field.
Since the electric fields are already of the order $c^{-2}$ in the Hamiltonian, there are no further relevant corrections.

\subsubsection{PZW transformation}
We perform the PZW transformation in Coulomb gauge where the vector potential commutes with itself and the magnetic field at any position.
Thus, only the transformation of minimally coupled momenta and the electric field is remaining.
As a result, we find the transformations~\cite{Baxter1993,Steck}
\begin{subequations}
\begin{align}
     \bars{P} \to& \vect{\hat{P}}_{\text{PZW}} =  \vect{\hat{P}}_Q + \vect{\hat{F}}^{(\text{cm})} \\
     \bars{p} \to& \vect{\hat{p}}_\text{PZW} = \vect{\hat{p}} + \vect{\hat{F}}^{(\text{rel})} \\
     \vect{\hat{E}}^\perp(\vect{y}) \to& \vect{\hat{E}}^\perp(\vect{y}) - \frac{1}{\varepsilon_0} \vect{\mathcal{P}}^\perp ( \vect{y}).
\end{align}
\end{subequations}
In the case of ions with $Q\neq 0$, the c.m. momentum couples minimally to a monopole evaluated at the c.m. position of the vector potential, \ie{} $\vect{\hat{P}}_Q = \vect{\hat{P}} - Q \vect{\hat{A}}(\vect{R})$.
Moreover, both c.m. and relative momenta are modified by generalized $\vect{r} \times \vect{\hat{B}}$ summarized by
\begin{subequations}
\begin{align}
    \vect{\hat{F}}^{(\text{cm})} &=  \int \dd[3]{y} \vect{\mathcal{P}} \left( \vect{y} \right) \times \vect{\hat{B}} \left( \vect{y} \right) \\
    \vect{\hat{F}}^{(\text{rel})} &= \sum_{j=e,n} q_j \frac{m_{\text{r}}^2}{m_j^2} \vect{r} \times \int \limits_{0}^{1} \dd{\rho} \rho \vect{\hat{B}} \left(\vect{R} +\rho (\vect{x}_j-\vect{R} ) \right),
\end{align}
\end{subequations}
where the c.m. momentum now includes the polarization field defined in Eq.~\eqref{eq:PZW}.
The electric fields in the single-coboson Hamiltonian are evaluated at positions $\vect{x}_i$.
Since the polarization field $\vect{\mathcal{P}} (\vect{x}_i) = 0$ vanishes, there is no electric field contribution from the single-coboson electric fields.

\subsubsection{Hamiltonian}
After we insert the PZW-transformed momenta, also into the corrections from Table~\ref{tab:relO}, the transformed single-coboson Hamiltonian $\hat{h}_\text{Cb}$ from Eq.~\eqref{eq:trafoa} resolves to
\begin{align} \notag
     \hat{h}_\text{MpCb}^\prime =& Mc^2 + \frac{\vect{\hat{P}}^2_Q}{2 M} \left( 1 + \frac{\hat{h}_\text{rel}^{(0)}}{Mc^2} \right) - \frac{\vect{\hat{P}}^4_Q}{8M^3c^2} + \hat{h}_\text{rel}^{(0)} + \hat{h}_\text{rel}^{(1)}\\ 
    &+ \hat{h}_\text{IB}^{(0)} + \hat{h}_\text{IB}^{(1)}.
\end{align}
The prime indicates that electric field contributions that arise from the transformation of the EM Hamiltonian are not yet included. 
The relative Hamiltonian is identical to Eq.~\eqref{eq:Ham_int1} and $\hat{h}_\text{IB}^{(0)}$ contains the magnetic field contribution of lowest-order light-matter interaction.
It is listed in Table~\ref{tab:Ham_I}.
The additional part, magnetic field contributions to light-matter interaction at the order of $c^{-2}$, are collected in
\begin{widetext}
\begin{align} \notag
    \hat{h}_\text{IB}^{(1)} =& - \left( \frac{\vect{\hat{P}}_\text{PZW}^2 \hat{h}_\text{rel,PZW}^{(0)} - \vect{\hat{P}}_Q^2 \hat{h}_\text{rel}^{(0)} }{4M^2c^2} + \text{H.c.} \right) - \frac{\vect{\hat{P}}_\text{PZW}^4- \vect{\hat{P}}_Q^4}{8M^3c^2} - \frac{m_n^3 + m_e^3}{M^3} \frac{\vect{\hat{p}}_\text{PZW}^4 - \vect{\hat{p}}^4}{8m_\text{r}^3c^2} \\ \notag
    &-\frac{\kappa}{\abs{\vect{r}}} \left( \vect{\hat{p}}_\text{PZW}^2 - \vect{\hat{p}}^2 \right) + \frac{\kappa}{\abs{\vect{r}}^3} \vect{r} \times \vect{\hat{F}}^{(\text{rel)}} \cdot \left( \alpha_{\ell S} \vect{\hat{S}} + \alpha_{\ell s} \vect{\hat{s}} \right) + \sum_i \left[ \overline{c}_\text{S}^{(i)} q_i \frac{\left( \frac{m_i}{M} \vect{\hat{P}}_\text{PZW} - \frac{q_i}{\abs{q_i}} \vect{\hat{p}}_\text{PZW} \right) \times \vect{\hat{E}} }{4m_i^2c^2} \cdot \vect{\hat{s}}_i + \text{H.c.} \right] \\ \notag
    &+ \sum_i \overline{c}_\text{W1}^{(i)} q_i \frac{\left\{ \Big( \vect{\hat{P}}_\text{PZW} - \frac{q_i}{\abs{q_i}} \frac{m_\text{r}}{m_i} \vect{\hat{p}}_\text{PZW} \Big)^2, \vect{\hat{s}}_i \cdot \vect{\hat{B}} (\vect{x}_i )\right\}}{4m^3_ic^2} - \sum_i \overline{c}_\text{A1}^{(i)} q^2_i \hbar^2 \frac{\vect{\hat{B}}^2(\vect{x}_i)}{8 m_i^3 c^2} + \frac{ \left\{ \vect{\hat{P}}_\text{PZW}^{(1)}, \vect{\hat{P}}_\text{PZW}\right\}}{2M} +  \frac{ \left\{ \vect{\hat{p}}_{\vect{B},\text{PZW}}^{(1)}, \vect{\hat{p}}_\text{PZW}\right\}}{2m_\text{r}} \\ \notag
    & - \sum_i \left( \vect{\hat{\mu}}_{i, \text{PZW}}^{(1)} \cdot \vect{\hat{B}} + \vect{\hat{\mu}}_i \cdot \vect{\hat{B}}_\text{PZW}^{(1)} \right) + \frac{\hbar^2}{16m_\text{r}M^2c^2} \! \left[ \vect{\hat{p}}_\text{PZW} \cdot \vect{\nabla} \times q_1 \vect{\hat{B}}_{q_1} - \left( \vect{\hat{P}}_\text{PZW} + 2 \frac{\Delta m}{m_\text{r}} \vect{\hat{p}}_\text{PZW} \right) \cdot \vect{\nabla} \times q_2 \vect{\hat{B}}_{q_2}  + \text{H.c.} \right]  \\ \label{eq:HIB_1} 
   &+ \frac{\hbar^2}{4m_\text{r}M^2c^2} \left[  q_2 \vect{\hat{B}}_{q_2} \cdot \left( Q \vect{\hat{B}}_Q + \frac{\Delta m}{m_\text{r}} q_1 \vect{\hat{B}}_{q_1} \right) + q_1 \vect{\hat{B}}_{q_1} \cdot \left( q_1 \vect{\hat{B}}_{q_1} + \frac{\Delta m}{m_\text{r}} q_2 \vect{\hat{B}}_{q_2} \right) \right].
\end{align}
\end{widetext}
Note that $\kappa=2 \kappa_{en}$. Due to the PZW transformation and the fact that we keep light-matter interactions also in the order $c^{-2}$, we find for all $c^{-2}$ terms from the single-coboson Hamiltonian light-field contributions that are listed in the first three lines of $\hat{h}_\text{IB}^{(1)}$.
Every term that appears with a subscript ``PZW'' contains PZW-transformed momenta.
The relativistic corrections $\vect{\hat{\mu}}_{i, \text{PZW}}^{(1)}$, $\vect{\hat{B}}_{\text{PZW}}^{(1)}$, $\vect{\hat{P}}_{\text{PZW}}^{(1)}$, and $\vect{\hat{p}}_{\text{PZW}}^{(1)}$ are the ones from Table~\ref{tab:relO}, only with PZW-transformed momenta, where $\vect{\hat{\mu}}_{i}^{(1)} =  \overline{c}_\text{F}^{(i)}  q_i \vect{\hat{s}}_i^{(1)}/m_i$.
Moreover, we collect all PZW-transformed terms of $\vect{\hat{p}}^{(1)}_\text{PZW}$ directly proportional to $\vect{\hat{B}}$ in $\vect{\hat{p}}_{\vect{B},\text{PZW}}^{(1)}$ that includes light-field-induced corrections. 
The last line in Eq.~\eqref{eq:HIB_1} represents the remainder of combining c.m. and relative parts of the kinetic correction and the relativistic correction $\vect{\hat{p}}^{(1)}$ due to the noncommutativity of c.m. and relative momenta.
We see that in $c^{-2}$ the influence of the magnetic field becomes cumbersome, but the general structure is expressed through $\vect{r} \times \vect{\hat{B}}$ terms, $\vect{\hat{B}}^2$ terms and $\vect{\nabla} \times \vect{\hat{B}}$-type terms in various combinations with other operators.

\subsection{Coboson scattering potential}
In the following, we determine the transformation of the scattering potential
\begin{align}
    \hat{V}_\text{scatt} = \sum_{i,j} \left( \hat{V}_{\text{C}}^{(ij)} + \hat{V}_{\text{LL}}^{(ij)} +\hat{V}_{\text{LS}}^{(ij)} + \hat{V}_{\text{SS}}^{(ij)} \right).
\end{align}
To our order, only relativistic corrections to the Coulomb potential 
\begin{align}
    \hat{V}^{(ij)}_{\text{C}} = \frac{q_i q_j}{8 \pi \varepsilon_0 \abs{\vect{\chi}_{ij}} }
\end{align}
between fermion $i$ of coboson 1 and fermion $j$ of coboson 2 at a relative distance $\vect{\chi}_{ij} = \vect{x}_{1,i} - \vect{x}_{2,j}$ have to be considered.
Here, $\vect{x}_{k,i} = \vect{R}_k - \text{sgn}(q_i) m_\text{r} \vect{r}_k / m_i$ in NR c.m. and relative coordinates.
The transformation resolves to
\begin{align}
    \hat{u}^{(\text{rel}) \, \dagger}_{12} \frac{1}{\abs{\vect{\chi}_{ij}}} \hat{u}^{(\text{rel}) }_{12} = \frac{1}{\abs{\vect{\chi}_{ij}}} + \hat{\delta}_{1,i}^{(ij)} + \hat{\delta}_{2,j}^{(ij)} + \mathcal{O} \left( c^{-4} \right)
\end{align}
with corrections $\hat{\delta}_{k,t}^{(ij)} = - \ii \big[ \hat{\lambda}^{(\text{rel})}_k, 1/\abs{\vect{\chi}_{ij}} \big]/\hbar$ that take the explicit form
\begin{widetext}
\begin{align} \notag
    \hat{\delta}_{k,t}^{(ij)} =  - \frac{(-1)^k}{4M^2c^2 \abs{\vect{\chi}_{ij}}^3 } \Bigg\{& \vect{r}_k \cdot \vect{\chi}_{ij} \frac{\Delta m q_e q_n}{4\pi \varepsilon_0 \abs{\vect{r}_k}} +  \bars{\ell}_\beta^{(ij)} \cdot \left( \bars{L}_k + \frac{\Delta m}{M} \bars{\ell}_k\right) + \left(\vect{\chi}_{ij} \cdot \bars{P}_k \right) \vect{r}_k \cdot \left( \bars{p}_k - \frac{q_t}{\abs{q_t}} \frac{m_\text{r}}{m_t} \bars{P}_k  \right)   \\ \label{eq:Scatt_correction}
    &+ \left( \vect{\chi}_{ij} \cdot \bars{p}_k \right) \vect{r}_k \cdot \left( \frac{\Delta m}{M} \bars{p}_k + \left(1-2 \frac{\Delta m}{M} \frac{q_t}{\abs{q_t}} \frac{m_\text{r}}{m_t} \right) \bars{P}_k  \right) + \frac{M}{m_\text{r}} \left( \bars{\ell}^{(ij)} + \frac{q_t}{\abs{q_t}} \frac{m_\text{r}}{m_t} \bars{L}^{(ij)} \right) \cdot \vect{\hat{s}}_k + \text{H.c.}  \Bigg\}.
\end{align}
\end{widetext}
These relativistic corrections can be identified with a scalar correction to the Coulomb potential, orbit-orbit-like and spin-orbit-like potentials.
For instance, the orbit-orbit scattering  potential $\hat{V}^{(ij)}_{\text{LL}}$ describes orbit-orbit coupling between fermions of two different cobosons due to their respective angular momentum $\vect{\chi}_{ij} \times \vect{\hat{p}}_i$.
Similar terms appear also in $\hat{\delta}_{k,t}^{(ij)}$ that depend on the angular momentum $\bars{\ell}_k^{(ij)} = \vect{\chi}_{ij} \times \bars{p}_k$, which is the outer product of the distance between two fermions of different cobosons and the relative momentum of coboson $k$, where the bar indicates again minimally coupled momenta.
These angular momenta between cobosons couple in this case to the \textit{internal} total $\bars{L}_k = \vect{r}_k \times \bars{P}_k$ and relative $\bars{\ell}_k = \vect{r}_k \times \bars{p}_k$ angular momentum of coboson $k$.
Next, we find a spin-orbit coupling between angular momenta $\bars{L}^{(ij)}$ and $\bars{\ell}^{(ij)}$ to the relative spin $\vect{\hat{s}}_k$.

Applying the PZW transformation as well changes all momenta in the scattering potentials to the PZW-transformed ones, also in the correction term given in Eq.~\eqref{eq:Scatt_correction}.
Consequently, the transformed scattering potential reads
\begin{align}
\begin{split}
    \hat{\mathcal{V}}_\text{scatt} = \sum_{i,j}& \frac{q_i q_j}{8 \pi \varepsilon_0} \left( \frac{1}{\abs{\vect{\chi}_{ij}}} + \hat{\delta}_{1,i}^{(ij,\text{PZW})} + \hat{\delta}_{2,j}^{(ij,\text{PZW})} \right) \\
    &+ \hat{\mathcal{V}}^{(ij)}_{\text{LL}} + \hat{\mathcal{V}}^{(ij)}_{\text{LS}} + \hat{\mathcal{V}}^{(ij)}_{\text{SS}}.
\end{split}
\end{align}
The potentials $\hat{\mathcal{V}}^{(ij)}_{v}$ are the ones from Fig.~\ref{fig:potential} where we replace $\vect{r}$ by $\vect{\chi}_{ij}$ as well as $\vect{\hat{p}}_i \to  m_i \vect{\hat{P}}_{1, \text{PZW}}/M - \text{sgn} (q_i) \vect{\hat{p}}_{1, \text{PZW}}$ and $\vect{\hat{p}}^\prime_j \to m_j \vect{\hat{P}}_{2, \text{PZW}}/M - \text{sgn} (q_j) \vect{\hat{p}}_{2, \text{PZW}}$ for the momenta.

Scattering between two cobosons reduces to interactions between fermion $i$ and $j$ of two different cobosons via the Coulomb potential in lowest order together with magnetic moments associated with both spin and orbital motion.
By that, all magnetic moments couple, \ie{} spin to spin, spin to orbit, and orbit to orbit where the latter has an additional retardation correction.
In addition, corrections to NR c.m. and relative coordinates arise from the Coulomb term and modify the Coulomb potential, the LL coupling, as well as the LS coupling.

\begin{table*}
\begin{ruledtabular}
    \centering
        \caption{Clebsch-Gordan coefficients $ \alpha_{j,S,m_S}$ for the angular-momentum eigenbasis of total angular momentum $\vect{\hat{J}} = \vect{\hat{\ell}} + \vect{\hat{S}}$.
        Here, $j$ is the quantum number of the total angular momentum, $S$ the quantum number of the total spin, and $m_S$ its magnetic quantum number.
        Moreover, $\ell$ is the quantum number of orbital angular momentum and $m_j$ the magnetic quantum number of the total angular momentum. 
        }
    \begin{tabular}{cccc}
    \diagbox[]{$(j,S)$}{$m_S$}   & 1 & 0 & -1  \\ 
    \colrule\addlinespace
       $(\ell+1,1)$  & $\sqrt{\frac{(\ell+m_j)(\ell+m_j+1)}{2 \left(\ell+1 \right) \left( 2 \ell +1 \right)}}$  & $\sqrt{\frac{(\ell - m_j +1)(\ell+m_j+1)}{\left( \ell+1 \right) \left( 2\ell +1 \right)}}$ & $\sqrt{\frac{(\ell-m_j)(\ell-m_j+1)}{2 \left(\ell+1 \right) \left( 2\ell +1 \right)}}$ \\ \addlinespace
       $(\ell,1)$  & $\sqrt{\frac{(\ell+m_j)(\ell-m_j+1)}{2 \ell \left( \ell +1 \right)}}$  & $-\frac{m_j}{\sqrt{\ell \left( \ell +1 \right)}}$ & $-\sqrt{\frac{(\ell-m_j)(\ell+m_j+1)}{2 \ell \left( \ell +1 \right)}}$  \\ \addlinespace
       $(\ell-1,1)$  & $\sqrt{\frac{(\ell-m_j)(\ell-m_j+1)}{2 \ell \left(2 \ell +1 \right)}}$  & $- \sqrt{\frac{(\ell-m_j)(\ell+m_j)}{ \ell \left(2 \ell +1 \right)}}$ & $\sqrt{\frac{(\ell+m_j)(\ell+m_j+1)}{2 \ell \left(2 \ell +1 \right)}}$  \\ \addlinespace
        $(\ell,0)$ & 0   & 1 & 0 
    \end{tabular}
    \label{tab:Clebsch}
\end{ruledtabular}
\end{table*} 
\subsection{EM Hamiltonian}
Finally, the transformation of the EM Hamiltonian requires the direct computation of the second-quantized unitary $\hat{U}$ as it contains no cobosonic field operators.
Hence, we determine
\begin{align}
    \hat{U}^\dagger_\text{PZW} \hat{U}^\dagger_\text{rel} \hat{H}_\text{EM} \hat{U}_{\text{rel}} \hat{U}_{\text{PZW}}.
\end{align}
Note that momentum operators contained in field-theoretical unitaries do not act on the variables of integration in $\hat{H}_\text{EM}$ such that the transformation is solely determined through EM fields.
Moreover, $\hat{U}_\text{rel}$ and $\hat{U}_\text{PZW}$ contain only the vector potential $\hat{\vect{A}}$ that commutes with itself and with the magnetic field $\vect{\hat{B}}$ in Coulomb gauge, so only the electric field gives rise to additional terms.
The relevant commutator between the vector potential and the electric field
\begin{align} \label{eq:commEA}
    \left[ \hat{A}^{(\ell)} (\vect{x}) , \hat{E}^{(m)} (\vect{y}) \right] = - \frac{\ii \hbar}{\varepsilon_0} \delta^{\ell m, \perp} ( \vect{x} - \vect{y} ) 
\end{align}
is defined through the transverse delta function~\cite{Belifante1946}.
The transformation generating the relativistic corrections gives rise to the form
\begin{align} \label{eq:relEM}
   \hat{U}^\dagger_\text{rel} \hat{H}_{\text{EM}} \hat{U}_\text{rel} =  \hat{H}_\text{EM} + \int \limits_{C_1} \dd[6]{\mathcal{R}_1} \hat{\varphi}^\dagger_1 \, \hat{\overline{ \!h}}_\text{IE}^{\, (1)} \hat{\varphi}_1.
\end{align}
The relativistic corrections can be written as $\hat{\overline{ \! h}}_{\text{IE}}^{\,(1)} = \frac{1}{2} \sum_i \left( \vect{\hat{E}}^\perp_i \cdot \bars{d}_i + \bars{d}_i \cdot \vect{\hat{E}}^\perp_i \right)$ and appears with $\bars{d}_i$ defined in Eq.~\eqref{eq:Rel_Ecorr}, but the momenta are still the minimally coupled ones.
Applying the PZW transformation to the first term in Eq.~\eqref{eq:relEM} results in
\begin{align}
\begin{split}
   \hat{U}_\text{PZW}^\dagger \hat{H}_\text{EM} \hat{U}_\text{PZW} =& \hat{H}_\text{EM} + \int \limits_{C_1} \dd[6]{\mathcal{R}_1} \hat{\varphi}^\dagger_1 \hat{h}_\text{IE}^{(0)} \hat{\varphi}_1 \\
    &+ \int \limits_{C_1} \dd[6]{\mathcal{R}_1} \int \limits_{C_2} \dd[6]{\mathcal{R}_2} \hat{\varphi}^\dagger_1 \hat{\varphi}^\dagger_2 \hat{\mathcal{V}}_\text{self} \hat{\varphi}_2 \hat{\varphi}_1.
\end{split}
\end{align}
Here, $\hat{h}_\text{IE}^{(0)}$ contains the electric multipole moments and the self-energy from Table~\ref{tab:Ham_I} and $\hat{\mathcal{V}}_\text{self}$ is the scattering self-energy known from Table~\ref{tab:Scattering}.

The PZW transformation of the second term in Eq.~\eqref{eq:relEM} reduces now, due to the coboson field operators, to the first-quantized PZW transformation of $\hat{\overline{ \! h}}_{\text{IE}}^{\,(1)}$ only, such that the electric field $\vect{\hat{E}}_i$ is again not affected as $\vect{\mathcal{P}} ( \vect{x}_i) = 0$ vanishes.
The momenta are exchanged by the PZW-transformed ones, \ie{} $\hat{\overline{ \! h}}_{\text{IE}}^{\,(1)} \to \hat{h}_\text{IE}^{(1)}$.
This Hamiltonian is the electric part of $c^{-2}$ corrections to light-matter interaction from Eq.~\eqref{eq:Rel_Ecorr}, where we replace canonical momenta with the PZW-transformed ones.

By taking into account the contributions from the EM Hamiltonian, the single-coboson Hamiltonian is modified in the light-matter interaction part to $\hat{h}^{(k)}_\text{I} = \hat{h}^{(k)}_\text{IB} + \hat{h}^{(k)}_\text{IE}$ with $k=0,1$.
The scattering potential gets an additional self-energy $\hat{\mathcal{V}}_\text{self}$.

\section{First-order energy shift} \label{App:E}

To determine first-order energy shifts, the actual form of the wave function of relative modes of hydrogenlike cobosons $\psi_\beta$ is required and is given in terms of quantum numbers $n,j,m_j,\ell, S$.
In particular, the wave function
\begin{align} \label{eq:cb_mot}
    \begin{split}
        \psi_{\beta} =&  \alpha_{j,S,1} \psi_{n,\ell,m_j-1} \chi_{S,1} + \alpha_{j,S,0} \psi_{n,\ell,m_j} \chi_{S,0} \\
        &+ \alpha_{j,S,-1} \psi_{n,\ell,m_j+1} \chi_{S,-1}
    \end{split}
\end{align}
consists of $\psi_{n,\ell,m}$, the standard spatial part of the solution to hydrogenlike Schrödinger equation.
The spin wave function $\chi_{S,m_S}$ is the eigenbasis of operators $\vect{\hat{S}}^2$ and $\hat{S}_z$ of the total spin $\vect{\hat{S}} = \vect{\hat{s}}_e + \vect{\hat{s}}_n$.
By that, the total spin of a coboson formed by two spin-1/2 fermions can either be $S=0$ or $S=1$, while its projection onto the $z$ axis can take magnetic spin numbers $m_S=0$ or $m_S = -1,0,1$, respectively.
In the superposition of Eq.~\eqref{eq:cb_mot} together with Clebsch-Gordan coefficients~\cite{Krainov1997} (detailed in Table~\ref{tab:Clebsch}), we find the eigenbasis of the operators $\vect{\hat{J}}^2, \hat{J}_z, \vect{\hat{\ell}^2}$, and $ \vect{\hat{S}^2}$, where $\vect{\hat{J}} = \vect{\hat{\ell}} + \vect{\hat{S}}$ is the total angular momentum.
With this explicit wave function in angular momentum eigenbasis we determine the first-order energy shift $E^{(1)}_\beta = \int \dd[3]r \psi_\beta^* \hat{h}_\text{rel}^{(1)} \psi_\beta$ and arrive at
\begin{widetext}
\begin{align}
    \begin{split}
        E_\beta^{(1)}=  \frac{m_{\text{r}}^2c^2(Z\alpha)^4}{M} \Bigg \{& \frac{m_e^3+m_n^3}{8 m_\text{r}M^2} \left( 3- \frac{8n}{2\ell+1}\right) \frac{1}{n^4} +\left( 1 - \frac{3n}{2 \ell+1} \right) \frac{1}{n^4}  + \left( \alpha_\text{D} - \frac{3}{4} \alpha_\mathrm{ss} + \alpha_\mathrm{ss} \delta_{S,1} \right) \frac{\delta_{\ell,0}}{n^3} + \frac{(\delta_{\ell,0} -1) \delta_{S,1}}{\ell (\ell+1) (2 \ell +1)} C_{j,\ell}\Bigg \},
    \end{split}
\end{align}  
that has been obtained before~\cite{Pineda1998b,Peset2015a}.
The individual terms correspond to the kinetic correction (first), orbit-orbit coupling (second), Darwin and contact interaction (third) where $\vect{\hat{s}}_e \cdot \vect{\hat{s}}_n = ( \vect{\hat{S}}^2 - \vect{\hat{s}}_e^2 - \vect{\hat{s}}_n^2)/2$ was exploited such that $\vect{\hat{s}}_i^2$ takes also a Darwin-like spin-independent form.
The last term combines both spin-orbit terms and the magnetic dipole-dipole potential in $\hat{S}_{ne}=- \hat{\vect{s}}_n \cdot \hat{\vect{s}}_e^\prime + 3 ( \vect{r} \cdot \hat{\vect{s}}_n ) (\vect{r} \cdot \hat{\vect{s}}_e^\prime)/\abs{\vect{r}}^2$ with 
\begin{align}
\begin{split}
    C_{j,\ell} = \left\{ \begin{array}{ll}
        \frac{\ell}{2\ell +3} \left[ 2(2 \ell +3) \left( \alpha_{\ell S} + \frac{\Delta m}{2M}  \alpha_{\ell s} \right) - \overline{c}_\text{F}^{(e)} \overline{c}_\text{F}^{(n)}  \right], & \text{for} \,  j=\ell+1 \\ \addlinespace
        -2 \left(  \alpha_{\ell S} + \frac{\Delta m}{2M}  \alpha_{\ell s} \right) + \overline{c}_\text{F}^{(e)} \overline{c}_\text{F}^{(n)}, & \text{for} \, j=\ell \\ \addlinespace
        - \frac{\ell +1}{2 \ell -1} \left[ 2 (2 \ell -1) \left( \alpha_{\ell S} +  \frac{\Delta m}{2M}  \alpha_{\ell s} \right) + \overline{c}_\text{F}^{(e)} \overline{c}_\text{F}^{(n)} \right], & \text{for} \, j=\ell-1
    \end{array} \right.
\end{split}
\end{align}
and the low-energy Wilson coefficients $\alpha_v$ are given in Table~\ref{tab:Wilson}.
Hence, the first-order energy shift depends on quantum numbers $n,j,\ell$, and $S$, but not on $m_j$.
\end{widetext}
\end{appendix}

\bibliography{Literatur}
\end{document}